\documentclass[amsmath,amssymb,amsfonts,twocolumn,english,showpacs,superscriptaddress,prb]{revtex4-1}
\pdfoutput=1

\usepackage[T1]{fontenc}
\usepackage[latin9]{inputenc}
\usepackage{times}
\usepackage{graphicx}
\usepackage{color}
\usepackage{mathrsfs}
\usepackage{float}
\usepackage{indentfirst}
\usepackage{babel}
\usepackage{wasysym}
\usepackage{tikz}
\usepackage{amsthm}
\usetikzlibrary{calc}
\usepackage{verbatim}
\usepackage{tabularx}
\usepackage{tabulary}
\usepackage{mwe}
\usepackage[normalem]{ulem}
\newcommand{\ket} [1] {\vert #1 \rangle}
\newcommand{\bra} [1] {\langle #1 \vert}

\usepackage{booktabs}
\usepackage{array, multirow}
\usepackage{subfigure}

\usepackage[colorlinks,bookmarks=false,citecolor=blue,linkcolor=blue,urlcolor=blue]{hyperref}
\setcitestyle{numbers,square}
\usepackage{array}

\graphicspath{{.}}

\begin{document}

\title{Edge states of 2D time-reversal-invariant topological superconductors with strong interactions and disorder: A view from the lattice}

\author{Jun Ho Son}
\affiliation{Department of Physics, Cornell University, Ithaca, New York 14853, USA}
\author{Jason Alicea}
\affiliation{Department of Physics and Institute for Quantum Information and Matter, California Institute of Technology, Pasadena, CA 91125, USA}
\author{Olexei I.~Motrunich}
\affiliation{Department of Physics and Institute for Quantum Information and Matter,
California Institute of Technology, Pasadena, CA 91125, USA}

\begin{abstract}
Two-dimensional time-reversal-invariant topological superconductors host helical Majorana fermions at their boundary.  We study the fate of these edge states under the combined influence of strong interactions and disorder, using the effective 1D lattice model for the edge introduced by Jones and Metlitski [Phys.~Rev.~B {\bf 104}, 245130 (2021)].  We specifically develop a strong-disorder renormalization group analysis of the lattice model and identify a regime in which time-reversal is broken spontaneously, creating random magnetic domains; Majorana fermions localize to domain walls and form an infinite-randomness fixed point, identical to that appearing in the random transverse-field Ising model.  While this infinite-randomness fixed point describes a  fine-tuned critical point in a purely 1D system, in our edge context there is no obvious time-reversal-preserving perturbation that destabilizes the fixed point. Our analysis thus suggests that the infinite-randomness fixed point emerges as a stable phase on the edge of 2D topological superconductors when strong disorder and interactions are present.
\end{abstract}

\maketitle

\section{Introduction} 
  
 Among many intriguing features of topological insulators and superconductors, their nontrivial boundary states yield the most direct experimental signatures. Whereas the bulk of these phases is either fully gapped (in the clean limit) or localized (when randomness is present and the notion of spectral gap is less meaningful), their interfaces with the trivial vacuum can host states that are gapless and delocalized. In the canonical examples of 2D and 3D topological insulators protected by time-reversal symmetry \cite{KaneMele2005_1,KaneMele2005_2,Bernevig2006,Fu2007}, band structure calculations famously reveal that their boundaries feature 1D and 2D massless Dirac fermions, respectively. Gapless, delocalized boundary states are not always guaranteed, however.   Notably, transcending band theory by adding disorder and/or interactions opens up possibilities for other interesting types of boundary states to emerge. Recent works in this direction include the construction of gapped, symmetry-preserving topologically ordered surface states of strongly interacting 3D topological insulators and superconductors \cite{surface1,surface2,surface3,surface4,surface5,surface6,surface7,surface8,surface9,surface10,surface11,surface12,surface13} and fully localized boundary states of 2D quantum spin Hall insulators \cite{Altshuler2013,Chou2018,Kimchi2020}. 
 
 The theoretical concept that unites such distinct boundary states is the quantum anomaly: $(d-1)$-dimensional boundary states of $d$-dimensional topological insulators and superconductors possess anomalies that preclude their appearance in purely $(d-1)$-dimensional systems with the same set of symmetries acting in the same manner.  The bulk of such topological phases precisely cancels these anomalies, thereby enabling the existence of anomalous boundary states \cite{Callan1985}. Reversing this logic, one can in principle envision a menagerie of possible boundary states for each type of topological insulator and superconductor---all sharing the same quantum anomaly cancelled by the bulk. This viewpoint is especially relevant when the boundary evades a  band-theoretic description.
 
In this paper, we explore edge states of 2D time-reversal-invariant topological superconductors \cite{Qi2009} subjected to  both strong interaction \emph{and} disorder. At the level of a non-interacting Bogoliubov-de Gennes treatment, one can view such a phase as a $p+ip$ superconductor for spin-up electrons composed with a $p-ip$ superconductor for spin-down electrons.  The clean, non-interacting edge accordingly hosts helical Majorana fermions that are perturbatively stable against time-reversal-symmetric interaction and disorder \cite{Chou2021}. This stability can be contrasted with the critical point in the clean quantum Ising chain where weak disorder is a relevant perturbation, even when ferromagnetic interactions and transverse fields are taken from the same distribution to maintain criticality, and the system flows to an infinite-randomness fixed point~\cite{Ma1979, Fishersdrg_1, Fishersdrg_2, Fishersdrg_3}. Stability of the clean gapless edge states of the 2D time-reversal invariant topological superconductors to weak disorder descends from the particular time reversal symmetry action in this case 
\footnote{We can loosely model the clean gapless edge of 2D time-reversal invariant topological superconductors by a critical Ising chain, with the time reversal invariance corresponding to the self-duality condition, which is required to hold for \emph{each disorder realization}. We can in turn model this condition by requiring that the random transverse field at site $j$ equals, say, the ferromagnetic interaction between spins at $j$ and $j+1$. We can argue that such a quantum Ising chain with perfectly correlated random local fields and interactions does not flow to strong disorder, unlike the Ising chain with uncorrelated random local fields and interaction with only ``statistical symmetry'' (self-duality) between the two. One way to see this is by mapping to extremal properties of an appropriate random walk~\cite{Eggarter1978, Igloi1998random, Igloi1998anomalous, Motrunich2001_2, Roberts2021}, which in the perfectly correlated disorder case essentially does not wander at all. Instances explored in this paper wherein disorder yields more dramatic consequences would likely need to start in a very different regime in the quantum Ising analogy, loosely requiring spontaneous breaking of self-duality but also other conditions, and it is not clear a priori if one can make such analogies precise.}. On the other hand, strong interaction and disorder of interest here may nevertheless drastically alter the edge physics.

To attack the problem, we exploit exactly solvable, commuting-projector Hamiltonians for 2D time-reversal invariant topological superconductors \cite{Wang2017}.  This starting point not only incorporates strong interactions at the outset, but also allows one to `peel off' a purely 1D lattice model \cite{Jones2019} that emulates the helical Majorana edge with an appropriate implementation of time-reversal symmetry; see also Refs.~\onlinecite{Tarantino2016,Ware2016,Bhardwaj2017}).  The purely 1D lattice model---which inherits strong interactions from the bulk---provides an efficient way of incorporating disorder via a strong-disorder renormalization group (RG) analysis \cite{Ma1979,Fishersdrg_3,Fishersdrg_1,Fishersdrg_2} that we adapt to this setting.  Our treatment thereby probes a \textit{completely different limit} from the usual continuum field theory considerations that generally start from clean fixed points.

\begin{figure}
     \centering
     \includegraphics[width=\linewidth]{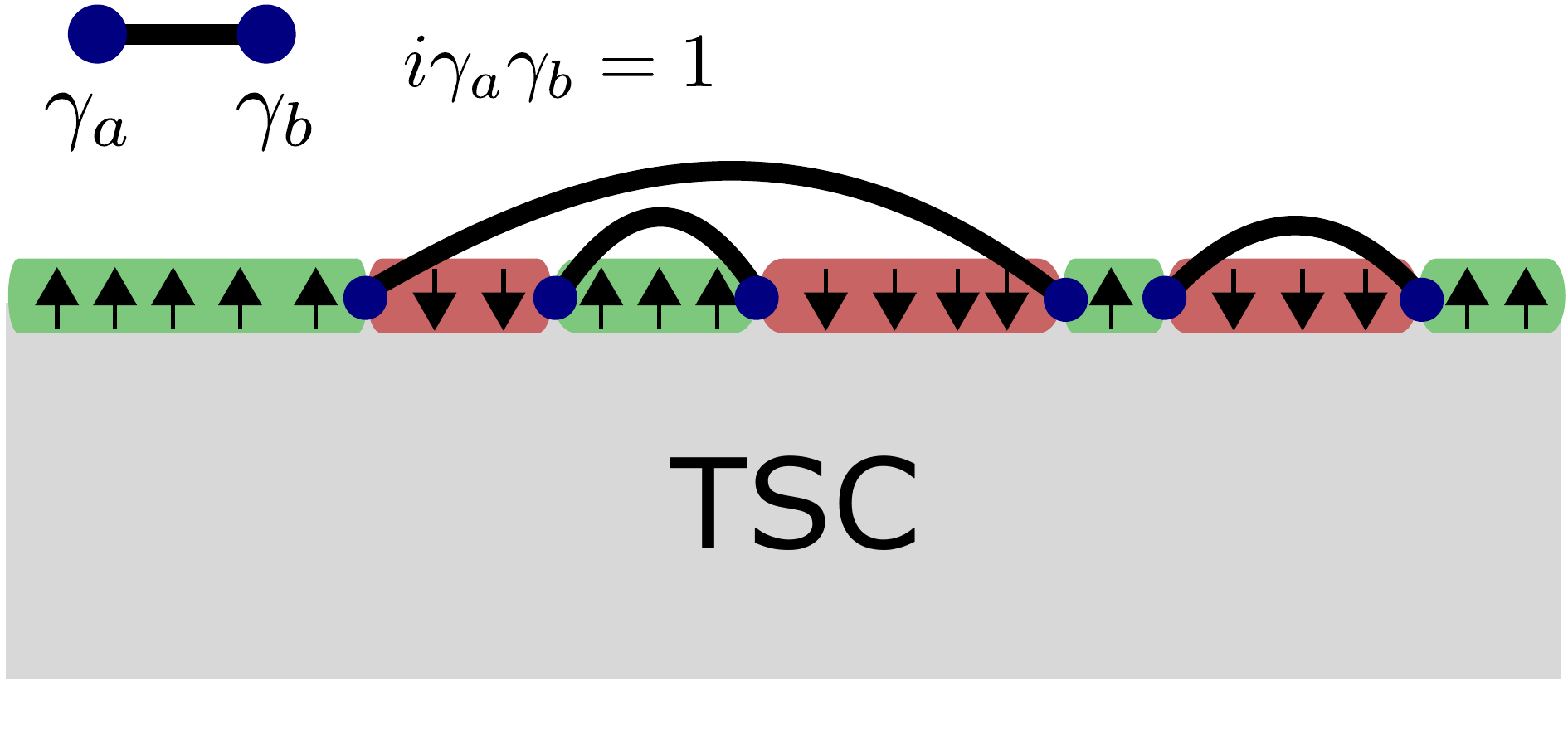}
     \caption{``Cartoon picture'' of edge states of 2D time-reversal invariant topological superconductors, assumed to be in the parameter range where the time-reversal is spontaneously broken on the edge. There is a 1D array of ferromagnetic domains with random lengths, two neighboring domains having opposite magnetizations; domains with up-magnetizations are colored green, while domains with down-magnetizations are colored red. Majorana fermions that live on the domain walls are illustrated as blue dots. These Majorana fermions form an infinite-randomness fixed point. To illustrate the infinite-randomness fixed point, we first define connecting two Majorana fermions $\gamma_{a}$ and $\gamma_{b}$ with a solid black line as representing a quantum state which is an eigenstate of $i\gamma_{a} \gamma_{b}$ with fixed eigenvalue $+1$ or $-1$ depending on specific situation. Then, we illustrate the infinite-randomness fixed point by connecting paired Majorana fermions with lines, lines not crossing with each other (the specific pairing is determined by energetics resulting from the original random couplings). }
     \label{fig:cartoon}
\end{figure}

Our strong-disorder RG analysis reveals a \textit{stable} edge phase enabled by a combination of strong disorder and interaction. This edge phase spontaneously breaks time-reversal symmetry; spin degrees of freedom form local ferromagnetic domains of random lengths, with neighboring domains exhibiting opposite magnetizations. Each domain wall hosts an unpaired Majorana fermion, which is the topological-superconductor analogue of the famous Jackiw-Rebbi $\pm \frac{e}{2}$ mode \cite{Jackiw1976} that appears on analogous ferromagnetic domain walls in a quantum spin Hall edge \cite{Qi2008}.  The collection of domain-wall Majorana fermions generically self-tunes to criticality and forms an infinite-randomness fixed point, identical to that found at the critical point of the 1D random transverse-field Ising model \cite{Fishersdrg_2,Fishersdrg_3} and at topology-changing critical points in 1D dirty superconductors without spin-rotation invariance~\cite{Motrunich2001}; see Fig.~\ref{fig:cartoon} for a cartoon picture. 

 The same edge physics was recently proposed by Chou and Nandkishore \cite{Chou2021} to arise when there is a statistical time-reversal symmetry, or equivalently, absence of long-range ferromagnetic order.  From the continuum-field-theory starting point utilized in that work, however, the physical mechanism and criteria for the emergence of statistical time-reversal symmetry remains unclear. The present approach sheds new light on this issue: Upon including a small amount of antiferromagnetic couplings in the effective edge Hamiltonian, our strong disorder RG demonstrates that local ferromagnetic domains form at intermediate length/energy scale; as the RG progresses, effect of antiferromagnetic interactions fail to become fully ``screened'' and dominate the ultimate IR physics.  In particular,  antiferromagnetic interactions that persist all the way to the IR preclude long-range order in conventional sense \footnote{Denote spin up/down at site $i$ as $\sigma_{i}^{z} = \pm 1$. The conventional disorder averaged correlator $\overline{\left\langle \sigma_{i}^{z} \sigma_{j}^{z}\right\rangle}$ vanishes as $i-j \rightarrow \infty$. However, a correlator $\overline{\langle \sigma_i \sigma_j \rangle  \langle \sigma_i \sigma_j \rangle }$, associated with a time-reversal odd Edward-Anderson order parameter, is long-ranged and may be used to diagnose spontaneous time-reversal symmetry breaking.}. Our analysis therefore suggests that this phase is a stable, generic edge state supported by time-reversal symmetric Hamiltonians for 2D topological superconductors with both strong disorder and interaction. We anticipate that the results developed here may sharpen the understanding of when similar novel boundary phases can emerge under the influence of disorder and interactions in topological phases more broadly.  
 
The paper is organized as follows: In Sec.~\ref{sec:Review}, we briefly review the purely 1D lattice model that mimics the edge state of 2D time-reversal invariant topological superconductors. In Sec.~\ref{sec:splim}, we take a closer look at the special limit in which spin degrees of freedom are strongly pinned by nearest-neighbor Ising interactions with random signs. Although this limit is arguably somewhat special, it makes many key properties of the edge states we are studying manifest.  In Sec.~\ref{sec:sdrg}, we carry out a full-fledged strong-disorder renormalization group analysis.  There we provide evidence that in a certain parameter regime, the RG flow is consistent with the physical scenario described in the previous paragraph, suggesting that the proposed edge state is a stable phase in this parameter regime. Concluding remarks appear in Sec.~\ref{sec:conc}.

\section{Review of the 1D Model}
\label{sec:Review}

 Here, we provide a brief review of the 1D model that mimics edge physics of 2D time-reversal-invariant topological superconductors, introduced by Jones and Metlitski \cite{Jones2019} and generalized to edges of 2D quantum spin Hall systems in Refs.~\onlinecite{Our2019,Metlitski2019}.
 
\subsection{Setup and Symmetries}

\begin{figure}
     \centering
     \includegraphics[width=\linewidth]{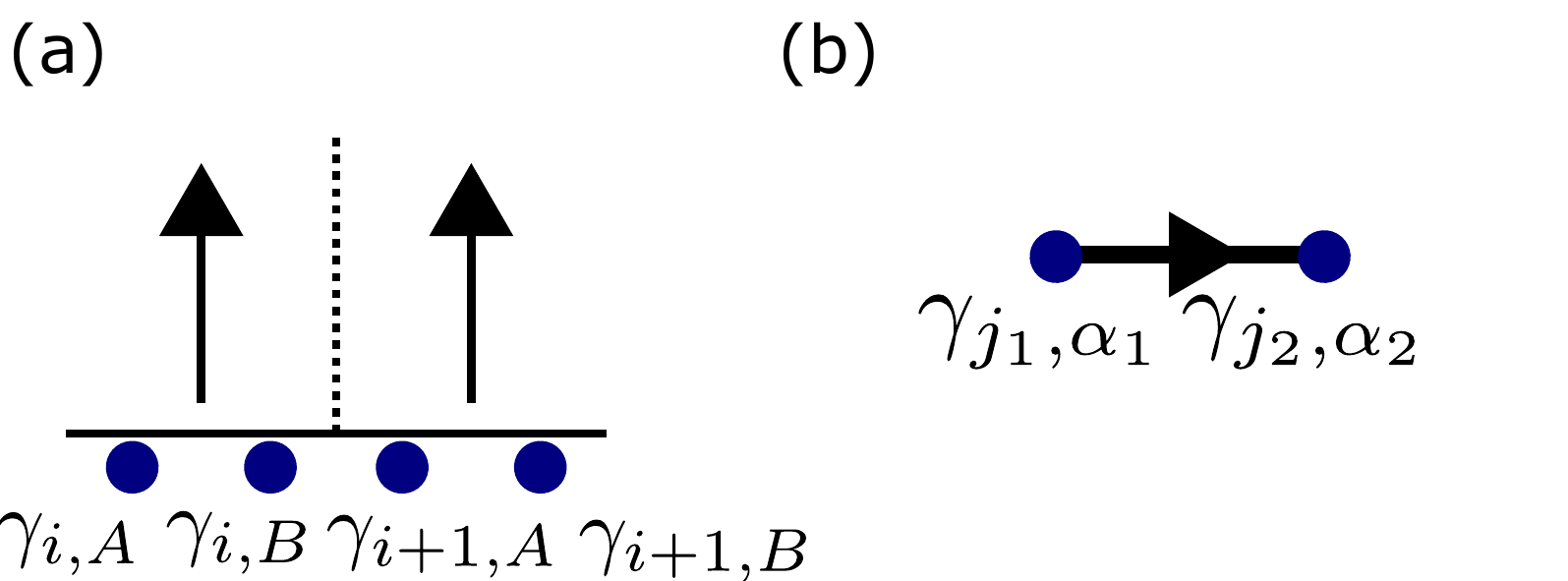}
     \caption{(a) Illustration of degrees of freedom in the 1D model we consider with Majorana fermion labelings. Arrows in the figure represent bosonic Ising spins, while blue dots represent Majorana fermions. (b) Illustration of the state with $ i\gamma_{j_{1},\alpha_{1}}\gamma_{j_{2},\alpha_{2}} = +1$, or equivalently, $P_{j_{1},\alpha_{1}}^{j_{2},\alpha_{2}}=1$} 
     \label{fig:review1}
\end{figure}

 In the 1D model of interest, each site $i$ hosts two Majorana fermions $\gamma_{i,A}$ and $\gamma_{i,B}$ and one spin-$\frac{1}{2}$ bosonic degree of freedom $\sigma^z_i$ that we will call an Ising spin. We will interchangeably denote the $\sigma^z_i$ eigenvalues by $+1,-1$ and $\uparrow,\downarrow$. Figure~\ref{fig:review1}(a) illustrates the setup, with arrows representing the bosonic spins and blue dots denoting Majorana fermions.
 
The Hilbert space for the bosonic and fermionic degrees of freedom is constrained in the 1D model. Before writing down this constraint explicitly, we introduce some mathematical and graphical notations that we will use throughout this paper. First, we define the projector 
\begin{equation}
 P_{j_{1},\alpha_{1}}^{j_{2},\alpha_{2}} = \frac{1+i\gamma_{j_{1},\alpha_{1}}\gamma_{j_{2},\alpha_{2}}}{2},
\end{equation}
where $\alpha_{1}, \, \alpha_{2} \in \{
A, B \} $; $P_{j_{1},\alpha_{1}}^{j_{2},\alpha_{2}}$ acts as the identity on states for which $ i\gamma_{j_{1},\alpha_{1}}\gamma_{j_{2},\alpha_{2}}$ has eigenvalue $+1$ but annihilates states with $-1$ eigenvalue. We often denote two Majoranas $\gamma_{j_{1},\alpha_{1}}$ and $\gamma_{j_{2},\alpha_{2}}$ as \textit{paired} on states with $P_{j_{1},\alpha_{1}}^{j_{2},\alpha_{2}}=1$. Graphically, we represent a pairing 
$ i\gamma_{j_{1},\alpha_{1}}\gamma_{j_{2},\alpha_{2}} = +1$ as a line connecting two blue dots representing the paired Majorana fermions, with the arrowhead directed from $\gamma_{j_{1},\alpha_{1}}$ to $\gamma_{j_{2},\alpha_{2}}$. See Fig.~\ref{fig:review1}(b) for an illustration.
 
The constraint term $R_{i}$ for each site $i$ is given as
\begin{equation}
\label{eq:ciconst}
\begin{split}
R_{i} = P_{i,A}^{i,B} \left( \ket{\uparrow_{i} \uparrow_{i+1}} \bra{\uparrow_{i} \uparrow_{i+1}} + \ket{\uparrow_{i} \downarrow_{i+1}} \bra{\uparrow_{i} \downarrow_{i+1}} \right) \\
+\ket{\downarrow_{i} \uparrow_{i+1}} \bra{\downarrow_{i} \uparrow_{i+1}} + P_{i,B}^{i+1,A} \ket{\downarrow_{i} \downarrow_{i+1}} \bra{\downarrow_{i} \downarrow_{i+1}}.
\end{split}
\end{equation}
Here $\ket{\uparrow_{i} \uparrow_{i+1}} \bra{\uparrow_{i} \uparrow_{i+1}}$ projects onto the state with Ising spins $\sigma_{i}^{z} = \sigma_{i+1}^{z} = +1 = \uparrow$; similarly, pairs of kets and bras involving different combinations of arrows in the above equation represent projections onto specific Ising configurations. Note that $R_i$'s commute with each other, and the constrained Hilbert space is defined by $R_i = 1$ for all $i$.
 
\begin{figure}
     \centering
     \includegraphics[width=\linewidth]{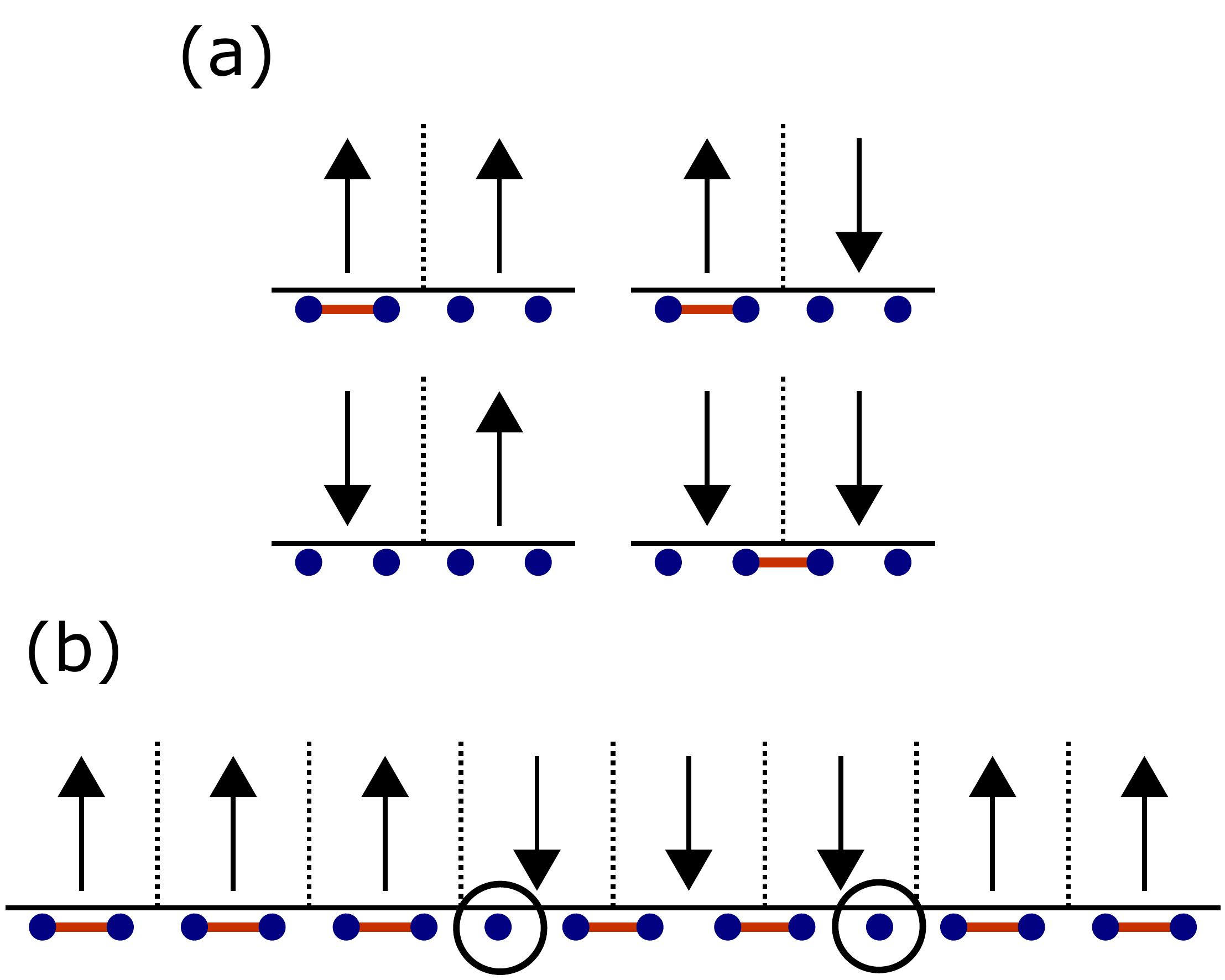}
     \caption{(a) Graphical illustration of $R_{i}$ of Eq.~\eqref{eq:ciconst}. (b) An example of state with the constraint $R_{i}=1$ satisfied for all $i$'s. The domain-wall Majorana fermions that remain unpaired by constraints are circled. In both (a) and (b), we suppressed arrowheads because they are all directed left to right. In the future figures, all red lines between Majorana fermions denote pairings enforced by the constraint $R_{i}=1$, and the arrowheads for them will be omitted. } 
     \label{fig:review2}
 \end{figure}

 To give some intuition about this constraint, we first examine Fig.~\ref{fig:review2}(a) where we illustrate the constraint $R_{i}=1$ for a single $i$. In Fig.~\ref{fig:review2} and all other figures in the paper, we will use red lines exclusively for Majorana fermion pairings enforced by the constraint $R_{i}=1$; the arrowheads on the red lines always point from the left to the right and will be suppressed. The two key features are:
\begin{itemize}
    \item If $\sigma_{i}^{z} =\,  \uparrow$, then $\gamma_{i,A}$ and $\gamma_{i,B}$, the two Majorana fermions at site $i$, are paired.
    \item If $\sigma_{i}^{z} = \sigma_{i+1}^{z} =\, \downarrow$, then $\gamma_{i+1,B}$ and $\gamma_{i,A}$, the two Majorana fermions that neighbor each other but belong to two different sites, are paired.
\end{itemize}
 Having the above in mind, it is straightforward to construct states which satisfy the constraint $R_{i}=1$ for all $i$'s. We illustrated one example of such states in Fig.~\ref{fig:review2}(b). In these states, along spin-up domains, Majorana fermions are paired within their own unit cells, but along spin-down domains, two nearest-neighbor Majorana fermions in two different unit cells are paired. The pairing patterns on up-domains and down-domains resemble cartoon pictures of trivial and topological states of Kitaev chains, respectively.
 
 The constraints we discuss leave one Majorana fermion on each Ising spin domain wall unpaired ($\gamma_{i,B}$ when $\sigma_{i}^{z}=\,\downarrow$ and $\sigma_{i+1}^{z}=\,\uparrow$, $\gamma_{i+1,A }$ when $\sigma_{i}^{z}=\,\uparrow$ and $\sigma_{i+1}^{z}=\,\downarrow$). These unpaired domain wall Majorana fermions are fermionic low-energy degrees of freedom that ``survive'' after restricting the Hilbert space with the constraints. We will sometimes refer to these Majorana fermions as ``free''. Hence, in our restricted Hilbert space, each fixed configuration of Ising spins has a $2^{N_\text{dw}/2}$ Hilbert space dimension, where $N_\text{dw}$ is the number of domain walls in the Ising spin configuration. Half of the domain walls are $\uparrow-\downarrow$ type, and the other half are $\downarrow-\uparrow$ type (note that the 1D chain has the topology of a circle, being a boundary of a 2D region).
 
 These free Majorana fermions on the Ising spin domain walls have a close connection to the following physical setup: Given a topological superconductor, gap its edge out by introducing two different ferromagnets that have opposite magnetizations and hence are time-reversal parteners to each other. The domain wall then binds a single Majorana zero mode. The constraints $R_{i}=1$ precisely capture this phenomenology on the lattice model. The 1D model being reviewed here promotes these ``magnetizations'' to dynamical degrees of freedom, i.e., Ising spins, and as we will soon see, can capture other edge phases including the gapless helical Majorana edge state.
 
Finally, we present the implementation of time-reversal symmetry in this model. Time-reversal symmetry here acts inherently non-locally -- any incarnation of local time-reversal symmetry in this model would be a direct contradiction to the fact that our 1D model mimics the edge of a 2D topological superconductor. The time-reversal symmetry, denoted $\mathcal{T}$ from now on, is defined as the following set of operations, applied sequentially:
\begin{enumerate}
    \item Flipping all Ising spins
    \item Kramers-Wannier-like half-unit-cell translation of Majorana fermions, defined as:
    \begin{equation}
        \gamma_{i,A} \rightarrow \gamma_{i,B}, \, \gamma_{i,B} \rightarrow - \gamma_{i+1,A}.
    \end{equation}
    \item $U=\displaystyle \prod_{i} U_{i}$ with $U_{i}$ a local unitary transformation defined as
    \begin{equation}
    \begin{split}
        & U_{i} = \frac{1 + \gamma_{i,B}\gamma_{i+1,B}}{\sqrt{2}}  \ket{\downarrow_{i}\uparrow_{i+1}} \bra{\downarrow_{i}\uparrow_{i+1}} +  \ket{\downarrow_{i}\downarrow_{i+1}} \bra{\downarrow_{i}\downarrow_{i+1}} \\
        & \quad \quad \quad \quad +  \ket{\uparrow_{i}\uparrow_{i+1}} \bra{\uparrow_{i}\uparrow_{i+1}}  +  \ket{\uparrow_{i}\downarrow_{i+1}} \bra{\uparrow_{i}\downarrow_{i+1}}.
    \end{split}
    \end{equation}
    \item Complex conjugation.
\end{enumerate}
The above set of operations does not map $R_{i}$ to the same operator. However, one can easily show that the restricted Hilbert space defined by the condition $R_{i}=1$ and the restricted Hilbert space defined by $\tilde{R}_{i}=1$, the ``time-reversal partner'' of $R_{i}=1$, are identical. Hence, the above set of operations is indeed closed under the restricted Hilbert space of  interest. Additionally, due to the non-local half-unit cell transformation, one might naively think that $\mathcal{T}^2$ implemented in this way is also a non-local operator. 
However, one can show that the free Majorana fermions on Ising spin domain walls are transformed into free Majorana fermions on the same domain walls under the above definition of $\mathcal{T}$.  Because of this property, $\mathcal{T}^{2} = (-1)^{F}$ (where $F$ denotes fermion parity) in the restricted Hilbert space, satisfying all the defining properties of time-reversal symmetry.

\subsection{Symmetry-respecting local terms}
\label{sec:localterms}

 Having discussed the Hilbert space and the time-reversal symmetry action of the model, we now present simple local time-reversal symmetric terms that one can add to the edge Hamiltonian and that will appear throughout this paper.
 
\subsubsection{Flip term}
 
\begin{figure}
     \centering
     \includegraphics[width=\linewidth]{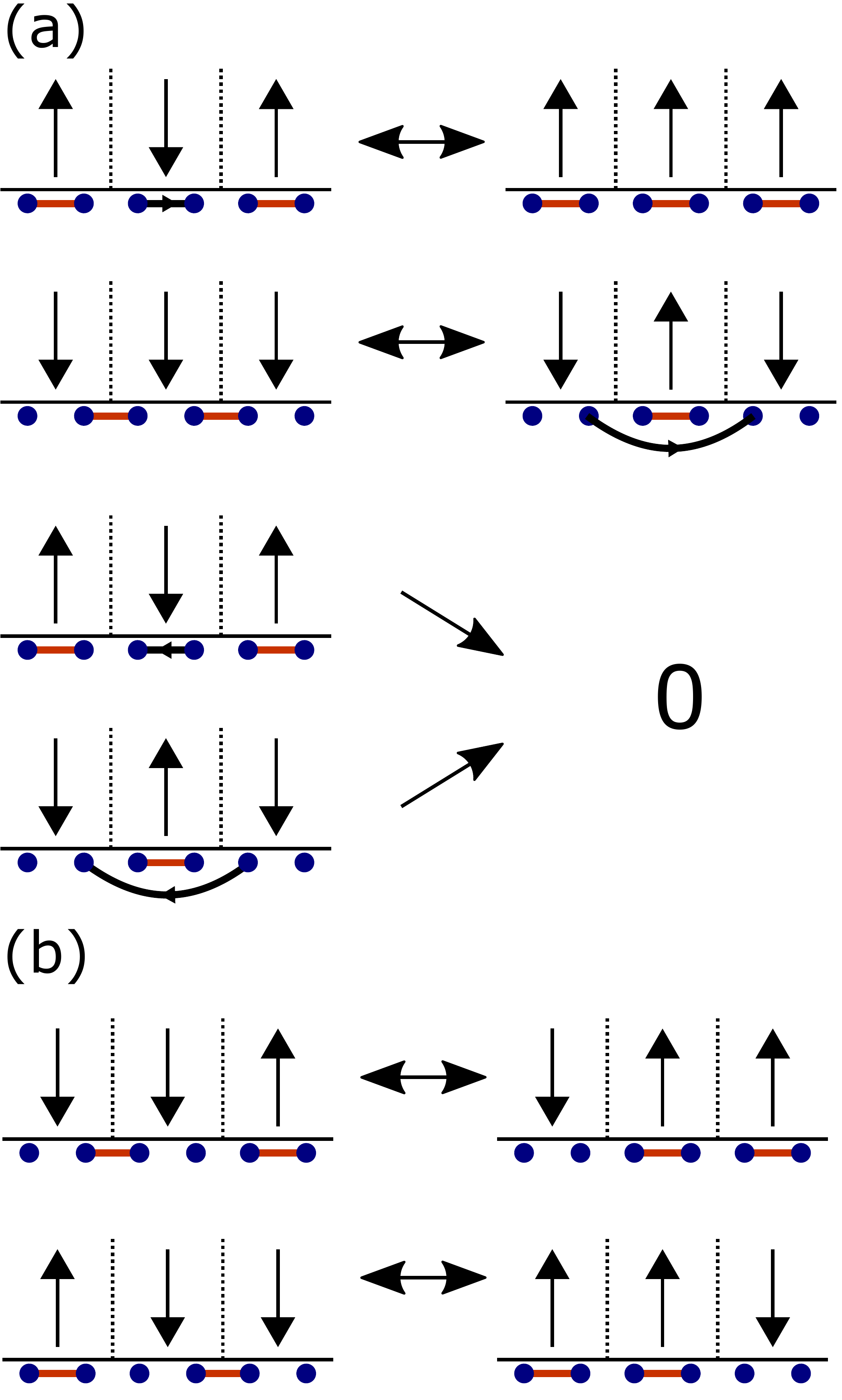}
     \caption{Action of the flip term $F_{i}$, when there are (a) even  or (b) odd number of domain walls in between site $i-1$, $i$, $i+1$} 
     \label{fig:reviewflip}
\end{figure}

 We will define a ``flip term'' $F_{i}$ at site $i$ to be a Hermitian term that is compatible with time-reversal symmetry and flips an Ising spin at site $i$ when the action is non-trivial. It turns out that these  conditions are restrictive enough to specify the flip term $F_{i}$ with just one complex parameter $p_{i}$ and one U(1) phase parameter $\phi$, up to a real constant which can be thought as a ``magnitude'' of the flip term when this term is added to the Hamiltonian. Specifically, $F_{i}$ is defined as:
 \begin{equation}
 \label{eq:flipdef}
 \begin{split}
 & F_{i} =  e^{i \phi}  P_{i,A}^{i,B} P_{i-1,A}^{i-1,B}\ket{\uparrow_{i-1} \uparrow_{i} \uparrow_{i+1}} \bra{\uparrow_{i-1} \downarrow_{i} \uparrow_{i+1}} \\ 
 & +  \frac{e^{i \phi}}{\sqrt{2}}P_{i,A}^{i,B} P_{i-1,B}^{i+1,A} P_{i-1,B}^{i,A} P_{i,B}^{i+1,A} \ket{\downarrow_{i-1} \uparrow_{i} \downarrow_{i+1}} \bra{\downarrow_{i-1} \downarrow_{i} \downarrow_{i+1}} \\
 & + p_{i} \frac{1}{\sqrt{2}}P_{i,A}^{i,B} P_{i-1,B}^{i,A}  \ket{\downarrow_{i-1} \uparrow_{i} \uparrow_{i+1}} \bra{\downarrow_{i-1} \downarrow_{i} \uparrow_{i+1}} \\
 & + p_{i} \frac{1}{\sqrt{2}}P_{i,A}^{i,B} P_{i,B}^{i+1,A} \ket{\uparrow_{i-1} \uparrow_{i} \downarrow_{i+1}} \bra{\uparrow_{i-1} \downarrow_{i} \downarrow_{i+1}} \\
 & + \, \text{H.c}.
 \end{split}
 \end{equation}
We illustrate the action of this flip term in Fig.~\ref{fig:reviewflip}. When there are zero or two domain walls between site $i-1$ and site $i+1$, the relevant terms in Eq.~\eqref{eq:flipdef} are the first two lines and their Hermitian conjugates. These terms create or annihilate two domain walls and reconfigure Majorana fermions accordingly -- see Fig.~\ref{fig:reviewflip}(a). For the Ising spin configurations with two domain walls, there are two free Majorana fermions. These two free Majorana fermions should be paired in a specific way for the action to be non-trivial due to the fermion parity conservation. For states with the wrong pairings [the bottom two rows in Fig.~\ref{fig:reviewflip}(a)], there is no local operation to flip a spin at site $i$ and reconfigure Majorana fermion pairings according to the constraint $R_{i}=1$. Hence, $F_{i}$ annihilates these states.
  
Figure~\ref{fig:reviewflip}(b) illustrate action of $F_{i}$ on the states with a single domain wall between site $i-1$ and site $i+1$. These states are acted upon by terms with coefficient $p_{i}$ or $p_{i}^{*}$ in Eq.~\eqref{eq:flipdef}, i.e., the third and fourth lines and their Hermitian conjugates. In this case, $F_{i}$ always acts non-trivially and simply shifts the location of the domain wall and the corresponding unpaired Majorana fermions. 
 
We note that a local time-reversal symmetric gauge transformation $e^{i \frac{\phi}{2} \sigma_{i}^{z}}$ absorbs the U(1) phase parameter $\phi$ into argument of $p_{i}$. In other words, one can always perform a gauge transformation to set $\phi=0$, at the cost of modifying $p_{i}$. For the rest of the paper, without loss of generality we therefore set $\phi =0$; $F_{i}$ is then only parameterized by a complex number $p_{i}$. 

From a naive viewpoint, $F_{i}$ plays a role similar to that of a Pauli operator $\sigma_{i}^{x}$ in the sense that both operators flip an Ising spin at site $i$. However, the key difference is that for $F_{i}$ to be compatible with the restricted Hilbert space we are working in, $F_{i}$ should non-trivially modify Majorana fermion degrees of freedom as well. Due to this feature, the commutator between two neighboring flip terms is generally nonzero, i.e., $[F_{i},F_{i+1}] \neq 0$. Hence, the simple time-reversal symmetric Hamiltonian $H= \displaystyle - \sum_{i} F_{i}$ already exhibits non-trivial quantum dynamics. Metlitski and Jones showed through numerics that the low-energy physics of this Hamiltonian with all $p_{i}=1$ is described by a central-charge $c=\frac{1}{2}$ Ising conformal field theory \cite{Jones2019}.  The 1D model we review here thus indeed captures the familiar helical Majorana fermion edge state of 2D time-reversal invariant topological superconductors!

In this paper, we find that if the parameter $p_{i}$ is assumed to be spatially uniform, the natural value is $p_i = \pm 2^{-1/4}$.  It turns out that $p_{i} = \pm 2^{-1/4}$ corresponds to the only two attractive fixed-point values of $p_{i}$ under the strong-disorder RG transformations we introduce (see Appendix~\ref{app:fpp} for the proof). Additionally, a gauge transformation $\exp\left[i \frac{\pi}{2}\left(N_{\text{dw}}+\sum_{i} \sigma_{i}^{z} \right) \right] $ maps all $p_{i}$'s to $-p_{i}$, implying that the models where $p_{i}= +2^{-1/4}$ versus $p_{i}=-2^{-1/4}$ are identical. In most cases we simply fix $p_{i}=2^{-1/4}$. However, we also emphasize that even if $p_{i}$'s are spatially random or are set to be a different value from $\pm 2^{-1/4}$ (for example, $p_{i}=1$ to make comparison to the disorder-free system studied in Metlitski and Jones), deep in the RG flow, the effective Hamiltonian will have $p_{i}$'s very close to $\pm 2^{-1/4}$. 

\subsubsection{Ising interaction}
A simple Ising interaction term $\sigma_{i}^{z}\sigma_{i+1}^{z}$ is time-reversal symmetric and hence can be added to the Hamiltonian.

\subsubsection{Domain-wall Majorana fermion bilinears}
Consider the following operator:
\begin{equation}
\label{eq:Tidef}
\begin{split}
& T_{i} = i\gamma_{i,A}\gamma_{i,B} \ket{\uparrow_{i-1} \downarrow_{i} \uparrow_{i+1}} \bra{\uparrow_{i-1} \downarrow_{i} \uparrow_{i+1}} \\
& +  i\gamma_{i-1,B}\gamma_{i+1,A} \ket{\downarrow_{i-1} \uparrow_{i} \downarrow_{i+1}} \bra{\downarrow_{i-1} \uparrow_{i} \downarrow_{i+1}}.
\end{split}
\end{equation}
The two ket-bras project onto Ising spin configurations which have two domain walls between site $i-1$ and $i+1$. Two Majorana fermions in front of each Ising spin projector are precisely free Majorana fermions corresponding to the two domain walls. Hence, one may interpret the above term as a nearest-neighbor hopping between free Majorana fermions. Especially, for the second row of Eq.~\eqref{eq:Tidef}, the constraints $R_{i}=1$ freeze $\gamma_{i,A}$ and $\gamma_{i,B}$ that lie between $\gamma_{i-1,B}$ and $\gamma_{i+1,A}$, so the term that appears in the second line is indeed nearest-neighbor hopping between two ``free" domain wall Majoranas. One can readily show that the above term is Hermitian and invariant under the time-reversal symmetry. We will see that this term governs the low-energy behavior of most of the phases of matter that appear in this paper.

Similarly, one can think of the following term that mediates next-nearest neighbor hoppings between free Majorana fermions, in the same sense as in the preceding paragraph:
\begin{equation}
\label{eq:Sidef}
\begin{split}
& S_{i} = i\gamma_{i,A}\gamma_{i+1,B} \ket{\uparrow_{i-1} \downarrow_{i} \downarrow_{i+1} \uparrow_{i+2}} \bra{\uparrow_{i-1} \downarrow_{i} \downarrow_{i+1} \uparrow_{i+2}} \\
& + i\gamma_{i-1,B}\gamma_{i+2,A} \ket{\downarrow_{i-1} \uparrow_{i} \uparrow_{i+1} \downarrow_{i+2}} \bra{\downarrow_{i-1} \uparrow_{i} \uparrow_{i+1} \downarrow_{i+2}}.
\end{split}
\end{equation}
The bosonic parts of the operator are projections onto spin configurations in which there is a domain wall between sites $i-1$ and $i$ and another between sites $i+1$ and $i+2$.  The operator $S_{i}$ is also a valid local term for our Hamiltonian and makes occasional appearances in this paper.

\section{Infinite-randomness fixed point from Ising-interaction-dominated limits}
\label{sec:splim}

In this section, we study the Hamiltonian 
\begin{equation}
\label{eq:Hamparent}
H = \sum_{i} - h_{i} F_{i} - J_{i} \sigma_{i}^{z} \sigma_{i+1}^{z}
\end{equation}
in two analytically tractable limits. The two limits are:
\begin{itemize}
    \item $J_{i} = -J$ with  positive $J \gg |h_{i}|$, i.e., $J_{i}$ is chosen to be antiferromagnetic and uniform with a magnitude much stronger than the flip terms.
    \item $J_{i}$ is randomly chosen to be either positive or negative, and when $J_{i} >0$, its magnitude is strongly random. The magnitude of any Ising interaction coefficient $J_{i}$ is nevertheless always much larger than $h_{i}$. We relegate the precise description of this limit to Sec.~\ref{sec:sec3rule}, which provides relevant technical details.
\end{itemize}
In either limit, the spin degrees of freedom are effectively pinned by strong nearest-neighbor Ising interactions, and the low-energy degrees of freedom are Majorana fermions that appear on spin domain walls. One can systematically derive an effective low-energy Hamiltonian consisting of domain wall Majorana fermions in both cases and show that these Majorana fermions form infinite-randomness fixed points. 

The first limit is the simplest case imaginable where such an infinite-randomness fixed point physics can arise.  Since each ``ferromagnetic domain'' has length $1$, however, it does not correspond to the edge phase we described in the introduction (Fig.~\ref{fig:cartoon}), yet serves as a useful stepping stone for understanding more complicated cases. The second limit allows some interactions to be ferromagnetic, and realizes the scenario we described in the introduction in which ferromagnetic domains have random lengths. While the Hamiltonian is fine-tuned here too, in the sense that Ising interactions are set to be much stronger than flip terms, this limit provides the most analytically well-controlled Hamiltonian that realizes the essential edge physics we are envisioning in this paper and makes key properties of the edge phase transparent. Also, the way we analyze the second limit serves as a primer to the more complicated strong-disorder RG we will develop in the next section.

\subsection{Preliminary example: 1D model with uniform antiferromagnetic interaction}\label{sec:3prelim}

\begin{figure}
\includegraphics[width=1.0\linewidth]{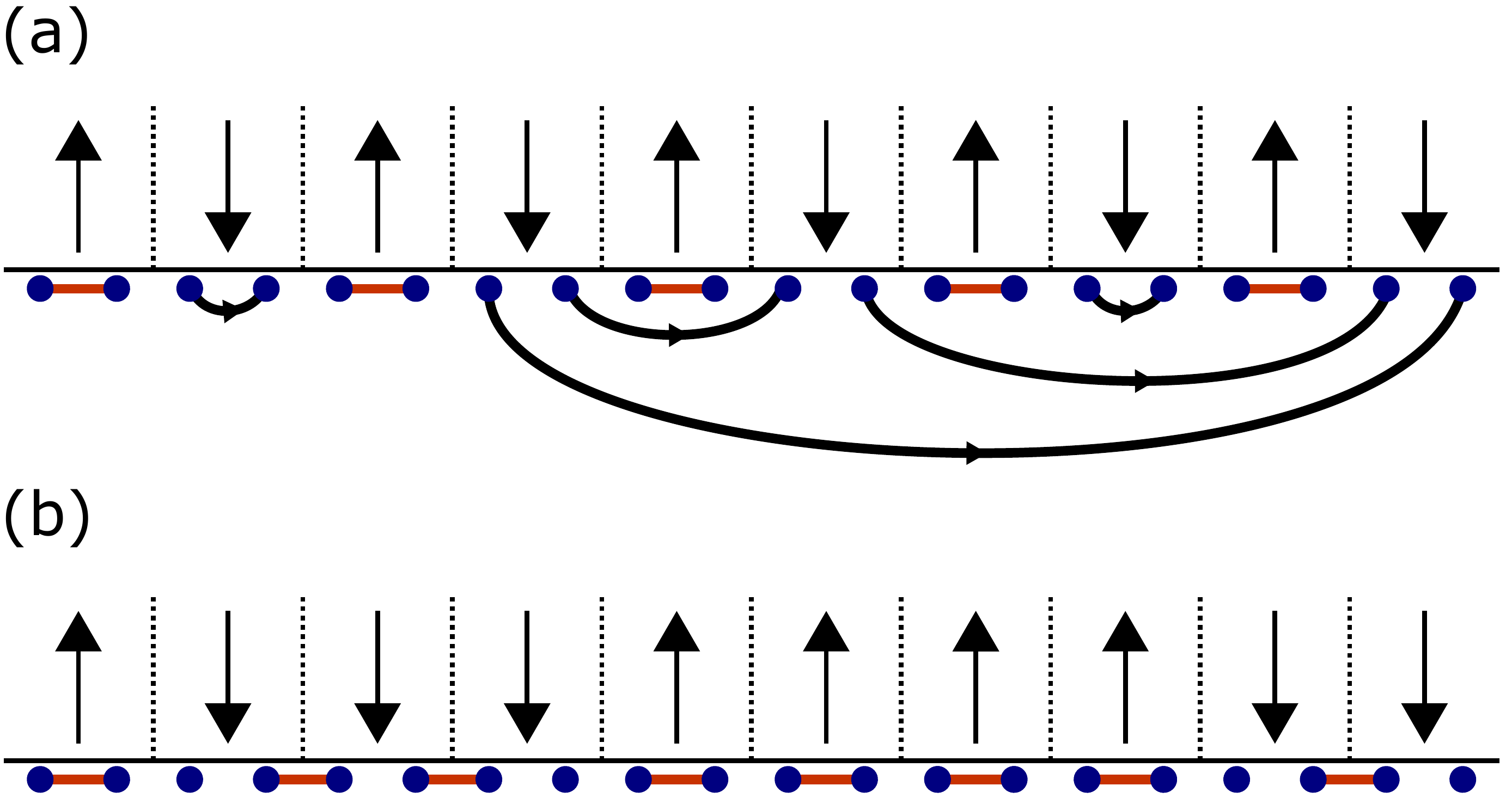}
\caption{(a) Illustration of the infinite-randomness fixed point that arises when Ising spins are antiferromagnetically aligned due to strong Ising interactions. Majorana fermions on the domain walls can pair with other Majorana fermions arbitrarily far away (albeit such events are rare), and these Majorana pairings form random-singlet-like infinite-randomness fixed points. (b) When Ising interactions have mixed sign, assuming Ising interactions are much larger than flip terms, the interactions spin configurations in IR, and the low-energy physics is governed by domain-wall Majorana fermions, which are shown in the figure as blue dots with no red line attached to them.}
\label{fig:spinfixed}
\end{figure}

As a preliminary and illustrative exercise, we will examine the Hamiltonian in Eq.~\eqref{eq:Hamparent} with $J_{i} = -J$ and $J > 0$. First, let us assume that $h_{i} =0$ for all $i$. In this case, the strong Ising interaction breaks time-reversal symmetry and induces antiferromagnetic ordering of the Ising spins. While Ising spins are then essentially frozen, Majorana zero modes arise between any two neighboring Ising spins. The Ising interaction contains no information on how these Majorana fermions couple with each other, and the ground state of the Hamiltonian correspondingly exhibits a massive degeneracy. (In stark contrast, when the Ising interactions are purely ferromagnetic, i.e., $J_{i} = -J$ and $J < 0$, the Ising spins fix the pairing of Majorana fermions and produce a unique ground state.)

Now we allow $h_{i}$ to be non-zero but take $J \gg h_{i}$. In this limit, one may use degenerate perturbation theory to study how flip terms generate Majorana-fermion couplings that lift the aforementioned degeneracy. Specifically, the unperturbated Hamiltonian $H_{0}$ contains the Ising interactions $J \sum_{i}  \sigma_{i}^{z} \sigma_{i+1}^{z}$, and the perturbation $H_{p}$ is the sum of flip terms $- \sum_{i} h_{i} F_{i}$. The degenerate eigenstates of $H_{0}$ to which we apply perturbation theory are states with perfectly antiferromagnetically ordered Ising spins. 
 
Recall that a flip term $F_{i}$ either annihilates a state or flips an Ising spin. Consequently, any first-order correction in degenerate perturbation theory vanishes. At second order, all terms generated from perturbation theory correspond to the following process: One may flip an Ising spin at site $i$ with a flip term $F_{i}$ to create an excited state with respect to $H_{0}$ and apply $F_{i}$ again to return to the original spin state. Hence, the second-order correction is given by
\begin{equation}
H^{(2)} = -\sum_{i} \frac{h_{i}^{2}}{4J} F_{i}^{2}.
\label{H2}
\end{equation} 

When acted on states with antiferromagnetically aligned Ising spins, either $F_{i}^{2} =1$ or $0$ depending on the Majorana degrees of freedom. On the states with $\sigma_{i}^{z} =\, \downarrow$ and $\sigma_{i-1}^{z} = \sigma_{i+1}^{z} =\, \uparrow$, one can show that  $F_{i}^{2} \equiv \frac{1+i\gamma_{i,A}\gamma_{i,B}}{2}$. Similarly, on states with $\sigma_{i}^{z} =\, \uparrow$ and $\sigma_{i-1}^{z} = \sigma_{i+1}^{z} =\, \downarrow$, one obtains $F_{i}^{2} \equiv \frac{1+i\gamma_{i-1,B}\gamma_{i+1,A}}{2}$. 
Equation~\eqref{H2} may then be expressed as
\begin{equation}
\label{eq:uafh2}
H^{(2)} = - \sum_{i} t_{i} T_{i} + \text{const.}, \, ~~~t_{i} = \frac{h_{i}^{2}}{8J},
\end{equation}
where $T_{i}$ corresponds to nearest-neighbor Majorana hopping terms we introduced in Eq.~\eqref{eq:Tidef}. Note that we are working on top of one of the antiferromagnetic ground states, i.e., assuming a frozen antiferromagnetic spin pattern. The free Majorana fermions are $\gamma_{i,A}, \gamma_{i,B}$ for sites with $\sigma_i^z = \downarrow$ (blue dots not connected by red lines in Fig.~\ref{fig:spinfixed}(a)). Each $T_i$ in the effective Hamiltonian retains a single fermion bilinear term that contains nearest-neighbor free Majoranas to the left and to the right of site $i$. When $h_{i}$'s are random, the low-energy domain-wall Majorana fermions are expected to flow to an infinite-randomness fixed point---the same one governing the crtical point in the random transverse-field Ising model \citep{Fishersdrg_2,Fishersdrg_3,Motrunich2001}. Figure~\ref{fig:spinfixed}(a) illustrates this phase.

This infinite-randomness fixed point, which represents a critical state and is hence delocalized in some sense, can be driven to a localized phase upon adding some perturbations. The most straightforward way to generate localization is to give a dimerization to the distribution of $h_{i}$, i.e., $\overline{h_{2k}}  > \overline{ h_{2k+1}}$, with the overlines denoting disorder averages here and below. Another, more indirect way is to add a small but uniform Zeeman field $ - B \sum_{i} \sigma_{i}^{z}$ to the Hamiltonian, with $B \ll J$. Then, one may similarly use degenerate perturbation theory to find that the coefficient $t_{i}$ is modified to
\begin{equation}
t_{i} = \begin{cases}
\frac{h_{i}^{2}}{8J+4B}, & \sigma_{i}^{z} =\, \uparrow , \\
\frac{h_{i}^{2}}{8J-4B}, & \sigma_{i}^{z} =\, \downarrow .
\end{cases}
\end{equation}
In stark contrast to the $B=0$ case where Majorana fermion couplings across  up-spins and down-spins are statistically the same, adding a $B$ field makes Majorana fermion hopping across down-spins statistically stronger than the hopping across up-spins. The resulting effective dimerization to $t_{i}$'s drives the Majorana fermions into a localized state.
 
For later purposes, it is useful to understand how Zeeman field terms and spatial modulations of flip terms localize Majorana fermion degrees of freedom from a symmetry perspective. The Hamiltonian we consider has two symmetries: time-reversal symmetry $\mathcal{T}$ and statistical translation symmetry $\overline{T}_{x}$. The \textit{composite} symmetry $\mathcal{T}\overline{T}_{x}$ guarantees the Majorana fermions to be critical. Spatial modulation of the flip terms or the Zeeman field terms break one of the two symmetries and hence break the composite symmetry. The explicit violation of $\mathcal{T}\overline{T}_{x}$  tunes the Majorana fermions away from criticality and drives localized behavior.
 
\subsection{The case with random Ising interactions: Overview}
\label{sec:sec3overview}

Now we will consider the same Hamiltonian Eq.~\eqref{eq:Hamparent}, with the Ising interaction coefficients allowed to be in general random---both in \textit{magnitude and sign}. We will still maintain the condition $ | J_{i'} | \gg h_{i} $ for any $i$ and $i'$ so that in the low-energy limit the Ising spins are pinned by the Ising interactions. However, note that due to the mixed sign of the Ising interactions, the magnetic ordering of the Ising spins is neither perfectly ferromagnetic nor antiferromagnetic. 
   
The low energy degrees of freedom, similar to the example in the previous subsection, are Majorana fermions living at the Ising spin domain walls; see Fig.~\ref{fig:spinfixed}(b) for an illustration. One may again employ perturbation theory to derive couplings between Majorana fermions. The key difference from the case with the uniform strong antiferromagnetic Ising interactions covered in the previous subsection is that one needs to go to higher order in perturbation theory to derive effective couplings between Majorana fermions. To understand this point, recall that in the previous example, nearest-neighbor couplings between Majorana fermions are generated by the second-order perturbation theory term that corresponds to a process of flipping a spin at one site and flipping it back. As a generalization, when treating the flip terms perturbatively, terms in the $2n$-th-order perturbation theory can be understood as a process of flipping $n$ spins back and forth. In the cases of interest in this subsection, between two consecutive antiferromagnetic bonds, there can be $n \geq 1$ spins that are ferromagnetically linked. Couplings between a pair of Majorana fermions located at two such antiferromagnetic bonds first appear in the $2n$-th order perturbation theory, corresponding to processes that flip $n$ spins between the two antiferromagnetic bonds one-by-one and then flip these spins back. 

There is no a priori obstruction to computing all terms in perturbation theory and deriving effective couplings between Majorana fermions that result.  Nevertheless, we will further restrict how the coefficients in the Hamiltonian of  Eq.~\eqref{eq:Hamparent} are chosen to make the analysis more concise; importantly, the case we consider can be naturally generalized to the strong-disorder RG analysis covered in Sec.~\ref{sec:sdrg}. We specifically assume that the parameters $\{h_{i} \}$, $\{ J_{i} \}$ in Eq.~\eqref{eq:Hamparent} are randomly selected according to the following rules:
 \begin{enumerate}
\item When $J_{i}$ is negative, we assume $J_{i}$ to take a uniform value $\Lambda_{A} \gg 1$. Additionally, we require that if $J_{i}<0$ for a certain $i$, then $J_{i'} >0$ for $i' = i-2,i-1,i+1,i+2$, i.e., within the next nearest neighbors of an antiferromagentic bond, one may not find another antiferromagnetic bond.
\item When $J_{i}$ takes a positive value, we assume $J_{i}$ to be \textit{infinitely random} in some limit. As a specific example, one can imagine drawing $J_{i}$ from the following probability distribution, with the limit $\Lambda_{F} \rightarrow 0$:
\begin{equation}
P(J) = \begin{cases}
\frac{1}{J \, \ln{\frac{1}{\Lambda_{F}}}} & \Lambda_{F} < J < 1 \\
0 & \text{otherwise}
\end{cases}.
\end{equation}
\item The parameter $h_{i}$ is chosen to be random and much smaller than $\Lambda_{F}$. For example, drawing $h_{i}$ uniformly from $(0, \Lambda_{H})$ with $\Lambda_{H} \ll \Lambda_{F}$ will achieve this condition.
\end{enumerate}
The analysis we perform in the next two subsections is expected to be asymptotically accurate in the limit $\Lambda_{A} \rightarrow \infty$, $\Lambda_{F} \rightarrow 0$, $\Lambda_{H} \rightarrow 0$ with $\frac{\Lambda_{H}}{\Lambda_{F}} \rightarrow 0$ (and is expected to be qualitatively correct for a range of parameters close to this limit). The key feature here is that all the Ising interactions are much larger than the flip terms and strongly random, approaching infinite-randomness in the aforementioned limit. We will see in the next subsection that this strong randomness condition allows us to derive couplings between Majorana fermions via sequence of \textit{local, real-space transformations}. These real-space transformations are simplified versions of RG transformations that appear in Sec.~\ref{sec:sdrg}. 

\subsection{The model with mixed-sign Ising interactions: Transformation rule}
\label{sec:sec3rule}

\begin{figure}
\includegraphics[width=\linewidth]{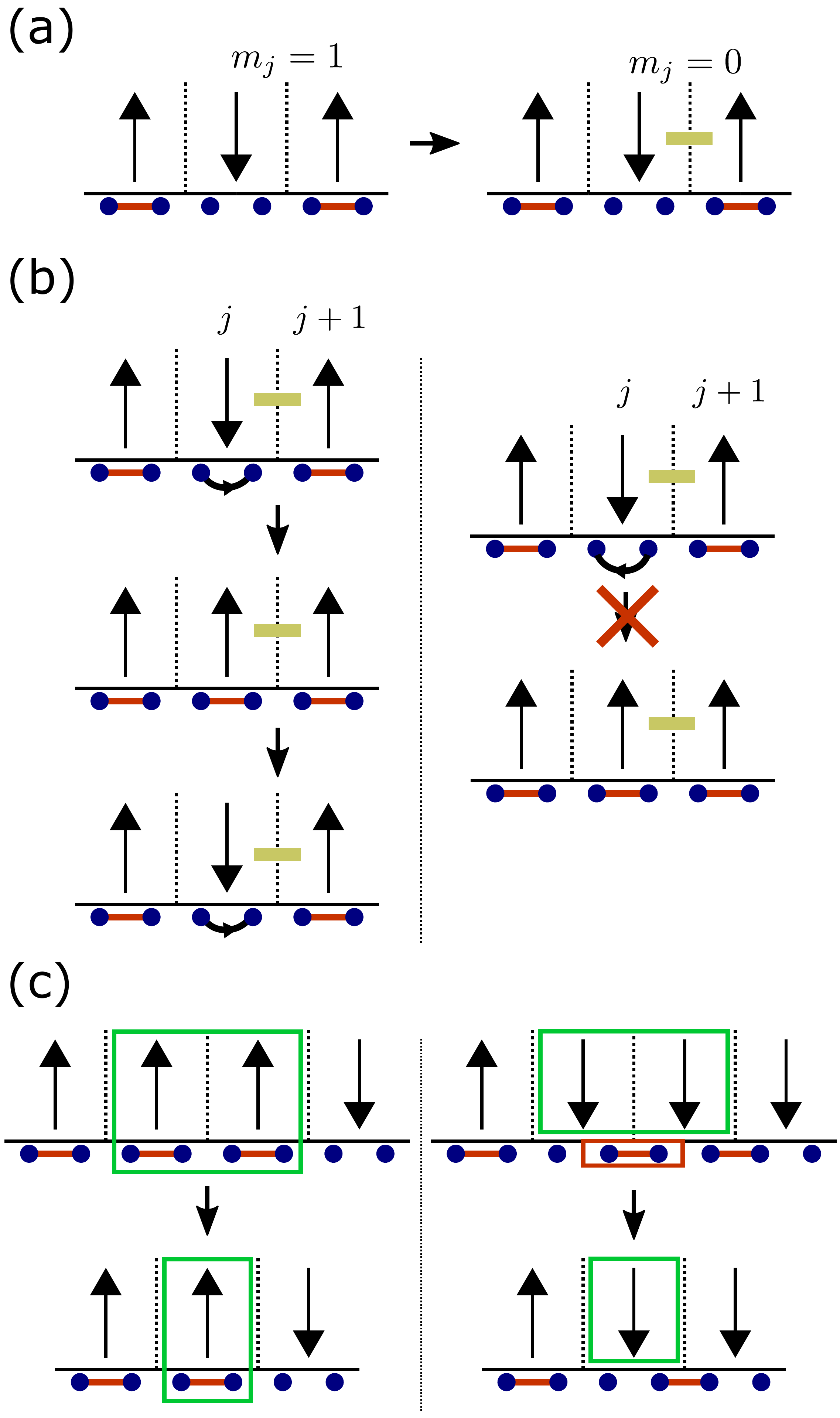}
\caption{(a) Graphical notation for the transformations. (b) Graphical illustration of the processes that give rise to $T_{j}$ term in the second-order perturbation theory. When starting with the state with $i\gamma_{j,A}\gamma_{j,B} = +1$ (left panel), one may flip a spin at site $j$ and and flip it back. However, this process is forbidden on the state  with $i\gamma_{j,A}\gamma_{j,B} = -1$ (right panel). Hence, the spin-flip process gives rise to the energy difference between the states in the left panel and the right panel, and this energy difference is encoded precisely by the effective $T_{j}$ term.
(c) Illustration of the transformation that decimates a strong ferromagnetic Ising interaction $-J_{j} \sigma_{j}^{z}\sigma_{j+1}^{z}$. In the right panel, one removes Majorana fermions in red squares to reduce number of Majorana fermions in the system by $2$.} 
\label{fig:tf1}
\end{figure}

Let us first sketch the big ideas behind real-space transformations that we will employ to derive the couplings between Majorana fermions. In the Hamiltonian we are considering, the largest local terms are given by strong antiferromagnetic nearest-neighbor Ising interactions. Remaining ferromagnetic Ising interactions and flip terms are assumed to be much smaller than the antiferromagnetic Ising interactions. Hence, to capture the low-energy physics, we imagine projecting our system onto the low-energy subspace in which two spins joined by an antiferromagnetic Ising interaction are always anti-aligned. In Fig.~\ref{fig:tf1}(a), we graphically depict the projection by marking the antiferromagnetic bond with a gold line---two spins neighboring the gold line are \emph{rigidly constrained} to be antiferromagnetically aligned with each other within the subspace. To encode the non-trivial effect of the flip terms which may locally change the Ising spin orientation and bring the system out of the low-energy subspace, we employ second-order degenerate perturbation theory and incorporate the terms generated from the perturbation theory in the Hamiltonian.

That is, we effectively ``integrate out'' or ``decimate'' strongly antiferromagnetic bonds and incorporate the local couplings generated from such procedures into the Hamiltonian. 
The key feature behind this transformation is that a process in the second order perturbation theory that flips a spin at site $i$ and flips the spin back generates $T_{i}$, via a similar mechanism we saw in the earlier subsection, see Fig.~\ref{fig:tf1}(b).

One can pursue similar ideas to treat ferromagnetic bonds as well. After decimating all strongly antiferromagnetic bonds, we then pick the strongest ferromagnetic bond corresponding to an Ising interaction $-J_{i}\sigma_{i}^{z}\sigma_{i+1}^{z}$. Due to the assumptions we made about parameters, all local terms in the Hamiltonian involving site $i$ or $i+1$ are much smaller in magnitude than $|J_i|$. Once again, we do a low-energy projection to the states where two spins $i$ and $i+1$ are ferromagnetically aligned. In this case, one may combine two ferromagnetically aligned spins into a single spin and also appropriately modify Majorana fermion degrees of freedom to combine two sites $i$ and $i+1$ into a single \textit{super-site} [Fig.~\ref{fig:tf1}(c)]. 

After decimating the strongest bond in the system, one may decimate the bond associated with the next strongest ferromagnetic interaction, or equivalently, the strongest ferromagnetic bond in the system after the first ferromagnetic decimation, and so on. One can continue this process until all ferromagnetic Ising interactions are decimated or integrated out. At the end of the procedure, all bonds in the system are marked with gold lines in our graphical illustration [recall Fig.~\ref{fig:tf1}(a) for the definition of gold lines]; each Ising spin represents a \textit{cluster} of original Ising spins linked by ferromagnetic Ising interactions. Since all the remaining Ising spins are hard-constrained to be antiferromagnetically aligned, the only meaningful local degrees of freedom at this point are Majorana fermions.  The low-energy fate of the system is controlled by the nearest-neighbor Majorana fermion hopping terms $T_{i}$ generated along the way via a sequence of transformations.

We now present these transformation rules in more formal, quantitative language (deferring precise technical details about the derivations of these rules to Appendix~\ref{app:df}). At any stage of the transformations, we keep track of the following information: non-negative parameters at each site/bond $\{h_{i}, J_{i}, t_{i}, s_{i} \}$, which determine coefficients of each term in the Hamiltonian, and $\{ m_{i}=0,1 \}$, a binary label on each bond. The label $m_{i} = 0$ indicates a bond with a gold line between site $i$ and $i+1$, as in Fig.~\ref{fig:tf1}(a), generated from integrating out the antiferromagnetic Ising interaction. 

The Hamiltonian at each stage takes the form 
\begin{equation}
\label{eq:tfham}
\begin{split}
&H(\{ m_{i}, h_{i}, J_{i}, t_{i}, s_{i} \} ) = \sum_{i} (-h_{i} m_{i-1}m_{i} F_{i} \\
& \quad \quad -   J_{i} \sigma_{i}^{z}\sigma_{i+1}^{z} - 2t_{i} T_{i}  - 2 s_{i} S_{i}) .
\end{split}
\end{equation}
We refer to Sec.~\ref{sec:localterms} for the definition of each term in the Hamiltonian. Here $\{ J_{i}, t_{i}, s_{i} \}$ directly specifies coefficients of local terms in the Hamiltonian, while $h_{i}$ only enters as a coefficient of the flip term $F_i$ when $m_{i-1} = m_{i} =1$. Physically, $m_{i-1} = m_{i} =1$ means that the Ising spin at site $i$ is not hard-constrained to be anti-aligned with respect to the Ising spin at site $i-1$ or $i+1$ and hence can be flipped without violating the constraint. Although the coefficient of the flip term at site $i$ vanishes when $m_{i-1}$ or $m_{i}$ is 0, we will see that the values of $h_{i}$ when the coefficient of the actual flip term vanishes plays an important role in some of the analysis in the appendices.  We will thus keep track of $h_{i}$'s for all $i$'s regardless of the value of $m_{i-1}$ and $m_{i}$.

At the beginning, $t_{i} = s_{i}=0$ for all $i$'s, and $m_{i}=1$ for all bonds $i$. Observe that the initial Hamiltonian has the same form as in Eq.~\eqref{eq:Hamparent}.
The transformation proceeds by integrating out strong antiferromagnetic bonds first, then ferromagnetic bonds. To integrate out an antiferromagnetic bond $j$ with $J_{j} = -\Lambda_{A}$, we perform the following transformation:
\begin{equation}
\label{eq:afrule}
\begin{split}
& m_{j}= 1\rightarrow m_{j}=0, \\
& t_{j}=0 \rightarrow t_{j} = \frac{h_{j}^{2}}{8 \Lambda_{A}}, \\
& t_{j+1}=0 \rightarrow t_{j+1} = \frac{h_{j+1}^{2}}{8\Lambda_{A}}.
\end{split}
\end{equation}
Assigning $m_{j}=0$ emphasizes that the strong antiferromagnetic bond enforces $\sigma_{j}^{z} = -\sigma_{j+1}^{z}$ as a hard constraint after the transformation. Changes in Hamiltonian parameters are derived from the second-order perturbation theory. As we saw earlier in the subsection and in Fig.~\ref{fig:tf1}(b), the second-order perturbation theory generates $t_{j}$ and $t_{j+1}$, whose precise values are given above.
 
 Decimation of ferromagnetic bonds proceeds from the largest to the smallest. Decimating a ferromagnetic bond between site $j $ and $j+1$ transforms the parameters $\{ m_{i}, h_{i}, J_{i},t_{i}, s_{i} \}$ into the new set of parameters $\{ m'_{i}, h'_{i}, J'_{i},t'_{i}, s'_{i} \}$ \ given by:
\begin{equation}
\label{eq:frule}
\begin{split}
& (m'_{i}, J'_{i}) = \begin{cases}
(m_{i}, J_{i}) & i<j \\
(m_{i+1}, J_{i+1}) & i \geq j
\end{cases}, \\
& h'_{i} = \begin{cases}
h_{i} & i < j \\
\frac{2^{-1/4} h_{j}h_{j+1}}{J_{j}} & i = j \\
h_{i+1} & i > j
\end{cases}, \\
& t'_{i} = \begin{cases}
t_{i} & i < j \\
t'_{j}  & i = j  \\
t_{i+1} & i > j
\end{cases}, \\
& s'_{i} = \begin{cases}
s_{i} & i < j-1 \\
m_{j-1}\frac{t_{j-1} t_{j+1} h_{j}^{2}}{2J_{j}^{3}}  & i=j-1  \\
m_{j+1}\frac{ t_{j} t_{j+2} h_{j+1}^{2}}{2J_{j}^{3}} & i = j \\
s_{i+1} & i > j
\end{cases}, \\
& t'_{j}  = m_{j+1} t_{j} \frac{h_{j+1}^{2}}{4 \sqrt{2} J_{j}^{2}} + m_{j-1} t_{j+1} \frac{h_{j}^{2}}{4 \sqrt{2} J_{j}^{2} }  + s_{j}.
\end{split}
\end{equation}
The transformation rule for $h'_{i}$ comes from the second-order degenerate perturbation theory, while keeping track of $t'_{i}$ and $s'_{i}$ requires going to the third order and fourth order, respectively. At first glance, it may appear unnatural that one should include terms from the higher-order perturbation theory -- non-trivial corrections appear already at the second order, and the higher-order terms are generally much smaller than the second-order terms. However, in our case, $T_{i}$'s control the eventual low-energy physics of the system, and keeping track of these hoppings requires going to higher-order in perturbation theory. \textit{The necessity of including the higher-order perturbation theory results to keep track of couplings between domain-wall Majorana fermions is a recurrent theme in this paper and plays an important role in understanding the strong-disorder RG flow presented in the next section.}

\begin{figure}
\includegraphics[width=\linewidth]{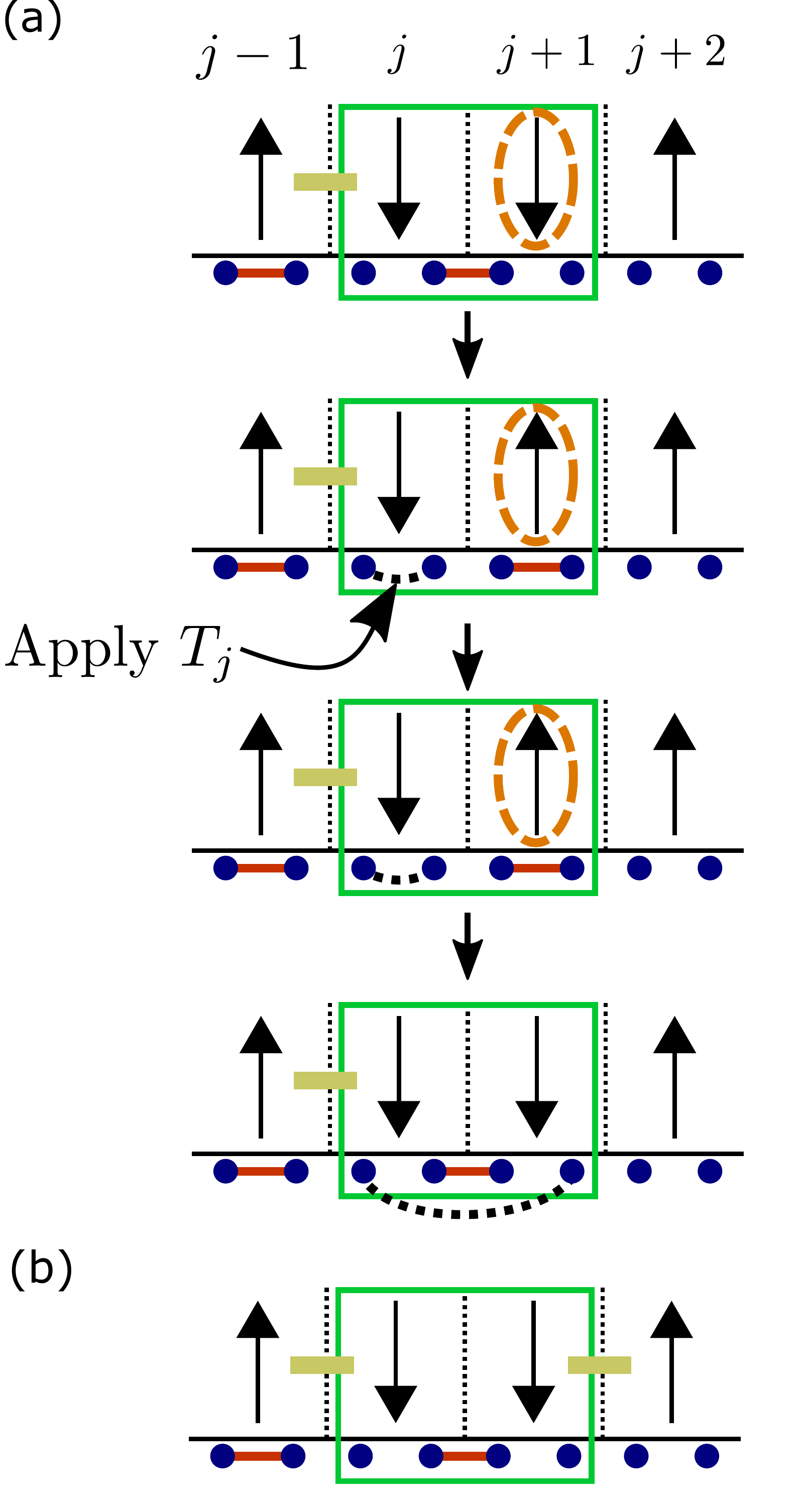}
\caption{
Decimation of a strong ferromagnetic bond leading to a superspin inside the green box [cf.\ Fig.~\ref{fig:tf1}(c)] when one of the spins is already antiferromagnetically locked with a neighbor (indicated by gold line).
(a) Illustration of a process that appears in the third-order in perturbation theory that renormalizes effective $T_{j}'$. (b) The ``problematic configuration'' in which a process illustrated in (a) is not available.}
\label{fig:third_sec3}
\end{figure}

Let us now pinpoint which processes in perturbation theory generate effective $t_{i}$ and $s_{i}$, relegating the detailed justification to Appendix~\ref{app:df}. The non-trivial transformation rule of $t_{i}$ under a ferromagnetic bond decimation is encoded in the last line of Eq.~\eqref{eq:frule}. As we mentioned earlier, there is no term in the second-order perturbation theory that gives a non-trivial $t'_{j}$. However, when $t_{j} \neq 0$, $m_{j+1} = 1$, the following process in the third-order perturbation theory contributes to a non-trivial $t'_{j}$: Flip the spin at site $j+1$, act $T_{j}$, and flip the spin at site $j$ back. We illustrated this virtual process in Fig.~\ref{fig:third_sec3}(a). In particular, the process of flipping the spin at site $j+1$ back and forth is crucial since $T_{j}$ acts as zero on the starting configuration -- it only acts non-trivially after flipping the spin at the site $j$. This contribution is encoded in the first term of the last line of Eq.~\eqref{eq:frule}. The second term on the last line of Eq.~\eqref{eq:frule} is relevant when $t_{j+1}\neq 0$, $m_{j-1} = 1$ and is simply the mirror-inverted version of the case we have just discussed.
 
\begin{figure}
\includegraphics[width=\linewidth]{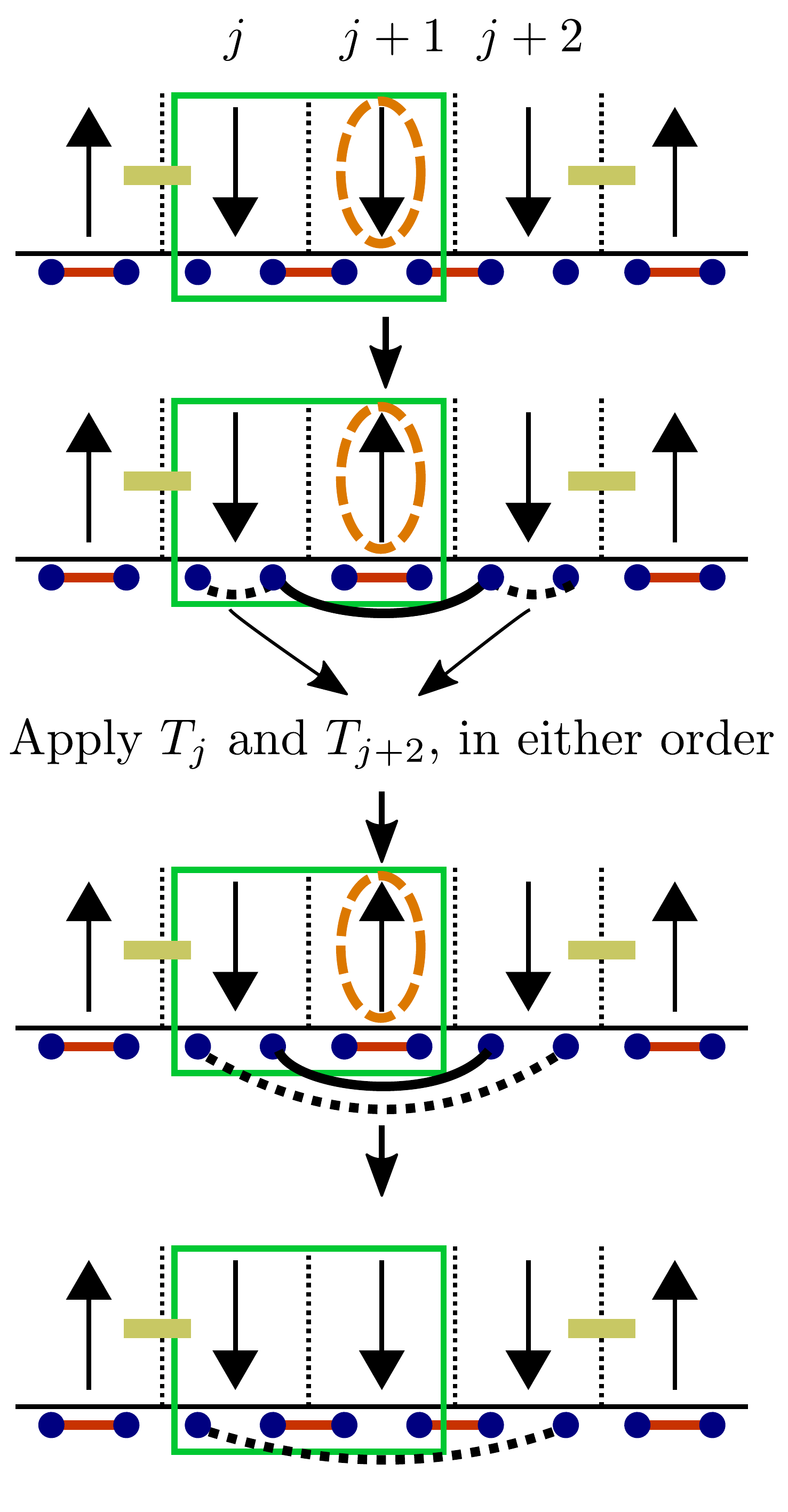}
\caption{RG step decimating strong ferromagnetic bond in the same setting as in Fig.~\ref{fig:third_sec3}. Illustration of a process that appears in the fourth-order in perturbation theory and generates $S_{j}$.
}
\label{fig:fourth_sec3}
\end{figure}

The situation is a little more complicated when $m_{j-1}=m_{j+1}=0$. In this case, as we illustrate in Fig.~\ref{fig:third_sec3}(b), both spins at site $j$ and $j+1$ are locked by gold lines, and there is no spin to flip and generate coupling between Majorana fermions akin to the process described in the previous paragraph. To see how our transformation rule bypasses this issue, we consider a configuration in which two bonds with $m_{i}=1$ label are surrounded by two bonds with $m_{i}=0$ label. Upon decimating one of the bonds with $m_{i}=1$ label, we end up with the ``problematic configuration'' we described in the beginning of the paragraph. To resolve the problem, we generate a term $s_{i}S_{i}$ when one of the two bonds with $m_{i} = 1$ label is decimated. Such a term first appears at the fourth-order in perturbation theory; we illustrate the process in Fig.~\ref{fig:fourth_sec3}. After generating a non-trivial $s_{j}S_{j}$ term whenever one ends up with a configuration $m_{j-1}=m_{j+1} = 0$ and $m_{j} =1$, $s_{j}$ directly enters into $t_{j}$ when the bond $j$ is decimated to form a configuration in which $m_{j}=m_{j+1}=0$, as encoded in the very last term of Eq.~\eqref{eq:frule}. The key feature here is that processes that generate terms like $s_{j}S_{j}$ appear in the higher-order perturbation theory and usually are not incorporated in our transformation rules. However, in the specific situation we illustrated here, $s_{j}S_{j}$ enter as the lowest-order couplings between Majorana fermions at the later stage, and one may incorporate this higher-order perturbation theory result only for this specific situation. 

\subsection{The model with mixed-sign Ising interactions: Results and stability}
\label{sec:sec3result}

 Here, we quantitatively analyze the general consequence of the successive transformations introduced in the previous subsection.
 
 In what follows we define a bond with index $i$ as a bond between sites $i$ and $i+1$.  Consider then an antiferromagnetic bond $i_1$ in the original Hamiltonian, and the closest antiferromagnetic bond $i_2$ to the right of $i_{1}$. Let us follow the transformation rules in Eqs.~\eqref{eq:afrule} and \eqref{eq:frule} to  understand the coupling generated between Majorana fermions at $i_{1}$ and $i_{2}$ at the end of the transformations. At the first step where antiferromagnetic bonds are treated, the transformation generates $t_{i_{1}+1} T_{i_{1}+1}$ with $t_{i_{1}+1} =\frac{h_{i_{1}+1}^{2}}{8\Lambda_{A}}$ and $t_{i_{2}} T_{i_{2}}$ with $t_{i_{2}} = \frac{h_{i_{2}}^{2}}{8\Lambda_{A}}$. After antiferromagnetic bond decimations, ferromagnetic bond decimations combine two sites into a single super-site and renormalize $t_{i_{1}+1}$ and $t_{i_{2}}$ according to the rule from the third-order perturbation theory we described earlier.  Ferromagnetic bond decimation continues until only two bonds between $i_{1}$ and $i_{2}$, say $i_{1,2,s}$ and $i_{1,2,ns}$, remain undecimated.  (Superscripts $s$ and $ns$ stand for smallest and next smallest, referring to the relative magnitude of ferromagnetic Ising interaction for bonds between two antiferromagnetic bonds $i_{1}$ and $i_{2}$; we will assume $J_{i_{1,2,s}} < J_{i_{1,2,ns}}$.) When $i_{1,2,ns}$ is decimated, it generates an $S_{i}$ term, and this term enters back into $T_{i}$ that mediates couplings between Majorana fermions at $i_{1}$ and $i_{2}$ when $i_{1,2,s}$ is decimated. Tracking this procedure, the coefficient $t_{i_{1},i_{2}}$ of the $T_{i}$ term---which at the very end of the transformation couples Majorana fermions at $i_{1}$ and $i_{2}$---takes the following form:
\begin{equation}
\label{eq:nmgeneral}
t_{i_{1},i_{2}} = \alpha(J_{i_{1}+1}, J_{i_{1}+2},\cdots,J_{i_{2}-1}) \frac{ J_{i_{1,2,s}}^{2} \prod_{i=i_{1}+1}^{i_{2}} h_{i}^{2} }{\Lambda_{A}^{2} J_{i_{1,2,ns}} \prod_{i = i_{1}+1}^{i_{2}-1} J_{i}^{2}}.
\end{equation}
 Here $\alpha(J_{i_{1}+1}, J_{i_{1}+2},\cdots,J_{i_{2}-1})$ is a numerical constant that depends on the number of bonds between $i_{1}$ and $i_{2}$ and the order by which the bonds are decimated.
 
 The right hand side in Eq.~\eqref{eq:nmgeneral} depends on the values of $h_{i}$ between $i_{1}$ and $i_{2}$, the values of ferromagnetic Ising interactions, the number of ferromagnetic bonds between $i_{1}$ and $i_{2}$, and the relative ordering of ferromagnetic interaction in terms of magnitude and spatial index. All of these dependencies are random in our setup, and therefore the derived $t_{i_{1},i_{2}}$ is strongly random. Similarly, for any two neighboring antiferromagnetic bonds $i_{n}$ and $i_{n+1}$, $t_{i_{n},i_{n+1}}$ is given by the same formula in Eq.~\eqref{eq:nmgeneral}, with $i_{1}$ and $i_{2}$ replaced by $i_{n}$ and $i_{n+1}$, and depends on same types of random variables. Hence $\overline{t_{i_{n},i_{n+1}}}$ is identical for any two neighboring antiferromagnetic bonds $i_{n}$ and $i_{n+1}$. In particular, there should be in general no dimerization tendency in the derived nearest-neighbor coupling between two Majorana fermions, and the low-energy physics should be described by an infinite-randomness fixed point which is a critical 1D \textit{phase}. 

 It is instructive to apply the symmetry perspective similar to the one we provided at the end of Sec.~\ref{sec:3prelim} to understand what symmetry protects the infinite-randomness fixed point. An Ising spin at the very end of the transformations contains a random number of the original ``UV'' spins, as opposed to the situation encountered in Sec.~\ref{sec:3prelim} in which there is exactly one UV spin between two domain wall Majorana fermions. Nevertheless, there is an IR average translation symmetry $\overline{T}_{x,\text{IR}}$ which together with exact time reversal symmetry $\mathcal{T}$ protects the critical 1D phase. This IR symmetry, albeit distinct from the UV average translation symmetry $\overline{T}_{x,\text{UV}}$ discussed in Sec.~\ref{sec:3prelim}, emerges naturally when the symmetry is present already in the UV (as is natural for condensed matter realizations of 2D topological superconductors). 

 Interestingly, due to the random nature of the signs of Ising interactions, even if one breaks $\overline{T}_{x,\text{UV}}$ by, for example, adding dimerization patterns to $h_{i}$'s, there is no straightforward way for this dimerization tendency to show up in the IR and break $\overline{T}_{x,\text{IR}}$. This observation leads us to conjecture that any spatial modulation in the UV quantities will not translate to spatial modulation in the IR description, and that $\overline{T}_{x,\text{IR}}$ will emerge even when there is no $\overline{T}_{x,\text{UV}}$. However, since $\overline{T}_{x,\text{UV}}$ is a natural one to assume, to avoid any subtle issues that might invalidate our conjecture, we assume $\overline{T}_{x,\text{UV}}$ throughout this paper.
 
One can infer from the above argument that, although there is no obvious way to break the average translation symmetry $\overline{T}_{x}$ in the IR, one may break time-reversal symmetry $\mathcal{T}$ to drive domain-wall Majorana fermions away from criticality and localize them. A straightforward way to break $\mathcal{T}$ is by adding a uniform Zeeman field (which without loss of generality we assume  energetically favors up spins). From a more microscopic viewpoint, one can envision that a Zeeman field yields the following two effects:
\begin{itemize}
    \item It generally costs more energy to flip spins from up to down than down to up when the Zeeman field is added. Additionally, recall that couplings between Majorana fermions are generated by processes of flipping spins back and forth. With a uniform Zeeman field, it is therefore generally expected that Majorana fermions couple more weakly across the up-spin domains than across down-spin domains---effectively dimerizing  the Majorana fermion couplings. This effect is analogous to what we saw at the end of Sec.~\ref{sec:3prelim} for the special case where all Ising interactions are uniformly antiferromagnetic.
    \item It is energetically favorable for some of the spins pointing down before adding the Zeeman field to reverse their alignments, producing a small net magnetization. Cast in a slightly different language, the up-spin domains are correspondingly longer on average than the down-spin domains. Notice that in Eq.~\eqref{eq:nmgeneral}, the couplings are on average weaker when the domain length is longer. Hence, this mechanism also makes couplings across the down-spin domains more dominant. This effect originates from the random nature of ferromagnetic couplings and does not have any analogue in the discussion from Sec.~\ref{sec:3prelim}.
\end{itemize}
These two effects both favor the Majorana fermion couplings across the down-spin domains, so we conclude that the explicit time-reversal symmetry breaking through a uniform Zeeman field destabilizes the infinite-randomness fixed point and leads to a localized phase.

\section{Strong-disorder RG}  
\label{sec:sdrg}
 
In the previous section, we pinned the spin degrees of freedom with random nearest-neighbor Ising interactions and studied the effective Hamiltonian governing Majorana fermions that live on the spin domain walls. We observed that these Majorana fermions in general form an infinite-randomness fixed point and a uniform Zeeman field that explicitly breaks the time-reversal symmetry does localize the Majorana fermions in our setup, reminiscent of gapping out topological edge/surface states by breaking symmetry. 

 These results suggest a tantalizing picture wherein the infinite-randomness fixed point of Majorana fermions can exist as \textit{a stable phase} on the edge of 2D time-reversal-invariant topological superconductors as long as the time-reversal symmetry is preserved in the microscopic Hamiltonian but is spontaneously broken in the phase. However, in the example we considered in the previous section, we imposed a rather arbitrary limit on how one chooses parameters, with the goal of clarifying the physics of the infinite-randomness fixed point. Here, we continue our journey of studying the Hamiltonian
\begin{equation}
\label{eq:prelimsdrg}
H = - \sum_{j} J_{j} \sigma_{j}^{z} \sigma_{j+1}^{z} - \sum_{i} h_{i} F_{i},
\end{equation}
 now without assuming $h_{i} \ll J_{j}$, by developing a strong-disorder RG analysis. The strong-disorder RG is simple when $J_{j}$ is forbidden to be antiferromagnetic---in this case, the only strong-disorder RG fixed points are associated with ferromagnetic Griffiths phases. Strong-disorder RG with $J_{j}$ allowed negative is in general complicated, but we provide an argument that one may truncate some of the terms that complicate the analysis in the limit when the bonds with antiferromagnetic Ising interactions are \textit{rare}. We provide theoretical arguments and numerical implementations of the strong disorder RG to show that introducing dilute antiferromagnetic bonds in the UV alters the IR physics completely; in this case, the IR physics is governed by the infinite-randomness fixed point of Majorana fermions. This result suggests that a strongly disordered ferromagnet on the edge of topological superconductor is unstable to introducing antiferromagnetic bonds and that the infinite-randomness fixed point of Majorana fermions appears as a stable, \textit{generic} edge phase when disorder is strong. 

\subsection{The case without antiferromagnetic Ising interaction}
\label{sec:prelimsdrg}

 We first take a look at the Hamiltonian Eq.~\eqref{eq:prelimsdrg} from the strong-disorder RG viewpoint when all $J_{j} \geq 0$, i.e., nearest-neighbor Ising interactions are forbidden to be antiferromagnetic. To study this Hamiltonian from the strong-disorder RG perspective, we envision the following set of transformations to the 1D chain, applicable when the parameters in the Hamiltonian are strongly random: At each step the energy scale is defined to be $\Omega = \max \{ J_{i}, h_{i}\}$. If $\Omega = J_{j}$ for some $j$, then under the strong randomness assumption, couplings around the bond $j$ except for the strongest Ising interaction $-J_{j} \sigma_{j}^{z} \sigma_{j+1}^{z}$ are expected to be weak. As an approximation that captures the low-energy physics, one can implement the ferromagnetic bond decimation transformation employed in the previous section, wherein we combine two sites $j$ and $j+1$ into a single super-site. Next one can employ degenerate second-order perturbation theory to derive new effective couplings of the Hamiltonian after the transformation (we fix $p_{i}$ in the definition of the flip term Eq.~\eqref{eq:flipdef} to be $p_{i}=2^{-1/4}$, as explained in Sec.~\ref{sec:localterms}): 
\begin{equation}
\label{eq:prelimsdrgrule1}
\begin{split}
& J'_{i} = \begin{cases}
J_{i} & i<j-1 \\
J_{j-1} + \frac{(\sqrt{2}-1) h_{j}^{2}}{4 \sqrt{2} J_{j-1}} & i=j-1 \\
J_{j+1} + \frac{(\sqrt{2}-1) h_{j+1}^{2}}{4 \sqrt{2} J_{j-1}}  & i=j \\
J_{i+1} & i >j
\end{cases}\\
& h'_{i} = \begin{cases}
h_{i} & i < j \\
\frac{2^{-1/4} h_{j}h_{j+1}}{J_{j}} & i = j \\
h_{i+1} & i > j
\end{cases}
\end{split}
\end{equation}

This transformation rule is very similar to that appearing in Eq.~\eqref{eq:frule} for $h_{i}$ and $J_{i}$; the only difference is the extra $ \frac{(\sqrt{2}-1) h_{j}^{2}}{4 \sqrt{2} J_{j-1}}$ and $\frac{(\sqrt{2}-1) h_{j+1}^{2}}{4 \sqrt{2} J_{j-1}}$ contributions to the effective Ising interactions in the second and the third lines. In the previous section, we dropped such terms due to the assumption $h_{i} \ll J_{i}$. We are not imposing this limit here, however, so one must now include these terms to capture the correct low-energy physics. We will later see that these contributions play a crucial role in determining the IR physics.

If $\Omega = h_{j}$, one may similarly invoke the strong-randomness assumption and project the system into eigenstates with the \textit{two} lowest eigenvalues of $-h_{j} F_{j}$ to capture the low-energy physics. Both of these states contain an equal superposition of $\sigma_{j}^{z} =\, \uparrow$ and $\sigma_{j}^{z} =\, \downarrow$, i.e., the corresponding Ising spin is disordered. This projection can be understood as effectively removing site $j$ and hence is dubbed site decimation. Zeroth and first-order perturbation theory yields the following transformation rule:
\begin{equation}
\label{eq:prelimsdrgrule2}
\begin{split}
& J'_{i} = \begin{cases}
J_{i} & i<j-1 \\
\frac{(1-2^{-1/4}) h_{j}}{2} & i=j-1 \\
J_{i+1} & i >j-1
\end{cases}\\
& h'_{i} = \begin{cases}
h_{i} & i < j-1 \\
\frac{1+2^{-1/4}}{2} h_{j-1} & i = j-1 \\
\frac{1+2^{-1/4}}{2} h_{j+1} & i = j \\
h_{i+1} & i > j
\end{cases}
\end{split}
\end{equation}
The $J'_{j-1} = \frac{(1-2^{-1/4}) h_{j}}{2}$ piece reflects the zeroth order term, i.e., the difference between the two lowest eigenvalues of $-h_{i}F_{i}$. The transformation rule for $h_{i}$ corresponds to the first-order correction. 

Successive transformations in which one finds the dominant local terms and performs site decimations or bond decimations define a real-space RG procedure where the energy scale $\Omega$ monotonically decreases. Studying how the energy scale, Hamiltonian, and distribution of couplings in the Hamiltonian evolve under the RG flow reveals the fate of the system in the IR. The RG transformations we introduce here are very similar to those used to study the 1D random transverse-field Ising model; in particular, the flip terms in our model play a similar role as  transverse-field terms in the random transverse-field Ising model in the sense that both induce a transformation which removes a site from the 1D chain. However, the nature of couplings generated/renormalized from our RG rules differs from the random-transverse-field Ising model, and therefore the IR physics differs as well. 

Let us heuristically see what happens under this RG transformation when $h_{i}$ and $J_{i}$ are strongly random. Decimation of a site with $h_i$ creates a ferromagnetic bond $J'_{i-1} = \frac{1-2^{-1/4}}{2} h_{i}$, which is still $O(h_{i})$, and under the standard strong randomness assumption, is likely to be a strong coupling as well. Hence, any site decimation is likely to be followed by another bond decimation, and it suffices to see what happens for the bond decimation. In our case, the bond decimation not only generates a super-site flip term of $O\left( \frac{h_{i}h_{i+1}}{J_{i}} \right)$, but also generates extra ferromagnetic Ising interactions of order $O\left( \frac{h_{i}^{2}}{J_{i}} \right)$ and $O\left( \frac{h_{i+1}^{2}}{J_{i}} \right)$. Strong randomness implies signifcant likelihood that either $h_{i} \gg h_{i+1}$ or $h_{i+1} \gg h_{i}$, i.e., coefficients of the two neighboring flip terms are likely very different. One of the generated ferromagnetic Ising interactions is therefore likely much stronger than the super-site flip term, so that this super-site Ising spin is likely to be flanked by a strong Ising interaction bond. This process continues, and deep in the RG flow, one will always find some neighboring Ising interactions that dominate the flip terms. The decimation procedure will thus be eventually dominated by ferromagnetic bond decimations, even if the UV Hamiltonian has $ \overline{\ln h_{i}} \gg \overline{\ln J_{j}}$. \textit{Thus, when couplings are strongly random and all Ising interactions are ferromagnetic, regardless of the initial conditions, the system is expected to flow to the ferromagnetic Griffiths phase, with the locally disordered spins giving rise to Griffiths effects.} In sharp contrast, the random transverse field Ising model additionally supports a disordered phase in which spin flip terms, or equivalently transverse-field terms, dominate.
 
We numerically implemented this RG procedure to confirm the above picture. In our implementation, we chose the couplings $h_{i}$ and $J_{i}$ from power-law distributions $P(h) =r_{h} h^{-1 + r_{h}}$ and $P(J) =r_{J} J^{-1 + r_{J}}$, with $0 < r_h, r_J < 1$ and $r_{h}, r_{J} \in \{ 0.05,0.1, 0.15, 0.2, 0.25\}$.  We started with system size $L = 5 \times 10^{5}$ and performed the RG transformation until $L=100$. 
In the Griffiths phase of our interest, the distribution of the log couplings is expected to follow
 \begin{equation}
 \label{eq:griffithscaling}
 \begin{split}
& P(\eta) = \alpha e^{-\alpha \eta}, \, P(\theta) = \beta e^{-\beta \theta} \, \\
& \alpha \rightarrow 0, \beta \rightarrow \frac{1}{z} \text{ as $\Omega \rightarrow 0$},
\end{split}
 \end{equation}
where $z$ is a non-universal dynamical exponent. The limiting behavior of $\alpha$ and $\beta$ can be probed by 
 \begin{equation}
 \delta_{h} =  \frac{\overline{\eta_{i}^{2}} - \overline{ \eta_{i} }^{2}}{\overline {\eta_{i}}}, \,  \delta_{J} = \frac{\overline{\theta_{i}^{2}} - \overline{\theta_{i} }^{2}}{\overline{ \theta_{i}}},
 \end{equation}
 which can be directly equated to $1/\alpha$ and $1/\beta$ assuming that $\eta_{i}$ and $\theta_{i}$ follow the exponential distributions of Eq.~\eqref{eq:griffithscaling}. Additionally, the number of remaining sites $L_{\Omega}$ and the energy scale $\Omega$ at a given RG step are expected to follow the scaling relation 
\begin{equation}
L_{\Omega}\sim \Omega^{\frac{1}{z}}
\end{equation}
in the IR. 

\begin{figure}  
    \includegraphics[width=\linewidth]{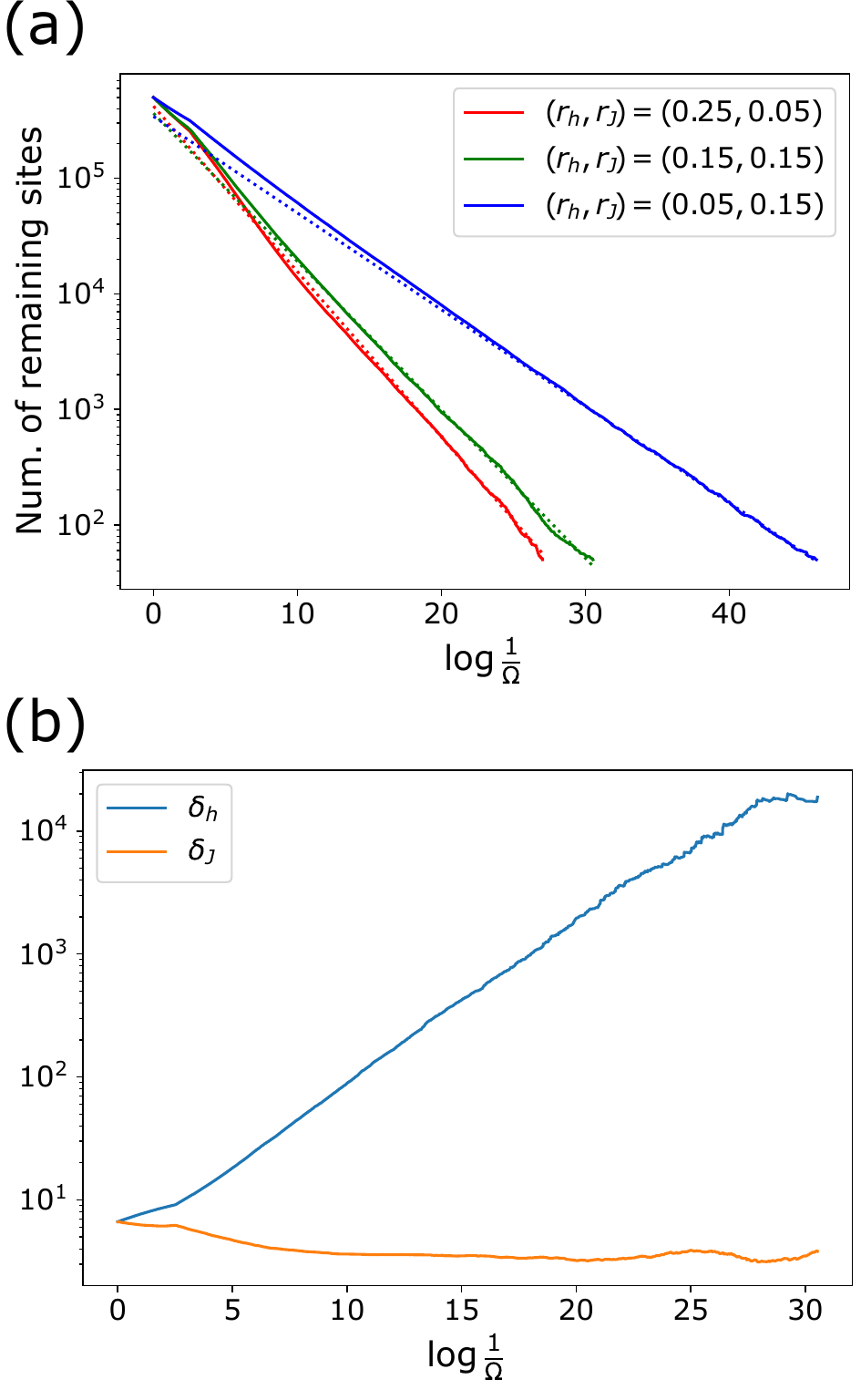}
    \caption{(a) Number of remaining sites versus $\log \frac{1}{\Omega}$ for three different initial conditions. (b) Evolution of $\delta_{h}$ and $\delta_{J}$ for the initial condition $(r_{h}, r_{J}) = (0.15,0.15)$.}
    \label{fig:onlyferro_graph}
\end{figure}

Figure~\ref{fig:onlyferro_graph}(a) plots the number of the remaining sites versus $\log(1/\Omega)$ for the selected initial distributions of $h_{i}$ and $J_{i}$. Dotted lines are obtained by performing the least linear squares fit on $\log L_{\Omega}$ and $\log(1/\Omega)$ for the last 1000 data points. One can clearly see that the scaling $L_{\Omega} \sim \Omega^{\frac{1}{z}}$ indeed emerges in the low-energy limit. 

Meanwhile, Fig.~\ref{fig:onlyferro_graph}(b) shows the evolution of $\delta_{h}$ and $\delta_{J}$ versus $\log(1/\Omega)$ for $(r_{h},r_{J}) = (0.15,0.15)$. Recall that $\delta_{h}$ and $\delta_{J}$ are proportional to $1/\alpha$ and $1/\beta$ in Eq.~\eqref{eq:griffithscaling}, and as the RG proceeds, $\alpha$  monotonically decreases to zero while $\beta$ saturates to some constant. One can see that $\delta_{J}$ retains relatively constant values, but $\delta_{h}$ rapidly increases, confirming the aforementioned expectations. We checked that these behaviors emerge in other initial conditions and in different realizations of disorder as well. Our numerical simulations thus strongly support our prediction that the ferromagnetic Griffiths phase always emerges in the strong-disorder limit of the Hamiltonian in Eq.~\eqref{eq:prelimsdrg} without antiferromagnetic Ising interactions, even when the ferromagnetic interactions are very weak in the UV. The physics is that such ferromagnetic interactions are always generated when decimating the flip terms and never renormalize down under subsequent RG steps and eventually become the dominant terms.

\subsection{Strong-disorder RG involving antiferromagnetic Ising interaction: Overview}
\label{sec:sec4sdrgoverview}

Building on the above result, we next explore the effect of introducing antiferromagnetic bonds into the model, beginning with some general features of the strong-disorder RG when one incorporates antiferromagnetic Ising interactions into the strong-disorder RG procedure. We can similarly define the strong-disorder RG procedure on the Hamiltonian by first setting the energy scale at each step to be $\Omega = \max \{ h_{i}, |J_{i}| \}$ (the absolute value is inserted because $J_{i}$ can take either sign).  Let us see what kind of transformations one should invoke when the energy scale is set by the antiferromagnetic Ising interaction, i.e., $\Omega = |J_{i}|$ with $J_{i}<0$. We proceed with the standard strong-disorder assumption and assume that one may project the system into the lowest-eigenvalue configurations of $+J_{i} \sigma_{i}^{z}\sigma_{i+1}^{z}$. In our case, the two Ising spins at site $i$ and $i+1$ are anti-aligned to each other always. Then, there is always one free Majorana fermion that lives on a bond between site $i$ and $i+1$, so in general, one is \textit{not} allowed to do any operations that combine a site $i$ and $i+1$ into a single site as one can do for ferromagnetic bond decimations. Instead, one can follow a similar procedure employed in Sec.~\ref{sec:sec3rule} to treat strong antiferromagnetic bonds: Introduce a binary label $m_{i}=0,1$ to each bond. If $m_{i} =0$ on a bond $i$, the spins at site $i$ and $i+1$ neighboring the bond $i$ are hard-constrained to be anti-aligned to each other. At the start of the RG, $m_{i}=1$ for all bonds, and all spins are ``free''. However, as one encounters strong antiferromagnetic bonds in the RG steps, some of the bonds will be transformed to carry a label $m_{i} =0$. This transformation associated with antiferromagnetic bonds does not reduce the system length but reduces the degrees of freedom in the system and fits into a standard paradigm of the RG transformation. 
 
After performing the transformation, one may incorporate couplings generated from the second-order degenerate perturbation theory to the Hamiltonian. First, assuming that all bonds near the antiferromagnetic bond $i$ on which we apply the transformation have label $m_{j} =1$, the second-order degenerate perturbation theory generates the following terms:
\begin{itemize}
    \item The process that flips the spin at site $i$ back and forth generates nearesst-neighbor ferromagnetic Ising interaction $-\tilde{J}_{i-1} \sigma_{i-1}^{z}\sigma_{i}^{z}$. A similar process involving the spin $i+1$ generates $-\tilde{J}_{i} \sigma_{i}^{z}\sigma_{i+1}^{z}$.
    \item The same process as the above also generates $-\tilde{t}_{i}T_{i}$ and $-\tilde{t}_{i+1}T_{i+1}$. This term represents a coupling between a Majorana fermion trapped by the bond with $m_{i}=0$ and the rest of the system.
    \item The process that flips the spin at site $i$ and then the spin at site $i+1$ (or vice versa) generates a term that flips two spins at site $i$ and $i+1$ simultaneously. This ``double-spin flip term'', in contrast to the term that flips the spin at site $i$ or $i+1$ individually, does not violate the constraint imposed by $m_{i} =0$ and is a legal term to add to the Hamiltonian after the transformation.
\end{itemize}
The nearest-neighbor Ising interactions are already in the Hamiltonian Eq.~\eqref{eq:prelimsdrg}; the two other terms generated are not, however.   Hence, one generally has to incorporate new types of couplings that are not present in the original Hamiltonian to capture the correct low-energy physics using the strong-disorder RG procedure.
 
The terms described in the second bullet point are relatively easy to incorporate into the RG but will be crucial in recovering the correct IR physics. On the other hand, the terms in the third bullet point hint at the difficulty of studying this RG procedure in full generality. In the putative strong-disorder RG procedure one might potentially develop, one would ``integrate out'' strong antiferromagnetic bonds at site $j$ by changing the label $m_{j}$ from 1 to 0, and in a later RG step generate a cluster consisting of multiple antiferromagnetically aligned spins---each bond linking two neighboring spins in this cluster carrying the label $m_{i} = 0$. Upon keeping all the second-order degenerate perturbation theory results at each step of the transformation, one then encounters a highly non-local term that flips the whole cluster of spins at once, the natural multi-site generalizationsof the double-spin flip terms in the third bullet point. Additionally, imagine, for example, that the current energy scale in the RG procedure is set by a strong ferromagnetic bond neighboring the cluster of spins we just mentioned before. Upon performing the ferromagnetic bond decimation and including terms generated from the second-order degenerate perturbation theory, one includes a term generated from a process where one flips the whole cluster of spins back and forth. This process generates non-local interactions between Majorana fermions that live on the domain walls of the multi-spin clusters. Hence, if one implements the strong-disorder RG through the naive generalization of keeping all the terms generated from the second-order perturbation theory, one has to keep track of an infinite number of non-local couplings, and the RG flow is intractable. 

While the naive RG procedure generates non-local couplings, it is conceivable that in some parameter regime, such non-local couplings are suppressed, and one may capture the correct physics by only keeping track of a limited set of local couplings. In the rest of this section, we explore the limit where antiferromagnetic bonds in the Hamiltonian Eq.~\eqref{eq:prelimsdrg} exist but are rare in the UV. Physically, this limit represents a scenario where we deform a dirty ferromagnet on the edge of the topological superconductor, whose physics from the strong-disorder RG is covered in the previous subsection, by introducing a small amount of antiferromagnetic bonds. We further argue that one may capture low-energy physics without keeping track of the aforementioned non-local couplings in Appendix~\ref{app:more}. 

 Thus we keep track of a finite number of local couplings to capture the low-energy physics in our RG procedure. Now the question is what is the low-energy fate of the putative strong-disorder RG procedure. Two natural scenarios arise: 
\begin{enumerate}
    \item As we will see in the next subsection, some of the decimation procedures that we introduce due to presence of additional types of local terms can remove bonds with the label 0. Hence, one can envision a scenario in which even though bonds with the labels $m_{i}=0$ are generated at some points in the strong-disorder RG, such bonds may get ``screened'', and in later steps of RG, the proportion of bonds with $m_{i}=0$ becomes vanishingly small. In this scenario we expect that the physics is governed by fixed points identical to the case where there are no antiferromagnetic bonds, i.e., ferromagnetic Griffiths fixed points.
    \item The bonds with $m_{i}=0$ labels survive down to low energies, and in the later steps of RG, the system is filled with $m_{i}=0$ bonds. This scenario represents a case in which spin degrees of freedom are frozen by Ising interactions with random signs, and the low-energy physics is governed by Majorana fermions living on the domain walls.
\end{enumerate}
In the next subsection, we will develop the strong-disorder RG in more detail and also present our numerical implementation.  There we will see that scenario 2 prevails. In particular, recall that even if the system is completely filled with $m_{i}=0$ bonds at a later step of the strong-disorder RG, each spin actually represents a random number of spins linked by strong ferromagnetic bonds. Hence, the symmetry argument at the end of Sec.~\ref{sec:sec3result} tells us that effective couplings between Majorana fermions have emergent average translation symmetry; correspondingly, whenever scenario 2 prevails, the Majorana fermions form an infinite-randomness fixed point.
 
\subsection{The RG procedure}

In the last subsection, we focused on a heuristic picture of how our strong-disorder RG works upon incorporating the effect of antiferromagnetic bonds and highlighted the physics that emerges. Here, we provide more precise prescriptions for the strong-disorder RG procedure. At each step, we keep track of $\{ m_{i} \}$, the binary labels on bonds mentioned earlier, and the couplings $\{ h_{i}, J_{i}, t_{i}, s_{i} \}$ that together parametrize the Hamiltonian according to
\begin{equation}
\label{eq:fullRGham}
\begin{split}
& H(\{ m_{i}, h_{i}, J_{i}, t_{i}, s_{i} \} ) = -\sum_{i} h_{i}m_{i-1}m_{i} F_{i} \\
& -\sum_{i} \frac{h_{i} h_{i+1}}{|J_{i}|} (1-m_{i}) m_{i-1} m_{i+1} \frac{F_{i}F_{i+1}+ F_{i+1}F_{i}}{2} \\
& - \left( \sum_{i} J_{i} \sigma_{i}^{z}\sigma_{i+1}^{z} + 2t_{i} T_{i}  +2 s_{i} S_{i} \right) .
\end{split}
\end{equation}
The first line represents the usual single-site flip terms, which are nonzero only when $m_{i-1} =m_{i} = 1$; recall that if $m_{i-1}$ or $m_{i}$ is $0$, site $i$ is hard-constrained to be anti-aligned with another neighboring spin, and a flip term that flips a single spin at site $i$ is not allowed. The second line is a double-site flip term that appears when $m_{i} =0$ but $m_{i-1} = m_{i+1}=1$. This pattern of $m$'s represents a situation in which two spins at site $i$ and $i+1$ are hard-constrained to be anti-aligned, but are not further constrained. Hence, we allow a term that flip two spins simultaneously. As we mentioned in the previous subsection, we omit terms that flip three or more spins simultaneously.
The remaining three terms in the Hamiltonian were already introduced and encode nearest-neighbor Ising interaction as well as first- and second-neighbor Majorana fermion hopping. We choose $t_{i} \neq 0$ only when $m_{i-1} =0$ or $m_{i} = 0$; also, as in Sec.~\ref{sec:sec3rule} $s_{i}\neq 0$ only for the specific configuration $m_{i}=1$ and $m_{i-1}=m_{i+1} = 0$. The precise reasoning behind this choice can be found in the bullet points in the first part of Appendix~\ref{app:rule}. Note that at the initial stage, $m_{i}=1$ for all bonds, so the Hamiltonian only contains single-site flip terms and nearest-neighbor Ising interactions---consistent with the UV Hamiltonian given in Eq.~\eqref{eq:prelimsdrg}. However, the three other terms in the above Hamiltonians are naturally generated under the RG when there are strong antiferromagnetic bonds.

\begin{figure}
\includegraphics[width=1.0\linewidth]{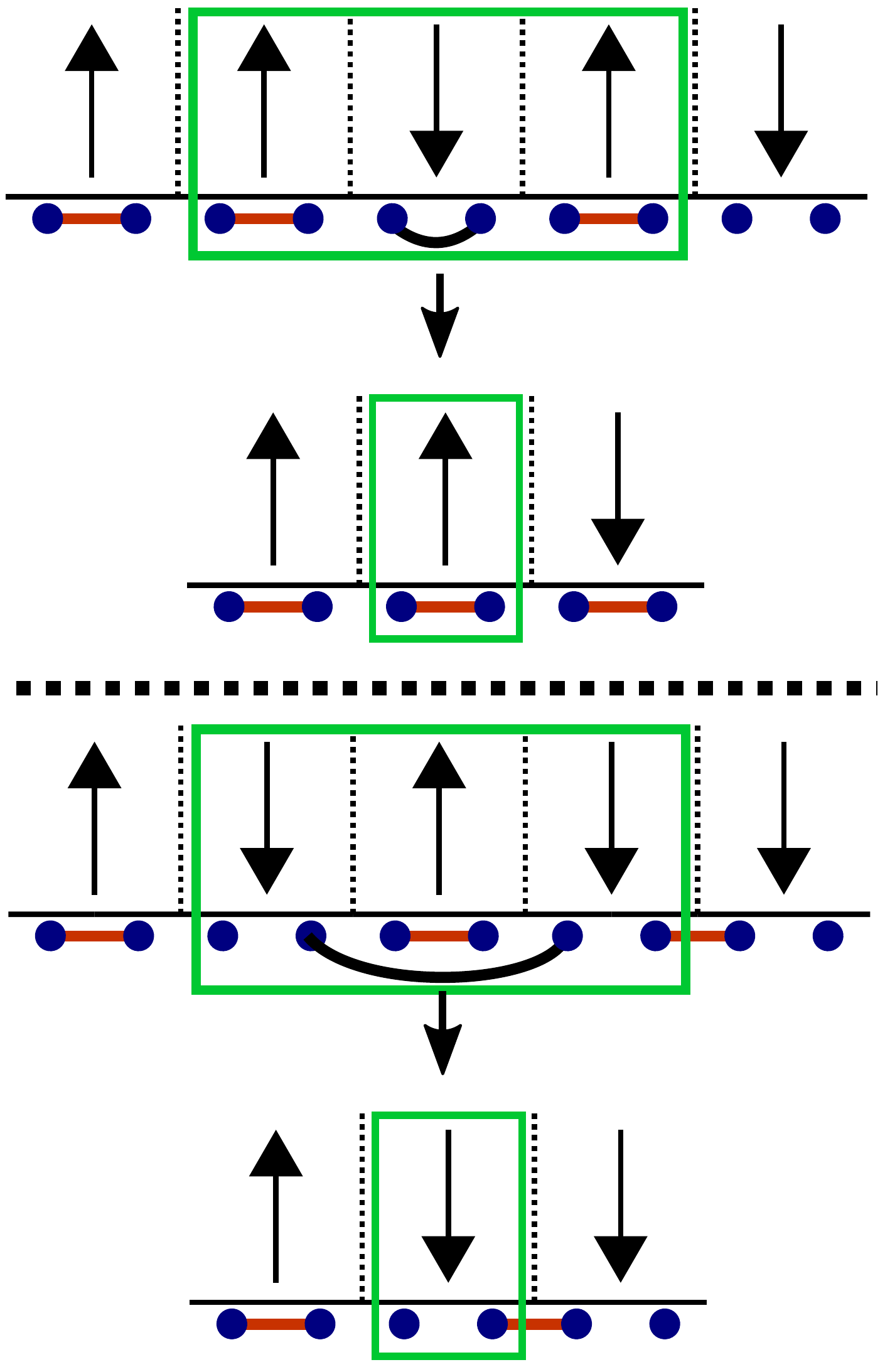}
\caption{Graphical illustration of the transformation we implement when $\Omega = t_{i}$. The three sites squared in green in the upper row, on which $T_{i}$ acts, are grouped into a single site after the transformation. Black lines denote specific Majorana fermion states one projects into upon performing the decimation.}
\label{fig:tf2}
\end{figure}

The energy scale at each RG step is given by 
\begin{equation}
\begin{split}
\Omega & = \max \{ m_{i} | J_{i} | \} \cup \{ t_{i} \} \cup \{ s_{i} \} \cup  \{ m_{i}m_{i-1} h_{i} \}\\
& \quad \quad \quad \quad \cup  \{ d_{i} = m_{i-1}(1-m_{i})m_{i+1} 
\frac{\sqrt{2} + 1}{2 \sqrt{2} }\frac{h_{i-1}h_{i}}{ |J_{i}|}  \},
\end{split}
\end{equation}
i.e., it is the largest coefficient for local terms among Ising interactions for $m_{i}=1$ bonds, $-t_{i}T_{i}$, $s_{i}S_{i}$, and single/double site flip terms. The numerical constant in the definition of $d_{i}$ is added so that each candidate energy scale is set by half of the difference between the lowest energy eigenvalue and the highest energy eigenvalue of each local term.
Similar to the procedure detailed in Sec.~\ref{sec:prelimsdrg}, between each RG step, we project our system into the lowest-energy eigenstate of the local term that sets the energy scale $\Omega$ and sequentially eliminate degrees of freedom. Perturbation theory then dictates how the couplings in the system renormalize after the transformation. The key difference from the earlier consideration in Sec.~\ref{sec:prelimsdrg} is that we have more types of local terms and correspondingly a broader set of RG transformations to invoke depending on the dominant term.  

When the energy scale is set by antiferromagnetic Ising interaction, i.e., $\Omega = |J_{i}|$ with $J_{i} < 0$, we change the bond label $m_{i}$  from $1$ to $0$, marking that now the Ising spins at site $i$ and $i+1$ are antiferromagnetically linked, as described earlier. If $\Omega=d_{i}$, we perform a similar projection to the case with $\Omega = h_{i}$, projecting out all states but those with the two lowest eigenvalues of the double-site flip term associated with $d_{i}$. In these eigenstates, the antiferromagnetically linked Ising spins at $i$ and $i+1$ remain disordered, so the projection may be equivalently thought as removing the two sites $i$ and $i+1$ and hence will be called a double-site decimation. 

 For $\Omega = t_{i}$ or $\Omega = s_{i}$, we project onto the states with the lowest eigenvalues of $-t_{i}T_{i}$ or $-s_{i} S_{i}$, respectively.  In the case of $-t_{i}T_{i}$, these lowest-eigenvalue states have $\sigma_{i-1}^{z} =\sigma_{i+1}^{z} =\, \uparrow$ and $\sigma_{i}^{z} =\, \downarrow$, or $\sigma_{i-1}^{z} = \sigma_{i+1}^{z} =\, \downarrow$ and $\sigma_{i}^{z} =\, \uparrow$; we group the three sites $i-1$, $i$, and $i+1$ into a single site whose Ising spin is given by $\sigma_{i-1}^{z} =\sigma_{i+1}^{z}$. Figure~\ref{fig:tf2} illustrates this transformation. Similarly, in the $-s_{i} S_{i}$ case, we group four sites involved in $S_{i}$ into a single site, whose Ising spin is given by $\sigma_{i-1}^{z} =\sigma_{i+2}^{z}$. These transformations may be viewed as removing two bonds (when $\Omega = t_{i}$) or three bonds (when $\Omega = s_{i}$) and grouping the surrounding sites into a single site. Hence, we will respectively refer to these transformations as double-bond and triple-bond decimations. 
 
In the main text, we skip how the parameters in the Hamiltonian are transformed after each RG step, instead relegating the full transformation rule and its derivation to Appendix~\ref{app:rule}. Here we simply point out key features of the transformation rules that give insight into the numerical result of our strong disorder RG procedure presented in the next subsection. 

First, in keeping track of $h_{i}$ and $J_{i}$, going to first order (if $\Omega = s_{i}$, $\Omega = h_{i}$, $\Omega = d_{i}$, or $\Omega = t_{i}$ with $m_{i-1}=m_{i}=0$) or second order in perturbation theory (if $\Omega = |J_{i}|$ or $\Omega = t_{i}$ with $m_{i-1}m_{i} =0$) suffices. However, to keep track of $s_{i}$ and $t_{i}$, one often needs to go to higher-order in perturbation theory. While keeping track of $s_{i}$ and $t_{i}$ involves higher-order corrections, they still provide the lowest-order route to the nearest-neighbor Majorana couplings within the spin cluster linked by $m_{i}=0$ bonds. Upon following the argument  presented in Appendix~\ref{app:more}, these nearest-neighbor couplings, although from higher-order corrections, dominate over non-local couplings generated from second-order perturbation theory involving multi-site flip terms in the limit we are considering in which antiferromagnetic bonds in the UV Hamiltonian are rare. 

  Also, the fact that $t_{i}$ and $s_{i}$ come from higher-order perturbation theory has an interesting implication on the expected RG flow. First, we observe that the double- and triple-bond decimation induced when $\Omega = t_{i}$ or $s_{i}$ removes some bonds with the label 0 and potentially screen strongly antiferromagnetic bonds. In contrast, single-site and single-bond decimations do not contribute to such screenings, and in fact increase the proportion of bonds with 0 label in the system. Meanwhile, the perturbatively generated couplings $h_{i}$ and $J_{i}$ will be much larger than $s_{i}$ and $t_{i}$ due to the fact that they originate from lower-order processes. The RG procedure is therefore dominated by single-site and single-bond decimations initially, and on the way, the proportion of bonds with label 0 increases. Scenario 2 mentioned at the end of Sec.~\ref{sec:sec4sdrgoverview} prevails via this mechanism.  
  
Second, our RG transformation rule \textit{does not} guarantee that the RG energy scale decreases monotonically, due to the following two processes: 
 \begin{enumerate}
     \item When decimating a single site, there is a chance that the generated $s_{i}$ is larger than the energy scale $\Omega$ -- see the second line of Eq.~\eqref{eq:hisi}.
     \item Assume that $m_{j-2} = m_{j+1} = 1$ and $m_{j-1}=m_{j}=0$, so that sites $j-1$, $j$, and $j+1$ are linked by $m_{i}=0$ bonds. Here there is a three-site flip term associated with this cluster of sites. While we do not choose to include this three-site flip term as a candidate for the RG energy scale, there can be occurrences where it exceeds $\Omega$. If this is the case, there is a possibility that after double-bond decimations involving the sites $j-1$ and $j$, or sites $j-2$ and $j-1$, this three-site flip term becomes a single-site or double-site flip term that enters the energy scale and is larger than the previous energy scale. One can envision a similar scenario involving four-site instead of three-site flip terms.
 \end{enumerate}
 For the first case, the generation of couplings larger than the original energy scale $\Omega$ does occur in different contexts (most notably, strong-disorder RG schemes for antiferromagnetic 1D spin chains with $S \geq 1$ \citep{Hyman1997,Monthus1997,Damle2002,Refael2002}), but in theses cases, it is understood that such events are suppressed near the strong-disorder fixed points. We will see that in our case as well, these events are very rare, and that our RG scheme also correctly captures the low-energy physics. 
 
 As for the second case, this possibility is associated with our choice to only keep track of single-site flip terms and double-site flip terms. Had we chosen to keep track of all possible flip terms that flip any number of spins linked by $m_{i}=0$ bonds, the second possibility would not appear. One interesting observation is that, if our approximation that assumes irrelevance of multi-site flip terms is valid, violation in the RG scale monotonicity due to the multi-site flip term contribution should be suppressed. Hence, whether the RG scale monotonically decreases or not also serves as an indirect test of the validity of our approximation. We will see from numerical implementations in the next subsection that the RG scale behaves more monotonically upon decreasing the proportion of antiferromagnetic bonds in the initial Hamiltonian, supporting our claim that our RG procedure is justified when antiferromagnetic bonds in the initial Hamiltonian are rare.  
 
\subsection{Numerical RG results}
 
  We turn now to  numerical implementation of the strong-disorder RG procedure described in the previous two subsections. In our simulations, we choose the flip term parameter $h_{i}$ in the Hamiltonian Eq.~\eqref{eq:prelimsdrg} randomly from the power-law distribution 
  \begin{equation}
      P_{h}(h_{i}) = \begin{cases}
      r_{h} h_{i}^{-1+r_{h}} & 0 < h_{i} < 1 \\
      0 & \text{otherwise}
      \end{cases},
  \end{equation} 
  as done earlier in Sec.~\ref{sec:prelimsdrg}. For the Ising interaction coefficients, with probability $p_{\text{AF}}$ we set $J_i$ to be antiferromagnetic ($J_{i}<0$), and choose its magnitude randomly from either the uniform distribution $(0,1)$ or from the same distribution as for the flip-term parameter $h_{i}$ with $r_{J} = r_{h}$ for simplicity. With probability $(1-p_{\text{AF}})$, we simply set $J_i = 0$. That is, in our numerical implementation, there are no ferromagnetic Ising interactions in the initial Hamiltonian: all ferromagnetic Ising interactions that appear in the middle of the RG procedure are dynamically generated. Recall also that our RG procedure is valid in the limit $p_{\text{AF}} \ll 1$. We benchmarked the RG procedure with the initial system size of $L =2 \times 10^{6}$, $r_{h} \in \{ 0.15,0.2,0.25 \}$ and $p_{\text{AF}} \in \{ 0.01,0.025,0.05,0.1 \}$. In this subsection, we primarily present results for the case $r_{h} = 0.2$, $p_{\text{AF}} = 0.05$, with the magnitudes of the antiferromagnetic Ising interactions chosen from the uniform distribution over segment $(0,1)$; however, we will also clarify the similarity and difference in the numerical data relative to other parameter choices. Also, the data shown or mentioned, unless stated otherwise, are not averaged over different initial couplings drawn from the same distribution, but we explicitly checked that different initial disorder realizations produce very similar cumulative measures of the RG flow described here.

 We will first see how the energy scale and remaining degrees of freedom $N_D$ in the system at each RG step evolve. We define the number of remaining degrees of freedom (effective qubits) as
\begin{equation}
    N_{D} = (L_\Omega - N_{m_{i}=0})\log{(1+\sqrt{2})} +  N_{m_{i}=0} \log \sqrt{2},
    \label{N_D}
\end{equation} 
where $N_{m_{i}=0}$ is number of bonds in the system with $m_{i}=0$, and $L_{\Omega}$ is the number of remaining sites at the current RG stage. Equation~\eqref{N_D} corresponds to the logarithm of the total Hilbert space dimensions at a given RG step, which can be seen as follows: A bond $i$ with the label $m_{i} = 1$ has quantum dimension $1+\sqrt{2}$, the $1$ coming from the spin configuration with no domain wall at bond $i$ and the $\sqrt{2}$ coming from the spin configuration with a domain wall at $i$, which hosts a Majorana fermion that contributes the $\sqrt{2}$ part. Meanwhile, a bond $i$ with the label $m_{i} =0$ only contributes a quantum dimension $\sqrt{2}$ since the spin configuration always exhibits a domain wall at that bond.

\begin{figure}
\includegraphics[width=\linewidth]{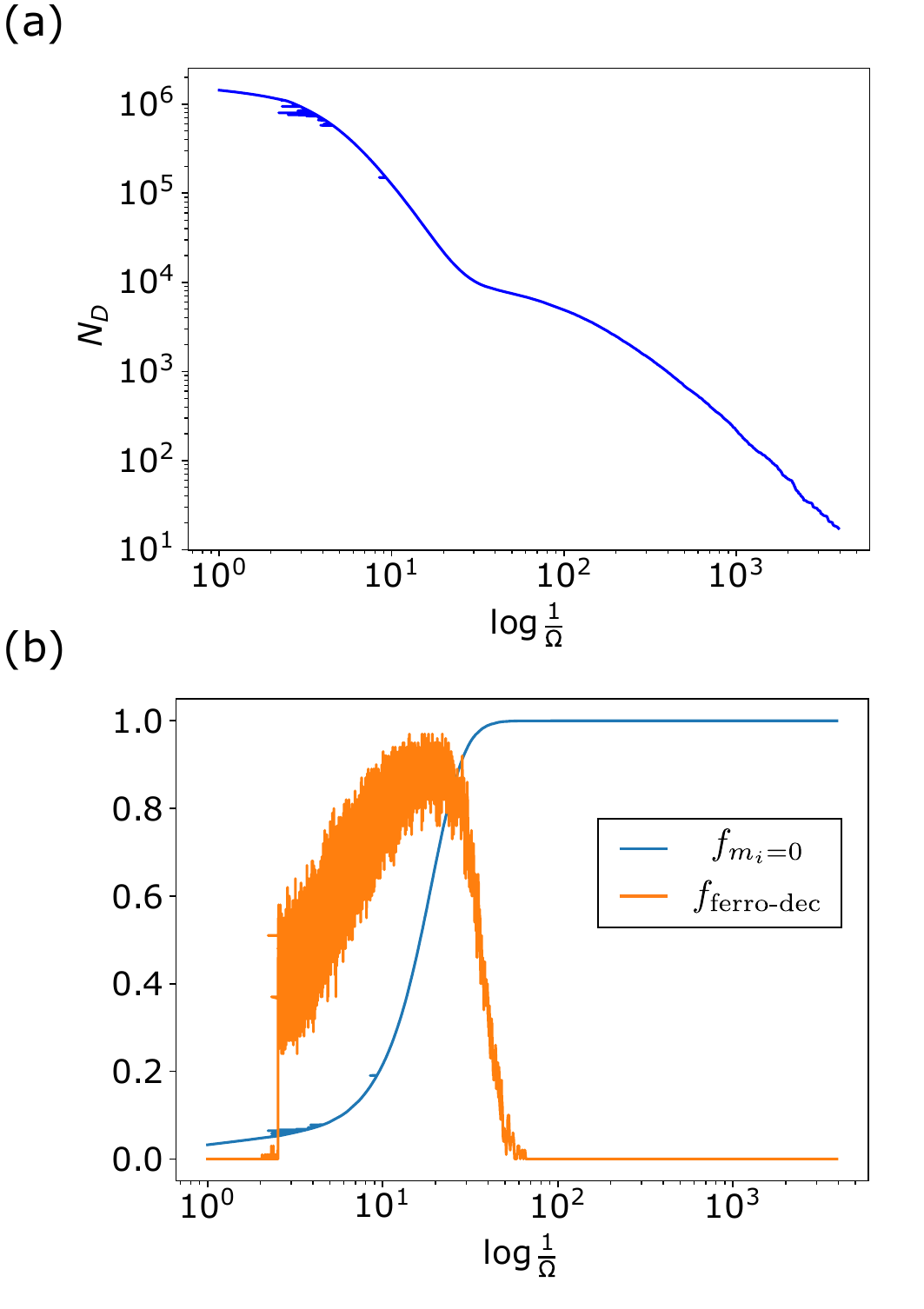}
\caption{(a) $N_{D}$ versus $\log \frac{1}{\Omega}$ graph, shown on log-log scale. (b)$f_{m_{i}=0} = \frac{N_{m_{i}=0}}{L_{\Omega}}$ (blue line) and $f_{\text{ferro-dec}}$ (orange line), defined as proportion of ferromagnetic bond decimations in the preceding 100 decimations prior to a given RG step, versus $\log \frac{1}{\Omega}$.
}
\label{fig:fullrggraph1}
\end{figure}

Figure~\ref{fig:fullrggraph1}(a) plots $N_D$ versus the log of the RG energy scale $\log \frac{1}{\Omega}$. One can clearly see that there are two regions where $N_{D}$ decreases monotonically with the energy scale; in between, there is a transient region where the slope becomes fairly flat. Comparing Fig.~\ref{fig:fullrggraph1}(a) with Fig.~\ref{fig:fullrggraph1}(b) provides a clearer picture behind this behavior. In Fig.~\ref{fig:fullrggraph1}(b), the blue line shows $f_{m_{i}=0} = \frac{N_{m_{i}=0}}{L_{\Omega}}$ versus the RG energy scale and tracks down the portion of bonds with the label $m_{i} =0$, while the orange line plots the proportion of ferromagnetic bond decimations in the preceding 100 decimations prior to a given RG step, defined as $f_{\text{ferro-dec}}$. Once past the very early stage in which the orange line is completely flat at zero due to the fact that dynamically generated ferromagnetic Ising interactions are too small to enter as the dominant RG energy scale, the orange line quickly increases to be near 1, and while this increase transpires, $f_{m_{i}=0}$ also quickly rises. This behavior indicates that dynamically generated ferromagnetic Ising interactions dominate this stage of the strong-disorder RG and lock the spin degrees of freedom in the system. This trend continues until the blue line saturates near 1; when $f_{m_{i}=0} \approx 1$, most of the spin degrees of freedom are locked up, and the RG enters the regime where the low-energy physics is primarily controlled by Majorana fermions that live on $m_{i} = 0$ bonds.  

Thus the two regions in Fig.~\ref{fig:fullrggraph1}(a) correspond first to a regime in which ferromagnetic Ising interactions dominate followed by a second regime in which the Majorana fermions govern the IR physics; the location of the intermediate transient region in Fig.~\ref{fig:fullrggraph1}(a) matches the energy scale in which the blue line in Fig.~\ref{fig:fullrggraph1}(b) saturates near $1$.  We observed similar behavior in all other initial conditions used in the RG procedure, indicating that the two-regime structure in Fig.~\ref{fig:fullrggraph1} is a generic feature of the RG for some extended range of initial conditions. Also, this behavior is consistent with the physical picture of the edge state we sketched in the introduction and Sec.~\ref{sec:sec4sdrgoverview} -- spin degrees of freedom are locked by Ising interactions, but due to the presence of the ``unscreened'' antiferromagnetic interaction, the true IR physics is governed by domain wall Majorana fermions.

As a final remark on Fig.~\ref{fig:fullrggraph1}(a) and (b), we comment on the the horizontal spikes in these plots. While blue lines in Fig.~\ref{fig:fullrggraph1}(a) and (b) mostly show monotonic behavior, we see some horizontal spikes, primarily in the region where $\log \frac{1}{\Omega} < 10$. These horizontal spikes represent RG steps at which the RG energy scale fails to decrease monotonically. As commented earlier, large number of these spikes would signal the breakdown of our strong-disorder RG. We point out the following: First, such points in Fig.~\ref{fig:fullrggraph1}(a) and (b) are concentrated near small $\log \frac{1}{\Omega}$. Hence, despite their appearance, the dangerous RG steps in which the RG energy scale does not decrease actually represent a very small portion of the decimation procedure. In the worst case among the parameters we studied in which $r_{h} = 0.25$ and $p_{\text{AF}} = 0.1$, there are $\sim 100$ such RG steps in our numerical implementations out of a far larger total of $O(L) \sim  10^{6}$ steps. These dangerous RG steps are further suppressed as one increases randomness in the initial couplings by choosing a smaller $r_{h}$ and a smaller $p_{\text{AF}}$; at $p_{\text{AF}}=0.01$ and $r_{h} = 0.15$, these events occur fewer than 10 times. This trend supports the expectation that the strong-disorder RG procedure works better when $p_{\text{AF}}$ is small and the randomness in couplings is large.
 
\begin{figure}
\includegraphics[width=\linewidth]{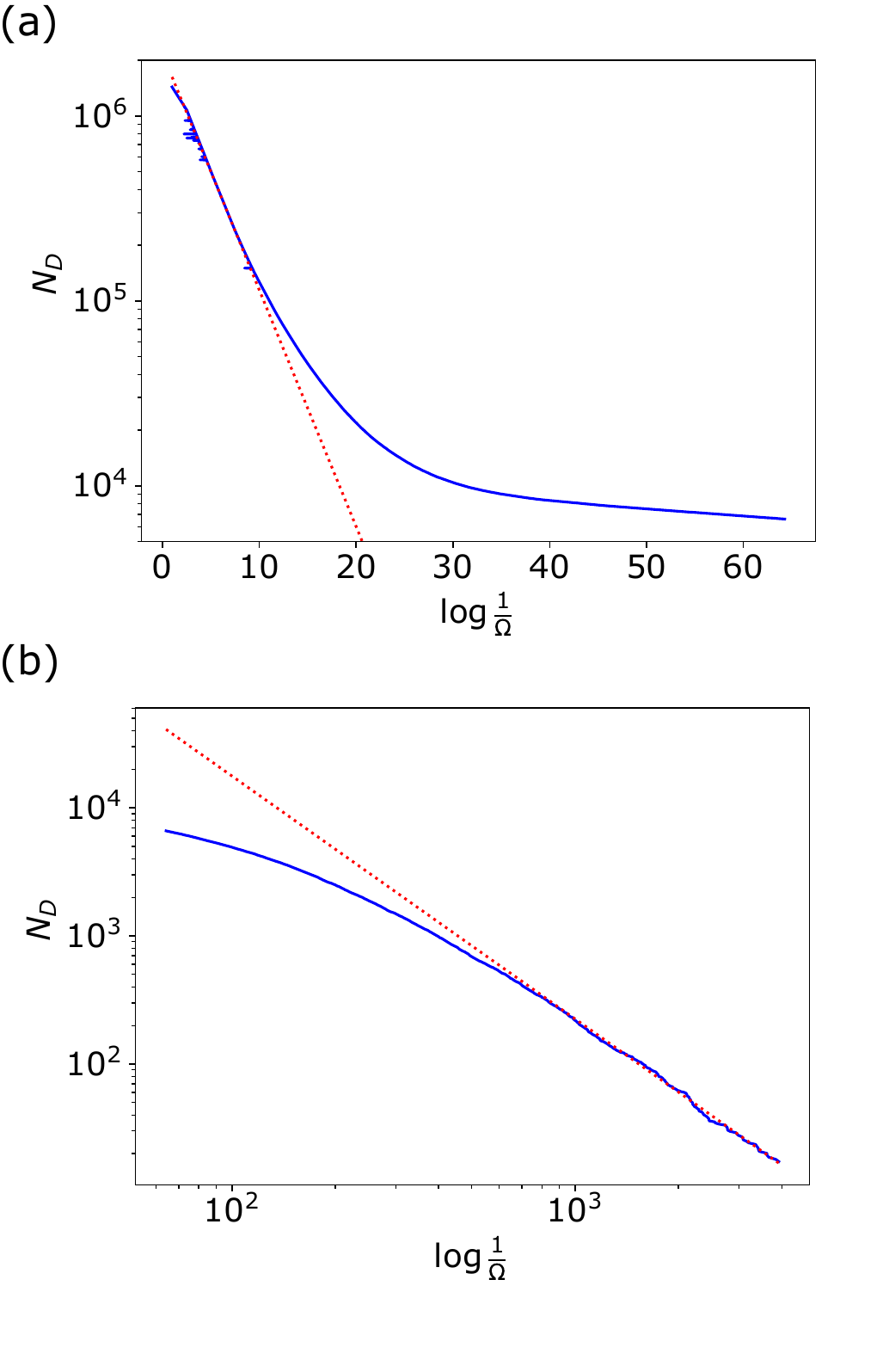}
\caption{The $N_D$ versus $\log \frac{1}{\Omega}$ plot, divided into the two regions (a) $f_{m_{i}=0} < 1$, using linear-log scale; and (b)$f_{m_{i}=0} \approx 1$, using log-log scale. The red dotted line in (a) is generated from the linear least-square fitting on the data in the region $5 < \log \frac{1}{\Omega} < 8 $, while the red dotted line in (b) is generated from the linear least-square fitting on the points from the last 200 RG steps.}
\label{fig:fullrggraph2}
\end{figure}

 Next we quantify the  scaling laws governing the two regimes in Fig.~\ref{fig:fullrggraph1}(a).  For clarity, Fig.~\ref{fig:fullrggraph2} presents subsets of the data from Fig.~\ref{fig:fullrggraph1}(a) corresponding to regimes with (a) $f_{m_{i}=0} < 1$ and (b) $f_{m_{i}=0} \approx 1$. We first discuss Fig.~\ref{fig:fullrggraph2}(a). As argued in Sec.~\ref{sec:sec4sdrgoverview}, at the early stage of the RG where two antiferromagnetic bonds are far from each other, the RG is ``unaware'' of the antiferromagnetic bonds, and the underlying scaling law will be that of the ferromagnetic Griffiths phase. In the ferromagnetic Griffiths phase, $N_{D}$ and the energy scale $\Omega$ are linked via the scaling relation
\begin{equation}
\label{eq:griffithantif}
    N_{D} \sim \Omega^{\frac{1}{z}}
\end{equation}
with non-universal exponent $z$.
In the log-linear plot of Fig.~\ref{fig:fullrggraph2}(a), the above scaling law appear as a straight line. At the early stage of the RG, indeed the red dotted line (obtained from least-square fitting) appropriately describes how $ N_{D}$ scales with $\Omega$, as expected. At later stages, however, significant deviations from the scaling law in Eq.~\eqref{eq:griffithantif} appear---indicating that antiferromagnetic bonds on average become close enough to modify the scaling behavior in such a way that $\log N_{D}$ decreases much more slowly than in the ferromagnetic Griffiths phase. Both the early RG steps governed by the scaling law from the Griffiths phase and the deviation from the Griffiths scaling at later steps are observed with different choices of initial conditions.

We reserve the precise nature of the slowdown in the evolution of $N_{D}$ to future work. However, we would like to point out one tantalizing possibility: Re-examining Fig.~\ref{fig:fullrggraph1}(a), and focusing on the first region dominated by ferromagnetic Ising interaction, the line is almost linear on a log-log plot at the later steps of that regime. This behavior suggests that $ N_{D}$ might scale with some power of $\log \Omega$, rather than of $\Omega$. Such activated scaling is unnatural in clean systems but commonly emerges in infinite-randomness fixed points \cite{Fishersdrg_1,Fishersdrg_2,Fishersdrg_3,Monthus1997,Motrunich2001_2,Refael2002}. In our problem, activated scaling could potentially originate from the existence of another strong-disorder fixed point which is not a true IR fixed point of the system but nevertheless controls the physics over some intermediate energy scales. 
 
Figure~\ref{fig:fullrggraph2}(b) represents the IR regime where the physics is governed by Majorana fermions living on the domain walls. Here, we expect an infinite randomness fixed point to arise, at which the following scaling between RG energy and $N_{D}$ holds:
\begin{equation}
\label{eq:activatedscaling}
   N_{D} \sim \left( \log \frac{1}{\Omega} \right)^{-\alpha}.
\end{equation}
More precisely, we anticipate an infinite-randomness fixed point akin to that of the random transverse-field Ising model, for which $\alpha = 2$. To check whether our numerical results exhibit this behavior, we performed a least-square fitting of $\log N_{D}$ versus $\log \log \frac{1}{\Omega}$ on the data from the last 200 RG steps; the slope estimated in this way corresponds to the exponent $-\alpha$ in the above scaling law. The red dotted line in Fig.~\ref{fig:fullrggraph2}(b) shows the result of this linear fitting, which appropriately capture how $N_{D}$ and $\log \frac{1}{\Omega}$ scale at the later steps.  Moreover, the fit yields $\alpha_{\text{est}} \approx 1.9$, reasonably close to the expected value $2$.  These features are consistent with  infinite-randomness fixed point behavior. However, we would like to mention that $\alpha_{\text{est}}$ varies quite a bit in different disorder realizations in our numerics, presumably due to finite-size effects. For the case we are presenting at $r_{h} = 0.2$ and $p_{\text{AF}} =0.05$, we observe $\alpha_{\text{est}}$ ranging from $1.64$ to $2.05$, and averaging $\alpha_{\text{est}}$  over ten different disorder realizations gives $\alpha_{\text{est},\text{avg}} \approx 1.84$. We recover a number closer to the expected value of $2$ when $p_{\text{AF}} =0.1$, where $\alpha_{\text{est},\text{avg}} \approx 1.95$, but a number farther from $2$ when  $p_{\text{AF}}=0.01$, where $\alpha_{\text{est},\text{avg}} \approx 1.64$, both of $\alpha_{\text{est},\text{avg}} \approx 1.95$'s obtained by also averaging for ten different disorder realizations.  We also find that $\alpha_{\text{est}}$ deviates a bit further from $2$ when the antiferromagnetic Ising interaction coefficients are chosen from the same power law as in the flip term instead of a uniform distribution.
 
The above result indicates that, within the system size we use for our numerical implementations, the scaling law Eq.~\eqref{eq:activatedscaling} with $\alpha=2$ is not manifest in all choices of parameters, but matches our simulation results increasingly well at larger values of $p_{\text{AF}}$. At larger values of $p_{\text{AF}}$, due to closer distance between antiferromagnetic bonds, fewer RG steps are needed to reach the IR regime governed by domain wall Majorana fermions. Hence, assuming that the RG flows to the same IR infinite-randomness fixed point for different choices of $p_{\text{AF}}$, choosing larger $p_{\text{AF}}$ gives more room for our system to flow close enough to IR fixed points to see the asymptotic scaling behavior. Hence, we believe that seeing $\alpha_{\text{est}}$ closer to $2$ as one chooses larger $p_{\text{AF}}$ constitutes a meaningful consistency check that the strong-disorder RG we presented eventually flows to the infinite-randomness fixed point with $\alpha=2$. We expect that numerical implementations with larger system sizes will make the IR scaling more manifest as well.

To summarize this subsection, we numerically implemented the strong-disorder RG procedure we developed and clearly resolved two energy regimes: the higher-energy regime in which dynamically generated ferromagnetic Ising interactions freeze spin degrees of freedom, and the IR regime in which the physics is governed by an infinite-randomness fixed point of the Majorana fermions residing on domain walls that exist in the magnetically ordered state due to the presence of antiferromagnetic interactions. A more quantitative analysis reveals a regime at the early stage of the RG there wherein the system scales as in the ferromagnetic Griffiths phase, followed by a regime in which ferromagnetic Ising interactions still dominate but $ N_{D}$ decrease much slower than one would expect in the ferromagnetic Griffiths phase due to the presence of antiferromagnetic interactions. We also investigated how $ N_{D}$ scales in the IR regime and found evidences that the scaling is consistent with the expected infinite-randomness fixed point in the random transverse-field Ising model at criticality.
 
\section{Conclusion}
\label{sec:conc}

In this paper, we studied the edge state of 2D time-reversal topological superconductor when both strong interaction and disorder are present. Instead of starting from the continuum field theory as in most studies of the edge/surface states of topological phases of matter, we exploited the recently derived 1D lattice model that mimics the helical edge states of a 2D topological superconductor. Our strategy was to write down model Hamiltonians within the framework of the 1D lattice model, consisting of nearest-neighbor Ising interaction, and ``flip terms'' that do not commute with the Ising interactions and 
thus provide non-trivial quantum dynamics. This 1D Hamiltonian allowed us to understand the edge physics from the standpoint of strong-disorder RG---which is more naturally understood in terms of lattice models than continuum field theories.

In Sec.~\ref{sec:splim}, we explored through an exactly solvable limit of the 1D lattice model how spin degrees of freedom frozen by Ising interactions, random in both sign and magnitude, give rise to a novel edge state in which time-reversal symmetry is spontaneously broken but Majorana fermions on spin domain walls form a critical infinite-randomness fixed point. In Sec.~\ref{sec:sdrg}, we extended the analysis of Sec.~\ref{sec:splim} to the full-fledged strong-disorder RG on the 1D lattice model to see how this edge state with infinite-randomness-fixed-point physics can arise generally when randomness in the couplings is strong. Due to the limitation of the strong-disorder RG method and approximations we made, we will not claim that our strong-disorder RG analysis definitively addresses the ultimate low-energy physics. Nonetheless, our analysis strongly suggests that the edge state featuring randomly oriented local ferromagnetic domains (frozen by the spontaneous time-reversal symmetry breaking) and infinite-randomness fixed point of domain wall Majorana fermions can arise generally when both strong interaction and disorder are present.
 
 We close this paper with some comments on future directions that our work illuminates:
\begin{enumerate}
    \item The numerical implementation of the strong-disorder RG scheme utilized in Sec.~\ref{sec:sdrg} seems to capture an extended phase that corresponds to the edge state with infinite-randomness fixed point physics. Can this strong-disorder RG capture different edge phases as well?  Our RG scheme breaks down when the concentration of antiferromagnetic bonds in the UV Hamiltonian is too large, as then non-local terms we chose to truncate become important. Turning this logic in reverse, extending our RG scheme to incorporate these non-local terms properly opens up the possibility to capture different edge phases and the transitions between them. 
    \item What are experimental signatures of the edge state with domain-wall Majorana fermions forming an infinite randomness fixed point? Apart from SQUID/scanning tunneling microscopy to detect local ferromagnetic domain formations and scaling of thermal conductance with respect to the system length discussed in Chou and Nandkishore \cite{Chou2021}, our RG scheme provides a systematic way to think about how these 1D edge states contribute to thermodynamic quantities such as spin susceptibility $\chi(T)$ and heat capacity $C(T)$ as a  function of temperature $T$. It is known, for instance, that the infinite-randomness fixed point that governs low-energy physics of our edge state contributes to the heat capacity via the scaling $C(T) \sim \frac{1}{\left( \log \frac{1}{T}\right)^{3}}$\citep{Fishersdrg_2,Fishersdrg_3}; as $T$ approaches to $0$, this 1D contribution decreases much slower than 2D bulk contributions that follow power laws, e.g., due to phonons. However, such a 1D contribution scales as linearly with respect to system size $L$, while 2D bulk contributions scale as $L^{2}$. Hence, it is not entirely clear whether there is a measurable window in which one may see the characteristic scaling from the 1D infinite-randomness fixed point.
    \item It is interesting to compare the edge state we proposed to arise generally in this paper to the localized edge state of quantum spin Hall insulators in \citep{Chou2018,Kimchi2020}. In such edge of quantum spin Hall insulators, Ising spins are similarly frozen with random orientations.  The key difference is that the domain-wall degrees of freedom, now full electrons instead of Majorana fermions, can undergo localization---in sharp contrast to the 2D topological superconductor situation, for which critical infinite-randomness fixed point behavior appears inevitable. It would be interesting to develop a deeper theoretical understanding of this difference. This understanding may lead to a generalization of novel edge states enabled by disorder and interaction to dirty surface states of higher-dimensional SPTs. 
    
    What we believe to lie at the heart of this question is that the group-cohomological classification of symmetry-protected topological phases can be obtained by assuming that local unitary transformations which are non-on-site in the case of non-trivial SPTs implement symmetry actions on the edge/surface \citep{Else2014}. It is also known that 2D time-reversal invariant topological superconductors are beyond-cohomology SPTs; the non-local time-reversal symmetry action of the 1D model we saw in Sec.~\ref{sec:Review} can be regarded as a manifestation of its beyond-cohomology nature. Meanwhile, the lattice 1D model that mimics the edge of quantum spin Hall insulators  do have  local unitary symmetry actions considered in \citep{Else2014}, and quantum spin Hall insulators may be regarded as ``cohomological'' SPTs \citep{Metlitski2019}. We speculate that the localization property of the edge/surface states when subjected to strong disorder and interaction is related to whether the SPT bulk is ``cohomological'' or not.
\end{enumerate}

\acknowledgments

We thank Sri Raghu for many comments and encouragement, as well as Pavel Nosov, Chaitanya Murthy, Rahul Nandkishore, Akshat Pandey, Josephine Yu, and Yue Yu for discussion.  We acknowledge support from the Caltech Institute for Quantum Information and Matter, an NSF Physics Frontiers Center with support of the Gordon and Betty Moore Foundation through Grant GBMF1250. The U.S. Department of Energy, Office of Science, National Quantum Information Science Research Centers, Quantum Science Center partially supported the manuscript preparation. OIM acknowledges support by the National Science Foundation through grant DMR-2001186.
\appendix
 
\section{More on the justification of the decimation rule}
\label{app:df}

 Here, we give technical details on how to derive the decimation rules described in Sec.~\ref{sec:sec3rule} for integrating out ferromagnetic and antiferromagnetic bonds.
 
Let us first introduce some notation. We define the following binary operation between flip terms:
\begin{equation}
\label{eq:symp}
F_{i_{1}} \odot F_{i_{2}} =  \frac{F_{i_{1}} F_{i_{2}} + F_{i_{2}}F_{i_{1}}}{2}.
\end{equation} 
 One may understand $\odot$ as a symmetrized product of two operators. Using this symbol, one may also define:
\begin{equation}
\label{eq:symp2}
\bigodot_{i=1}^{n} F_{i} = (\cdots((F_{1} \odot F_{2}) \odot F_{3}) \cdots \odot F_{n}).
\end{equation}
At the each step of the transformation, the Hamiltonian is written as
\begin{equation}
\label{eq:tfhamapp}
\begin{split}
&H(\{ m_{i}, h_{i}, J_{i}, t_{i}, s_{i} \} ) = -\sum_{ \{ \tilde{i}_{a} \} } c_{\tilde{i}_{a}} C_{\tilde{i}_{a}} \\
& \quad \quad - \left( \sum_{i} J_{j} \sigma_{j}^{z}\sigma_{j+1}^{z} + 2t_{i} T_{i}  +2 s_{i} S_{i} \right) \\
& C_{\tilde{i}_{a}} = \bigodot_{i=\tilde{i}_{a}+1}^{\tilde{i}_{a+1}} F_{i}, \quad c_{\tilde{i}_{a}} =  \frac{1}{\Lambda_{a}^{\tilde{i}_{a+1} - \tilde{i}_{a} -1}} \prod_{i=\tilde{i}_{a}+1}^{\tilde{i}_{a+1}} h_{i}.
\end{split}
\end{equation}
 In the above notation, $\{ \tilde{i}_{a} \}$ is defined to be a set of all indices with $m_{\tilde{i}_{a}} =1$, integer subscripts $a$ assigned with ascending order, i.e., $\tilde{i}_{1} < \tilde{i}_{2} < \tilde{i}_{3} < \cdots$. The basic intuition behind $C_{\tilde{i}_{a}}$ is that this term simultaneously flips all Ising spins in a cluster linked by bonds with $m_{i}=0$. 
 
Compared to the Hamiltonian given in Eq.~\eqref{eq:tfham} of the main text, Eq.~\eqref{eq:tfhamapp} is almost identical except that the flip terms $-h_{i}m_{i-1}m_{i}F_{i}$ are superseded by the terms $c_{\tilde{i}_{a}} C_{\tilde{i}_{a}}$, which generalize single-site flip terms in Eq.~\eqref{eq:tfham} in the sense that they now allow spins linked by $m_{i}=0$ bonds to be simultaneously flipped as well. We will see in this Appendix that the multi-site flip terms encoded by $c_{\tilde{i}_{a}} C_{\tilde{i}_{a}}$ do not play an important role in determining IR physics and hence are dropped in the main text. However, we include them here for completeness.
 
 \subsection{Integrating out antiferromagnetic bonds}
 
 To keep track of how parameters of the Hamiltonian should transform when an antiferromagnetic bond is integrated out, we can use the second-order degenerate perturbation theory on the local Hamiltonian around the antiferromagnetic bonds:
\begin{equation}
\label{eq:afpt}
\begin{split}
& H_{loc} = H_{0} + H_{p} \\
& H_{0} = \Lambda_{A} \sigma_{j}^{z} \sigma_{j+1}^{z}, \quad H_{p} = -h_{j} F_{j} - h_{j+1} F_{j+1},
\end{split}
\end{equation}
treating $H_{p}$ as a perturbation. The organizing principle behind the perturbation theory is almost identical to the one in Sec.~\ref{sec:3prelim}: On any state with $\sigma_{j}^{z} = -\sigma_{j+1}^{z}$, either $F_{j}$ or $F_{j+1}$ flips one Ising spin and brings that state to an excited state with $\sigma_{j}^{z} = \sigma_{j+1}^{z}$. Acting $F_{j}$ or $F_{j+1}$ once more sends the excited state back to the original state. Hence, all terms generated from the second-order perturbation theory involve flipping Ising spins twice, either by $F_{j}$ or $F_{j+1}$, and are given by:
\begin{equation}
\label{eq:afpt2nd}
H^{(2)} = - \frac{h_{j}^{2}}{2\Lambda_{A}} F_{j}^{2} - \frac{h_{j+1}^{2}}{2\Lambda_{A}} F_{j+1}^{2} - \frac{h_{j}h_{j+1}}{\Lambda_{A}} (F_{j} \odot F_{j+1}).
\end{equation}
 One can show that the term $\frac{h_{j}^{2}}{2\Lambda_{A}} F_{j}^{2}$, in our context, can be rewritten as:
\begin{equation}
\label{eq:afptsquare}
\frac{h_{j}^{2}}{2\Lambda_{A}} F_{j}^{2} \equiv \frac{h_{j}^{2}}{4 \Lambda_{A}} T_{j} + \frac{(\sqrt{2}-1) h_{j}^{2}}{8 \Lambda_{A}} \sigma_{j-1}^{z}\sigma_{j}^{z} + (\text{const.}),
\end{equation}
i.e., the second order perturbation theory generates $T_{j}$ and Ising interactions. However, due to our assumption that $\Lambda_{A} \gg J_{i} \gg h_{j}$, the generated Ising interaction is always much smaller than the Ising interaction already present in the Hamiltonian, and hence we only keep track of $T_{j}$, included in the second line of Eq.~\eqref{eq:afrule}. Similarly, one can show that the term $T_{j+1}$ is generated from $\frac{h_{j+1}^{2}}{2\Lambda_{A}} F_{j+1}^{2}$; this term corresponds to the third line in Eq.~\eqref{eq:afrule}.
As for $\frac{h_{j}h_{j+1}}{\Lambda_{A}} (F_{j} \odot F_{j+1})$, this term precisely corresponds to the multi-site flip term $C_{\tilde{i}_{a}}$ introduced in the first line of Eq.~\eqref{eq:tfhamapp} after the transformation.
 
\subsection{Integrating out ferromagnetic bonds}
\label{app:splimitf}

Now we derive the rule in Eq.~\eqref{eq:frule}. First, we take a look at terms generated from the first-order and the second-order perturbation theory result. As it turns out, one needs to incorporate higher-order perturbation theory results to keep track of $t_{i}$'s and $s_{i}$'s, the coefficients for Majorana-fermion hopping terms. We will explore what types of higher-order terms contribute to $s_{i}$'s and $t_{i}$'s as well.
  
Regarding $t_{i}$'s and $s_{i}$'s at each step in the middle of transformations according to the rule Eq.~\eqref{eq:afrule} and \eqref{eq:frule}, the following fact about these numbers will be useful in determining which terms to keep track of: When $t_{j}$ and $s_{j}$ are expressed in terms of the coefficients of the local terms in the original Hamiltonian (i.e., the Hamiltonian before any of the transformations given by the rules Eq.~\eqref{eq:afrule} and \eqref{eq:frule}), it has the following dependence on the strength of the Ising antiferromnagetic interaction $\Lambda_{A}$:
\begin{equation}
\label{eq:tjsjscale}
\begin{split}
   & t_{j} = \begin{cases}
    0 & m_{j-1}=m_{j} =1 \\
    O \left( \frac{1}{\Lambda_{A}}\right) & (m_{j-1} , m_{j}) = (0,1) \text{ or } (1,0) \\
    \left( \frac{1}{\Lambda_{A}^{2}}\right) & (m_{j-1} , m_{j}) = (0,0)
    \end{cases} \\
    & s_{j} = O \left( \frac{1}{\Lambda_{A}^{2}}\right),
\end{split}
\end{equation}
where dependence on all other coefficients is absorbed into the big-$O$ notation. On the other hand, in our assumption, we set $\Lambda_{A}$ to be the largest energy scale in the original Hamiltonian. Hence, terms whose denominators have higher powers of $\Lambda_{A}$ may be dropped in our limit.
   
\subsubsection{First-order and second-order contributions}
   The setup for the perturbation theory is similar to Eq.~\eqref{eq:afpt}. We work with the following, treating $H_{p}$ as a perturbation:
\begin{equation}
\label{eq:frulesetup}
\begin{split}
& H = H_{0} + H_{p}, \\
& H_{0} = - J_{j} \sigma_{j}^{z} \sigma_{j+1}^{z} \\
& H_{p} = - c_{l_{j}} C_{l_{j}} - c_{r_j} C_{r_j} - 2 t_{j} T_{j} - 2 t_{j+1} T_{j+1} - 2 s_{j}S_{j} + \cdots.
\end{split}
\end{equation}
 In $H_{p}$,  $c_{l_j} C_{l_j}$ and $c_{r_{j}} C_{r_{j}}$ correspond to the multi-site flip terms involving sites $j$ and $j+1$, respectively, with the symbols $l_{j}$ and $r_{j}$ denoting the bonds to the left and to the right of $j$. 
 
 The only terms that contribute at first order in $H_{p}$ are $ -2 s_{j}S_{j}$, which enters as $t'_{j}$ after the transformation. The first-order contribution from the $ -2 s_{j}S_{j}$ term is one of the three terms in Eq.~\eqref{eq:frule} on the expressions for $t'_{j}$ as well.
 
 The only two terms that contribute to the second-order perturbation theory are the Ising spin flip terms. The organization principle is almost identical to the one we saw in the antiferromagnetic bond decimation, and the terms generated at second order are the following:
 
\begin{equation}
\label{eq:f2ndfull}
H^{(2)} = -\frac{c_{l_{j}}^{2} }{2J_{j}} C_{l_{j}}^{2}-\frac{c_{r_j}^{2} }{2J_{j}} C_{r_j}^{2} - \frac{c_{l_{j}} c_{r_j} }{J_{j}} C_{l_{j}} \odot C_{r_j}.
\end{equation} 
Compared to the second-order correction from the antiferromagnetic bond decimation given in Eq.~\eqref{eq:afpt2nd}, the above equation can be simply obtained by substituting single-site flip terms $F_{j}$'s in Eq.~\eqref{eq:afpt2nd} with the corrsponding multi-site flip terms.

Let us see how the term $C_{l_{j}} \odot C_{r_j}$ becomes the transformation rule for $h_{i}$'s. We start by pointing out an important property of the symmetrized product we introduced in Eq.~\eqref{eq:symp} and \eqref{eq:symp2}. One can easily show that this product is commutative and associative. Hence, in the formula where multiple $\odot$'s are involved, the result is identical regardless of how you put parentheses on the formula or change the order of the product.

Since $C_{l_{j}}$ and $C_{r_j}$ are multi-site flip terms involving sites $j$ and $j+1$, respectively, one can without loss of generality write
\begin{equation}
\label{eq:f2ndflipsub}
\begin{split}
& C_{l_j} = \overline{C}_{l_j} \odot F_{j}\\
& C_{r_j} = F_{j+1} \odot \overline{C}_{r_j} \\
& c_{l_{j}} = \overline{c}_{l_{j}} h_{j} \\
& c_{r_{j}} = \overline{c}_{r_j} h_{j+1}.
\end{split}
\end{equation}
 Using the above, one may write
\begin{equation}
\label{eq:f2ndflip}
\begin{split}
& \frac{c_{l_{j}} c_{r_{j}} }{2J_{j}} C_{l_j} \odot C_{r_j} \\
& = \overline{c}_{l_{j}}  \overline{c}_{r_j}  \frac{h_{j}h_{j+1} }{J_{j}} \overline{C}_{l_j} \odot (F_{j} \odot F_{j+1}) \odot \overline{C}_{r_j}. 
\end{split}
\end{equation}
Now one can recast $ \frac{h_{j}h_{j+1} }{2J_{j}} F_{j} \odot F_{j+1}$ into a super-site flip term using
\begin{equation}
    \frac{h_{j}h_{j+1} }{J_{j}} F_{j} \odot F_{j+1} \equiv \frac{2^{-1/4} h_{j}h_{j+1} }{J_{j}} F'_{j} = h'_{j} F'_{j}
\end{equation}
with $F'_{j}$ denoting a normalized super-site flip term of amplitude $h_j'$, the prime added to emphasize that it is a flip term after the transformation. Note that the value of $h'_{j}$ above precisely matches that in Eq.~\eqref{eq:frule}.  One can plug the above redefinition into Eq.~\eqref{eq:f2ndflip} to show that it gives the correct form of the flip term in Eq.~\eqref{eq:tfhamapp} after the transformation as well. The key takeaway is that due to associativity and commutativity of the symmetrized product $\odot$ we defined, in decimating the ferromagnetic Ising bonds, one can renormalize the flip terms, even if they are multi-site, as if the flip terms at site $j$ and $j+1$ are single-site flip terms. We will use this idea in subsequent derivations as well.

Now we analyze how $-\frac{c_{l_j}^{2} }{2J_{j}} C_{r_{j}}^{2}$ and $-\frac{c_{l_j}^{2} }{2J_{j}} C_{r_j}^{2}$ in Eq.~\eqref{eq:f2ndflip} contribute to the transformation rule.  When $m_{j-1}=1$, $C_{l_j}$ is simply a single-site flip term $F_{j}$, and $c_{l_j} = h_{j}$. Then, one can show within the context of perturbation theory that 
\begin{equation}
\label{eq:f2ndsquare1}
\frac{c_{l_j}^{2} }{2J_{j}} C_{l_{j}}^{2} = \frac{h_{j}^{2} }{2J_{j}} F_{j}^{2} \equiv \frac{\sqrt{2}-1}{4\sqrt{2}} \frac{h_{j}^{2}}{J_{j}} \sigma_{j-1}^{z} \sigma_{j}^{z},
\end{equation}
i.e., it only contributes an Ising interaction. This Ising interaction is again much smaller than the one already present in the Hamiltonian due to our assumption that $h_{j}$ is much smaller than any other Ising interactions, and hence can be dropped for the purpose of Sec.~\ref{sec:splim}. If $C_{l_j}$ is a double-site flip term $F_{j-1} \odot F_{j}$, one can show that
\begin{equation}
\label{eq:f2ndsquare2}
\begin{split}
\frac{c_{l_j}^{2} }{2J_{j}} C_{l_j}^{2} = \frac{h_{j-1}^{2}h_{j}^{2}}{ 2\Lambda_{A}^{2} J_{j}} (F_{j-1} \odot F_{j})^{2} \\
\equiv \frac{h_{j-1}^{2}h_{j}^{2}}{ 16\sqrt{2}\Lambda_{A}^{2} J_{j}} \sigma_{j-1}^{z}\sigma_{j}^{z} + \frac{(1+\sqrt{2})h_{j-1}^{2}h_{j}^{2}}{ 16\Lambda_{A}^{2} J_{j}} T_{j-1}.
\end{split}
\end{equation}
 Due to similar logic we invoked before, we may drop the generated Ising interaction. For the $T_{j-1}$ term, we observe the following two facts: First, since we have a double-site flip term,  $m_{j-2} = 1$, while $m_{j-1}=0$ in this case. Second, at the beginning of the section---see also Eq.~\eqref{eq:tjsjscale}---we saw that the transformation rule in the main text gives $t_{j-1} = O(1/\Lambda_{A})$. However, the coefficient generated from the double-site flip terms is $O(1/\Lambda_{A}^{2})$ and constitutes only subleading corrections one may drop in our limit.
  
 Finally, when the number of sites involved in $C_{l_{j}}$ is more than two, $c_{l_{j}}$ contains two or higher power of $\Lambda_{A}$ in the denominator. Because of this property, any generated terms from $ -\frac{c_{l_{j}}^{2} }{2J_{j}} C_{l_{j}}^{2}$ have $\Lambda_{A}^{4}$ or higher power in the denominator. Meanwhile, all the terms generated in the transformation rule in the main text have at most $\Lambda_{A}^{2}$ in the denominator. Hence, when $\Lambda_{A}$ is very large, terms generated through the multi-site flip terms are also much smaller than those we retain in the main text, and we drop them in the transformation rule. 
  
 \subsubsection{Higher-order contributions}

 To capture the leading contribution to the nearest-neighbor coupling between Majorana fermions represented by $-t_{j}T_{j}$ terms, it is necessary to go to higher order in perturbation theory.
 Recall that within the context of degenerate perturbation theory, matrix elements corresponding to the third-order correction and the fourth-order correction are given by
\begin{equation}
\label{eq:3rdpgen}
    E^{(3)}_{ab} = \sum_{\substack{k_{1}, k_{2} \\ E_{k_{1}}, E_{k_{2}} \neq E_{0} }}\frac{\bra{a} H_{p} \ket{k_{1}} \bra{k_{1}} H_{p} \ket{k_{2}}\bra{k_{2}} H_{p} \ket{b}}{(E_{0} - E_{k_{1}})(E_{0} - E_{k_{2}})},
\end{equation}
\begin{equation}
\begin{split}
\label{eq:4thpgen}
    & E^{(4)}_{ab} = \\
     &\sum_{\substack{k_{1}, k_{2}, k_{3} \\ E_{k_{1}}, E_{k_{2}}, E_{k_{3}}\neq E_{0} }}\frac{\bra{a} H_{p} \ket{k_{1}} \bra{k_{1}} H_{p} \ket{k_{2}}\bra{k_{2}} H_{p} \ket{k_{3}} \bra{k_{3}} H_{p}\ket{b}}{(E_{0} - E_{k_{1}})(E_{0} - E_{k_{2}})(E_{0} - E_{k_{3}})},
\end{split}
\end{equation}
 where $H_{0}$ and $H_{p}$ are given in  Eq.~\eqref{eq:frulesetup}; $\ket{a}$ and $\ket{b}$ are degenerate eigenstates of $H_{0}$ with eigenvalue $E_{0}$. Here, we simply state which terms in the higher-order perturbation theory are responsible for the transformation rule in the main text. One may show that the remaining terms only generate interactions already accounted for in the lower-order perturbation theory, albeit with much smaller coefficients due to the fact that they appears at higher-order, and hence can be dropped.
 
 In the case with $m_{j-1} = 1$ and $m_{j+1} =0$, observe that in Eq.~\eqref{eq:frulesetup}, $c_{l_{j}} C_{l_{j}}$ is just a single-site flip term $h_{j} F_{j}$. The terms corresponding to choosing the first and the third $H_{p}$ in the numerator to be $-h_{j} F_{j}$ and the second $H_{p}$ to be $-2 t_{j+1} T_{j+1}$ decimation in Eq.~\eqref{eq:3rdpgen} generate the effective $t_{j}'$ given by 
 \begin{equation}
     t'_{j} = \frac{t_{j+1} h_{j}^{2}}{4\sqrt{2} J_{j}^{2}}.
 \end{equation}
 Similarly, one can show that if $m_{j-1} = 0$ and $m_{j+1} =1$, there is a term in the third-order perturbation theory that generates
  \begin{equation}
     t'_{j} = \frac{t_{j} h_{j+1}^{2}}{4\sqrt{2} J_{j}^{2}}.
 \end{equation}
 The preceding two terms precisely correspond to the first two terms for $t'_{j}$ in the last line of Eq.~\eqref{eq:frule}. These contributions do not add extra powers of $\Lambda_{A}$ to the coefficient owing to the fact that they involve single-site flip terms, and in the limit in the main text where $\Lambda_{A}$ is the largest energy scale, such contributions are the leading ones.
 
 The situation with $m_{j-1} = m_{j+1} =0$ is more complicated. In this case, both Ising spin flip terms covering sites $j$ and $j+1$ are multi-site flip terms, and analogues of the third-order terms introduced in the previous paragraph for the case we are considering will generally contain extra powers of $\Lambda_{A}$ in the denominator. In particular, since these third-order terms are linked to the physical processes of flipping spins back and forth, one should apply multi-site flip terms twice, resulting in introducing $\Lambda_{A}^{2}$ or higher powers to the denominators for Majorana fermion hopping terms. Also accounting for the $\Lambda_{A}$ in the denominator for the original $t_{j}$ and $t_{j+1}$ couplings, Majorana fermion hopping terms generated from the third order perturbation theory are $O(\frac{1}{\Lambda_{A}^{3}})$ in this case. This term is clearly much smaller than contributions given in the main text.
 
 It turns out that sticking to employing single-site flip terms only and keeping track of $s_{j}$ through employing the fourth-order perturbation theory in certain situations gives Majorana fermion hopping terms of $O(1/\Lambda_{A}^{2})$, which is the order being kept track of in the transformation rule of the main text. This $s_{j}$ directly enters as $t_{j}$ at first-order when decimating a ferromagnetic bond $j$ and constitutes the leading contributions over the terms involving multi-site flip terms discussed in the previous paragraph.
 
 To keep track of $s_{j}$ for our consideration, whenever we integrate out a ferromagnetic bond at $j$ and $m_{j-1} = m_{j+2} = 0,\, m_{j+1} = 1$ or $m_{j-2} = m_{j+1} =0 ,\, m_{j-1} = 1$, we use the fourth-order perturbation theory to generate non-zero $s'_{i}$'s. In both situations, in between two bonds with the label $m_{i}=0$, there are two bonds with $m_{i}=1$---one of which we are decimating. We first take a look at the case $m_{j-1} = m_{j+2} = 0,\, m_{j+1} = 1$. In this case, both bonds neighboring site $j+1$ have $m_{i}=1$. The terms that contributes to $-2 s'_{j} S_{j}$ after the transformation correspond to choosing the first and the fourth $H_{p}$ in Eq.~\eqref{eq:4thpgen} to be $-h_{j+1} F_{j+1}$, and the second and the third $H_{p}$ to be $-2 t_{j}T_{j}$ and $-2 t_{j+2}T_{j+2}$ or  $-2 t_{j+2}T_{j+2}$ and $-2 t_{j}T_{j}$. In this way one generates
 \begin{equation}
     s'_{j} =\frac{t_{j} t_{j+2} h_{j+1}^{2}}{2J_{j}^{3}}.
  \end{equation}
This term precisely corresponds to  the third line of the $s'_{i}$ value in Eq.~\eqref{eq:frule}. Similarly, one can show that there are fourth-order terms when  $m_{j-2} = m_{j+1} =0 ,\, m_{j-1} = 1$ that contribute to the second line of the $s'_{i}$ value in Eq.~\eqref{eq:frule}. Note that the above term only has dependence on $\Lambda_{A}$ from  $t_{j}$ and $t_{j+2}$, so $s'_{j} = O(1/\Lambda_{A}^{2})$ as we saw before.

Hereby we fully explained rationales on how each component of Eq.~\eqref{eq:frule} is derived.

\section{The full decimation rule}
\label{app:rule}

 This appendix presents the transformation rule for the full strong-disorder RG and its derivation. First, we briefly review the RG procedure and the protocol for deriving the decimation rules given in full form below.  At each RG step, the Hamiltonian reads
\begin{equation}
\label{eq:fullRGhamapp}
\begin{split}
&H(\{ m_{i}, h_{i}, J_{i}, t_{i}, s_{i} \} ) = -\sum_{ \{ \tilde{i}_{a} \} } c_{\tilde{i}_{a}} C_{\tilde{i}_{a}} \\
& \quad \quad - \left( \sum_{i} J_{j} \sigma_{j}^{z}\sigma_{j+1}^{z} + 2t_{i} T_{i}  +2 s_{i} S_{i} \right) \\
& C_{\tilde{i}_{a}} = \bigodot_{i=\tilde{i}_{a}+1}^{\tilde{i}_{a+1}} F_{i}, \quad c_{\tilde{i}_{a}} =  h_{\tilde{i}_{a+1}}  \prod_{i=\tilde{i}_{a}+1}^{\tilde{i}_{a+1}-1} \frac{h_{i}}{|J_{i}|} .
\end{split}
\end{equation}
The difference from the Hamiltonian given in the main text Eq.~\eqref{eq:fullRGham} is that we included multi-site flip terms linked by $m_{i}=0$ bonds here, similar to how we added multi-site flip terms to the analysis in Appendix~\ref{app:df}. It is also very similar to the Hamiltonian we encountered in Eq.~\eqref{eq:tfhamapp} -- the only difference is that now nearest-neighbor antiferromagnetic interaction can have different strength. Hence, the strength of antiferromagnetic bonds that appears in $c_{\tilde{i}_{a}}$, the coefficient for multi-site flip terms, have individual bond strengths $|J_{i}|$ instead of the uniform value $|\Lambda_{A}|$ as in Eq.~\eqref{eq:tfhamapp}. Although we mostly ignore contributions from multi-site flip terms, one may use values of these coefficients to track down how multi-site flip terms are renormalized to a very small magnitude.

Recall that at each step of the RG, we select a local term with the largest coefficient, reduce number of degrees of freedom by projecting the system onto the low-energy subspace of the local term, and derive the couplings in the effective Hamiltonian using perturbation theory. We identified six different types of decimations depending on which local term is chosen for the RG transformation in the main text: Ferromagnetic/antiferromagnetic bond decimations, single/double-site decimations, and double/triple bond decimations. In the main text, we did not give full details of how the couplings in the effective Hamiltonian change after each RG step. 

 Here we explicitly  write down and derive how the parameters $\{m'_{i}, h'_{i}, J'_{i}, t'_{i}, s'_{i} \}$ that determine effective couplings of the Hamiltonian after each RG step are given in terms of the parameters $\{m_{i}, h_{i}, J_{i}, t_{i}, s_{i} \}$ before the RG step of our interest. In each of the six subsections, we specify and justify the transformation rules for six different types of decimations. These rules allow one to implement the strong disorder RG iteratively on a computer and obtain the results presented in the main text.

 Before diving into details, we sketch here the general protocol for determining effective couplings of the Hamiltonian after a given transformation:
\begin{itemize}
    \item We incorporate all terms generated under first and second-order degenerate perturbation theory (for all bond decimations except for the case covered in Sec.~\ref{sec:tjmj0mj10}) and first-order perturbation theory (for site decimations and the bond decimations for the case in Sec.~\ref{sec:tjmj0mj10}). Incorporating these results suffices to provide leading order contributions on how $h_{i}$'s and $J_{i}$'s transform, and in some cases $t_{i}$'s as well.
    \item To track couplings between Majorana fermions with coefficients $t_{i}$ and $s_{i}$, we incorporate some  higher-order perturbation theory results compared to those specified in the previous bullet point. 
    
    The coupling $t_{i}$ is set to zero when $m_{i-1}=m_{i}=1$: In this case, none of the two bonds involved in the $-t_{i}T_{i}$ term is strongly antiferromagnetic.  One expects that because $t_{i}$ is generated from higher-order perturbation theory, it will be much smaller than the Ising couplings $J_{i}$ and flip terms $h_{i}$'s generated from the second-order perturbation theory, and that its effect is presumably negligible. However, within a spin cluster linked by strongly antiferromagnetic $m_{i}=0$ bonds, $t_{i}$ mediates nearest-neighbor Majorana fermion hoppings, and to capture the physics correctly within such a cluster, $t_{i}$ should be kept track of whenever $m_{i-1}$ or $m_{i}$ is 0. 

     Similarly, $s_{i}$ is only non-zero when $m_{i-1}=m_{i+1} = 0$ and $m_{i} = 1$ because, as we saw in Sec.~\ref{sec:sec3rule} and Appendix~\ref{app:df}, this term only plays a role of keeping track of the nearest-neighbor Majorana fermion hopping upon encountering this specific configuration of $m$'s 
     mid-RG. It involves incorporating higher-order perturbation results than keeping track of any other couplings, and consequently its effect should be negligible except for the specific case in which we set $s_{i}$ to be non-zero.

    \item We drop any contributions from multi-site flip terms involving three or more sites. We incorporate contributions from double-site flip terms when there is no bond with $m_{i}=0$ within the second nearest-neighbor of the double site-flip term of our interest,  on the grounds that when two antiferromagnetic bonds are close to each other, the flip terms are already renormalized to a very small magnitude. (See Appendix~\ref{app:more} for a more detailed argument.)
\end{itemize}
While the RG rules are admittedly complicated, having this protocol in mind helps one to understand the underlying organizing principles and its derivation.

 Finally, we introduce the following notation that will be useful for writing down the RG rules:
 \begin{equation}
\text{amax}\left( a,b\right)= \begin{cases}
a & |a| > |b| \\
b & |b| > |a|
\end{cases},
\end{equation}
i.e., the function above outputs the number with largest absolute value. We will use this notation to implement the ``max'' rule for the RG transformation in which, for example, a term with coefficient $a+b+c$ after the transformation receives a modified coefficient $\text{amax}(a,b,c)$, owing to the fact that with the strong-disorder assumption, the coefficients $a$, $b$, and $c$ are likely to have wildly different magnitudes.

\subsection{Decimating an antiferromagnetic bond at $j$} 

The transformation rule in this case reads:
\begin{equation}
m_{j} = 1 \rightarrow m_{j} =0 
\end{equation}
\begin{equation}
\label{eq:daf_2}
\begin{split}
& J_{j-1} \rightarrow \text{amax} \left( J_{j-1},  m_{j-1} \frac{(\sqrt{2} -1 ) h_{j}^{2}}{8|J_{j}|}\right) \\
& J_{j+1} \rightarrow \text{amax} \left( J_{j+1}, m_{j+1} \frac{(\sqrt{2} -1 ) h_{j+1}^{2}}{8|J_{j}|} \right)
\end{split}
\end{equation}
\begin{equation}
\begin{split}
& t_{j} \rightarrow \text{max} \left( t_{j}, m_{j-1} \frac{h_{j}^{2}}{8|J_{j}|} \right) \\
& t_{j+1} \rightarrow \text{max} \left(  t_{j+1}, m_{j+1}\frac{h_{j+1}^{2}}{8|J_{j}|} \right)
\end{split}
\end{equation}
\begin{equation}
\label{eq:afjisi}
\begin{split}
& s_{j} \rightarrow 0 \\
& s_{j-1} \rightarrow m_{j-1} \frac{1}{4\sqrt{2}}\frac{t_{j-2} h_{j}^{2}}{|J_{j}|^{2}} \\
& s_{j+1} \rightarrow m_{j+1} \frac{1}{4\sqrt{2}}\frac{t_{j+1} h_{j+1}^{2}}{|J_{j}|^{2}}.
\end{split}
\end{equation}
(For this case only, the system length does not change, and the Hamiltonian is determined by the same set of parameters before/after the transformation. Hence, for brevity, above we simply list parameters that undergo non-trivial transformations; we adopt different notation for the transformation rules in other cases.)
The transformation rule for $J_{i}$ and $t_{i}$ is captured by the second-order perturbation theory mostly identical to the one we saw in Eqs.~\eqref{eq:afpt}, \eqref{eq:afpt2nd} and \eqref{eq:afptsquare}. Key differences are:
\begin{itemize}
    \item In the fully general strong diorder RG we consider here, antiferromagnetic Ising interactions are also random. So $\Lambda_{A}$ in those equations should be $|J_{j}|$ for application to this case.
    \item In Eq.~\eqref{eq:afptsquare}, we dropped the contributions to the Ising interaction due to the special assumption on the energy scales of each term made for the setup in Appendix~\ref{app:df}. Here, however, one has to incorporate this nearest-neighbor Ising interaction as one can see in Eq.~\eqref{eq:daf_2}.
    \item $h_{j}$, $F_{j}$, $h_{j+1}$ and $F_{j+1}$ can be a part of multi-site flip terms and corresponding coefficients; meanwhile, in the setup of Sec.~\ref{sec:splim} and Appendix~\ref{app:df}, by assuming that antiferromagnetic Ising interactions 
    dominate and that antiferromagnetic bonds do not appear in the close neighborhood of each other, we avoided this situation. As explained in the main text and the beginning of this appendix, we mostly ignore terms generated from multi-site flip terms. Note that most coefficients generated in the above equations have $m_{j-1}$ or $m_{j+1}$ in the coefficients, denoting that if ($h_{j}$, $F_{j}$) or ($h_{j+1}$, $F_{j+1}$) is a part of a multi-site flip term, the corresponding coefficients vanish.
\end{itemize}
Also, we remark that the transformation rule $m_{j}=1 \rightarrow m_{j}=0$ suffices to keep track of the multi-site flip term generated from decimating an antiferromagnetic bond at $j$---which can be shown by employing associativity and commutativity of the $\odot$ operation introduced in Eqs.~\eqref{eq:symp} and \eqref{eq:symp2}. 

The transformation rule for $s_{i}$'s in Eq.~\eqref{eq:afjisi} is a new feature and is derived as follows. By marking $m_{j}=0$, the term $-2 s_{j}S_{j}$ cannot act non-trivially anymore because $S_{j}$ is only non-zero when $\sigma_{j}^{z} = \sigma_{j+1}^{z}$; yet by decimating the antiferromagnetic bond $j$, we are assuming $\sigma_{j}^{z} =-\sigma_{j+1}^{z}$ to be a hard constraint. Hence, we just set $s_{j}$ to be zero when decimating the strong antiferromagnetic bond $j$.
Meanwhile, if $m_{j-2} = 0$ and $m_{j-1} =1$, decimating the antiferromagnetic bond at site $j$ creates a configuration where we should assign non-zero $s_{j-1}$ values. We generate $s_{j-1}$ from the third-order perturbation theory. Observe that since $m_{j-1}=1$, the site $j$ has a single-site flip term before decimation. In the third-order formula in Eq.~\eqref{eq:3rdpgen}, one can show that selecting the first and the third $H_{p}$ to be $-h_{j}F_{j}$ and the second $H_{p}$ to be $-2t_{j-2}T_{j-2}$ generates the term we desire, corresponding to the second line in Eq.~\eqref{eq:afjisi}. One can similarly show that there is a third-order perturbation theory term that generates $s_{j+1}$ in the third line of Eq.~\eqref{eq:afjisi}.
 
\subsection{Decimating a ferromagnetic bond at $j$} 

In this case, the parameters $\{m'_{i}, h'_{i}, J'_{i}, t'_{i}, s'_{i} \}$ after the decimation, in terms of the parameters $\{m_{i}, h_{i}, J_{i}, t_{i}, s_{i} \}$ before the decimation, is given by:

\begin{equation}
m'_{i} = \begin{cases}
m_{i} & i < j \\
m_{i+1} & i \geq j
\end{cases}
\end{equation}
\begin{equation}
h'_{i} = \begin{cases}
h_{i} & i < j \\
\frac{2^{-1/4} h_{j} h_{j+1}}{J_{j}} & i = j \\
h_{i+1} & i> j
\end{cases} 
\end{equation}
\begin{equation}
J'_{i} = \begin{cases}
J_{i} & i < j-2 \\
\text{amax} \left( J_{j-2}, m_{j-2} (1-m_{j-1}) \frac{h_{j-1}^{2}h_{j}^{2}}{16\sqrt{2}J_{j}J_{j-1}^{2}} \right)  & i = j -2 \\
\text{amax} \left( J_{j-1}, m_{j-1} \frac{(\sqrt{2} - 1) h_{j}^{2}}{4\sqrt{2}J_{j}} \right)  & i = j -1 \\
\text{amax} \left( J_{j+1}, m_{j+1} \frac{(\sqrt{2} - 1) h_{j+1}^{2}}{4\sqrt{2}J_{j}}\right)  & i = j  \\
\text{amax} \left( J_{j+2}, m_{j+2} (1-m_{j+1}) \frac{h_{j+1}^{2}h_{j+2}^{2}}{16\sqrt{2}J_{j}J_{j+1}^{2}} \right) & i = j+1  \\
J_{j+1}  & j > i+1
\end{cases}
\end{equation}
\begin{equation}
\label{eq:fjiti}
t'_{i} = \begin{cases}
t_{i} & i < j-1 \\
\max \left(t_{j-1} , m_{j-2} (1-m_{j-1})\frac{(1 + \sqrt{2} ) h_{j-1}^{2}h_{j}^{2}}{32 J_{j}J_{j-1}^{2}} \right)  & i = j -1 \\
\max \left(m_{j+1}\frac{1}{4\sqrt{2}}\frac{t_{j} h_{j+1}^{2}}{ J_{j}^{2}}, m_{j-1}\frac{1}{4\sqrt{2}} \frac{t_{j+1} h_{j}^{2}}{ J_{j}^{2}}, s_{j} \right)  & i = j  \\
\max \left( t_{j+2}, m_{j+2} (1-m_{j+1}) \frac{(1 + \sqrt{2} ) h_{j+1}^{2}h_{j+2}^{2}}{32J_{j}J_{j+1}^{2}} \right)  & i = j +1 \\
t_{i+1} & i>j+1
\end{cases}
\end{equation}
\begin{equation}
\label{eq:fjisi}
s'_{i} = \begin{cases}
s_{i} & j < i-1  \\
m_{j-1} \frac{t_{j-1}t_{j+1} h_{j}^{2}}{2J_{j}^{3}} & i = j-1 \\
m_{j+1} \frac{t_{j}t_{j+2} h_{j+1}^{2}}{2J_{j}^{3}}  & i = j \\
s_{i+1} & j>i
\end{cases}.
\end{equation}
Notice that most of the coefficients are simply identical to those before the transformations -- only coefficients corresponding to the terms around the bond $j$ are non-trivially modified. Hereafter, for other types of decimation rules as well, we will simply focus on how the non-trivial transformation rules around the site/bond of interest are obtained.
 
 Appendix~\ref{app:splimitf} already covered the essential details regarding the derivation of the above rule, so here we only highlight the key differences:
\begin{itemize}
    \item The $\Lambda_{A}$'s in Appendix~\ref{app:splimitf} should be substituted for the Ising interaction coefficients $|J_{i}|$'s for the corresponding bonds.
    \item In Appendix~\ref{app:splimitf}, we ignored all contributions from $-\frac{c_{l_{j}}^{2} }{2J_{j}} C_{l_{j}}^{2}$ and $-\frac{c_{r_{j}}^{2} }{2J_{j}} C_{r_{j}+1}^{2}$ in Eq.~\eqref{eq:f2ndfull}. However, in the full strong-disorder RG, if these flip terms involve a single site or double site, we include these contributions in the decimation rule. Contributions from these terms are computed in Eq.~\eqref{eq:f2ndsquare1} and Eq.~\eqref{eq:f2ndsquare2}. We drop these contributions if flip terms involve three or more sites; in this case, these contributions may lead to non-nearest neighbor  Majorana fermion interactions. As mentioned in deriving the rule for the antiferromagnetic bond decimation, we argue in Appendix~\ref{app:more} that these terms are much smaller than $-2t_{i}T_{i}$'s and may be dropped.
\end{itemize}

\subsection{Single-site decimation at site $j$}

 For this decimation, we project onto eigenstates of $h_{j} F_{j}$ with the two lowest eigenvalues. These eigenstates exhibit a disordered Ising spin at $j$, i.e., $\langle \sigma_{j}^{z} \rangle = 0$, and the states with two different eigenvalues are distinguished by whether two Ising spins at $j-1$ and $j+1$ are aligned ($\sigma_{j-1}^{z} = \sigma_{j+1}^{z}$) or anti-aligned ($\sigma_{j-1}^{z} = -\sigma_{j+1}^{z}$). One may treat this ``projection'' as removing the disordered Ising spin $\sigma_{j}^{z}$ from the system---effectively decreasing the system size by 1. The new parameters are given as follows, incorporating first-order as well as some second-order corrections:
\begin{equation}
\label{eq:himi}
m'_{i} = \begin{cases}
m_{i} & i < j \\
0 & i = j  \\
m_{i+1} & i> j
\end{cases}
\end{equation}
\begin{equation}
\label{eq:hihi}
h'_{i} = \begin{cases}
h_{i} & i < j-1 \\
\frac{(1+2^{-1/4})h_{j-1}}{2} & i = j-1 \\
\frac{(1+2^{-1/4})h_{j+1}}{2} & i= j \\
h_{i+1} & i>j
\end{cases}
\end{equation}
\begin{equation}
\label{eq:hiji}
J'_{i} = \begin{cases}
J_{i} & i < j-1 \\
\frac{(1-2^{-1/4})}{2} h_{j} & i = j-1  \\
J_{i+1} & i> j-1
\end{cases}
\end{equation}
\begin{equation}
\label{eq:hiti}
t'_{i} = \begin{cases}
t_{i} & i < j-1 \\
\frac{t_{j-1}}{2} & i = j-1 \\
\frac{t_{j+1}}{2} & i = j \\
t_{i+1} & i>j
\end{cases}
\end{equation}
\begin{equation} 
\label{eq:hisi}
s'_{i} = \begin{cases}
s_{i} &   i < j-1  \\
\frac{2 t_{j-1}t_{j+1}}{h_{j}} & i = j-1 \\
s_{i+1} & i > j-1
\end{cases}.
\end{equation}

We set up the perturbation theory leading to this transformation by writing the Hamiltonian as
\begin{equation}
\label{eq:hipsetup}
\begin{split}
& H = H_{0} + H_{p}, \\
& H_{0} = - h_{j} F_{j} \\
& H_{p} = - c_{j-1} C_{j-1} - c_{j+1} C_{j+1} - 2 t_{j-1} T_{j-1} - 2 t_{j+1} T_{j+1}  + \cdots
\end{split}
\end{equation}
Here $- c_{j-1} C_{j-1}$ and $ - c_{j+1} C_{j+1} $ are multi-site flip terms associated with sites $j-1$ and $j+1$, respectively; terms that commute with $-h_{j}F_{j}$ and hence do not provide perturbative corrections are contained in the ellipsis. 
All the results except for Eq.~\eqref{eq:hisi} are derived from the first-order perturbation theory, i.e., by projecting the existing terms onto the low-energy eigenstates we specified. Equation~\eqref{eq:himi} simply reflects the fact that after removing site $j$, the bond connecting what was originally site $j-1$ and site $j+1$ should carry a label $0$. Equation~\eqref{eq:hiji} comes from the difference in the energy between states with $\sigma_{j-1}^{z} = \sigma_{j+1}^{z}$ and states with $\sigma_{j-1}^{z} = -\sigma_{j+1}^{z}$. 
 Equations~\eqref{eq:hihi} and \eqref{eq:hiti} are simply obtained by projecting the terms in $H_{p}$ of Eq.~\eqref{eq:hipsetup}. 

 After showing that Eq.~\eqref{eq:hihi} holds when $C_{j-1}$ and $C_{j+1}$ are single-site flip terms, one can show through the trick employed in Eqs.~\eqref{eq:f2ndflipsub} and \eqref{eq:f2ndflip} that this rule also correctly describes how multi-site flip terms evolve after the transformation. Generally, for any transformation rules for $h_{i}$'s in other types of decimations as well, using a similar trick shows that the same transformation rule applies regardless of whether $h_{i}$'s are parts of multi-site flip terms or single-site flip terms.

Equation~\eqref{eq:hisi} requires going to second-order perturbation theory. If $m_{j-2} = m_{j+1} = 0$, decimating site $j$ produces a configuration where $m'_{j-2}=m'_{j} = 0$ and $m'_{j-1} =1$.   Hence, one must go beyond first-order perturbation theory employed elsewhere in the rule to generate $-2 S_{j-1}$. Recall that degenerate perturbation theory gives the following energy correction at second order,
\begin{equation}
\label{eq:2ndog}
    E^{(2)}_{ab} = \sum_{k, E_{k} \neq E_{0}} \frac{\bra{a} H_{p} \ket{k}\bra{k} H_{p} \ket{b}}{E_{0}- E_{k} }.
\end{equation}
Choosing the first and second $H_p$ to be $-2 t_{j-1}T_{j-1}$ and $-2 t_{j+1}T_{j+1}$, respectively, and vice versa, yields the nonzero $s'_{j-1}$ in the second line of Eq.~\eqref{eq:hisi}.

\subsection{Decimating a double site covering $j$ and $j+1$}

This transformation rule is given by:
\begin{equation}
m'_{i} = \begin{cases}
m_{i} & i < j-1 \\
0 & i = j-1  \\
m_{i+2} & i> j-1
\end{cases}
\end{equation}
\begin{equation}
h'_{i} = \begin{cases}
h_{i} & i < j-1 \\
\frac{2^{-1/4}+\sqrt{\frac{1+\sqrt{2}}{2\sqrt{2}}}}{2} h_{j-1} & i = j-1 \\
\frac{2^{-1/4}+\sqrt{\frac{1+\sqrt{2}}{2\sqrt{2}}}}{2} h_{j+2} & i = j \\
h_{i+2} & i>j
\end{cases}
\end{equation}
\begin{equation}
J'_{i} = \begin{cases}
J_{i} & j < i-1 \\ \frac{1}{2} \left( 1- \frac{\sqrt{2}}{\sqrt{1+\sqrt{2}}} \right)  d_{j} & i = j-1 \\
J_{i+2} & j> i-1
\end{cases}
\end{equation}
\begin{equation}
\label{eq:diti}
t'_{i} = \begin{cases}
t_{i} & i < j-1 \\
\frac{1}{2\sqrt{2}}t_{j-1} & i = j-1 \\
\frac{1}{2\sqrt{2}}t_{j+2} & i = j \\
t_{i+2} & i>j
\end{cases}
\end{equation}
\begin{equation}
s'_{i} = \begin{cases}
s_{i} &   i < j-1 \\
\frac{t_{j-1}t_{j+2}}{d_{j}} & i = j-1 \\
s_{i+2} & i > j-1
\end{cases}.
\end{equation}
The derivation is almost identical to that presented for the single-site decimation case and is hence omitted.

\subsection{Double-bond decimation associated with $T_{j}$}

\subsubsection{$m_{j} = m_{j-1} = 0$}

\label{sec:tjmj0mj10}

For this case the rule is:
\begin{equation}
m'_{i} = \begin{cases}
m_{i} & i < j-1 \\
m_{i+2} & i \geq j-1
\end{cases}
\end{equation}
\begin{equation}
h'_{i} =  \begin{cases}
h_{i} & i < j-1 \\
\frac{(1+2^{1/2})h_{j-1} h_{j} h_{j+1}}{4J_{j}J_{j+1}} & i = j-1 \\
h_{i+2} & i>j-1
\end{cases}
\end{equation} 
\begin{equation}
J'_{i} = \begin{cases}
J_{i} & i \leq j-2 \\
J_{i+2} & i > j-2
\end{cases}
\end{equation}
\begin{equation}
\label{eq:ti1ti}
t'_{i} = \begin{cases}
t_{i} & j < i-1 \\
(1- m_{j-2} m_{j+1}) \frac{t_{j-1}t_{j+1}}{t_{j}} & i=j-1 \\
t_{i+2} & i>j-1
\end{cases}
\end{equation}
\begin{equation}
\label{eq:ti1si}
s'_{i} = \begin{cases}
s_{i} & i < j-1 \\
(1-m_{j+1})\frac{s_{j-2}t_{j+1}}{t_{j}} & i = j -2 \\
(1-m_{j-2})\frac{s_{j+1}t_{j-1}}{t_{j}}   & i = j-1  \\
s_{i+2} & i> j-1
\end{cases}.
\end{equation}
To this end we project onto the lowest-energy state of $H_{0} = -2 t_{j} T_{j}$. The rule for $m_{i}$'s and $J_{i}$'s corresponds to simply removing the two bond terms at $j-1$ and $j$, while the rule for $h_{i}$'s follows by projecting the multi-site flip terms involving sites $j-1$, $j$, and $j+1$ into the low-energy subspace, equivalent to applying the first-order perturbation theory. For the $h_{i}$'s, one may derive the rule when sites $j-1$, $j$, and $j+1$ are not linked by any other $m_{i}=0$ bonds, i.e., when $m_{j-2}= m_{j+2} = 0$; the usual trick can be employed to show that this rule applies in more general cases where $m_{j-2} \neq 0$ or $m_{j+2} \neq 0$ as well.

 To keep track of $t_{i}$ and $s_{i}$, we employ second-order perturbation theory. From the second-order formula in Eq.~\eqref{eq:2ndog}, choosing two $H_{p}$'s to be $-2 t_{j-1}T_{j-1}$ and $-2 t_{j+1}T_{j+1}$ (in either order) leads to the non-trivial rule for $t'_{j}$ in the second line of Eq.~\eqref{eq:ti2ti}. Note that when $m_{j-2}=m_{j+1} =1$, we assume $t'_{j-1}=0$ after the transformation -- hence the factor $(1-m_{j-2}m_{j+1})$.
 Similarly, when $m_{j-3} = 0$, $m_{j-2}=1$, and $m_{j+1} = 0$, after the decimation we encounter the situation in which $m'_{j-3} = m'_{j-1} = 0$ and $m'_{j-2}= 1$. Hence, one should keep track of $s'_{j-2}$. In this case, from Eq.~\eqref{eq:2ndog}, choosing two $H_{p}$'s to be $-2 s_{j-2}S_{j-2}$ (convince yourself that this term is non-zero in our case) and $-2 t_{j+1}T_{j+1}$ in either order leads to the non-trivial renormalization of $s'_{j-2}$ in the second line of Eq.~\eqref{eq:ti1si}. Similar considerations for the case $m_{j-1}= m_{j+2} =0$, $m_{j+1}=1$ lead to the third line of Eq.~\eqref{eq:ti1si}.
 
\subsubsection{$m_{j}=1$ and $m_{j-1} =0$}
\label{app:tib}

In this case, the rule is given by: 
\begin{equation}
m'_{i} = \begin{cases}
m_{i} & i < j-1 \\
m_{i+2} & i \geq j-1
\end{cases}
\end{equation}
\begin{equation}
\label{eq:ti2hi}
h'_{i} =  \begin{cases}
h_{i} & i < j-1 \\
\frac{(1+2^{1/2})h_{j-1} h_{j} h_{j+1}}{4t_{j}J_{j}} & i = j-1 \\
h_{i+2} & i>j-1
\end{cases}
\end{equation} 
\begin{equation}
\label{eq:ti2ji}
J'_{i} = \begin{cases}
J_{i} & i < j-2 \\
\text{amax} \left( J_{j-1}, m_{j-2}\frac{(\sqrt{2}+1)h_{j-1}^{2}h_{j}^{2}}{32\sqrt{2} t_{j} J_{j-1}^{2}}, -m_{j-2} \frac{t_{j-1}^{2}}{2 t_{j}} \right) & i = j -2 \\
\text{amax} \left( J_{j+2}, m_{j+1} \frac{(\sqrt{2}-1)h_{j+1}^{2}}{8t_{j}} \right)   & i = j-1  \\
J_{i+2} & i> j-1
\end{cases}
\end{equation}
\begin{equation}
\label{eq:ti2ti}
t'_{i} = \begin{cases}
t_{i} & i < j-1 \\
0 & i = j-1, m_{j-2}m_{j+1} = 1 \\
\frac{h_{j+1}^{2}}{4t_{j}^{2}} t_{j-1} & i = j-1, \, m_{j-2}=0, \, m_{j+1} =1 \\
\frac{t_{j-1}t_{j+1}}{t_{j}} & i = j-1, \text{all other cases}\\
t_{i+2} & i>j-1
\end{cases}
\end{equation}
\begin{equation}
\label{eq:ti2si}
s'_{i} = \begin{cases}
s_{i} & i < j-1 \\
(1-m_{j+1}) \frac{s_{j-2}t_{j+1}}{t_{j}}  & i = j -2 \\
m_{j+1} (1-m_{j-2}) \frac{3 t_{j-1}t_{j+2} h_{j+1}^{2}}{4\sqrt{2}t_{j}^{3}}   & i = j -1 \\
s_{i+2} & i> j-1
\end{cases}.
\end{equation}
We set up the degenerate perturbation leading to the above by writing
\begin{equation}
\begin{split}
    & H = H_{0} + H_{p}, \, H_{0} = - 2 t_{j}T_{j}, \\
    & H_{p} = - c_{l_{j}} C_{l_{j}} - c_{r_{j}} C_{r_{j}}  \\
    & \quad   -2 s_{j-2} S_{j-2} - 2 s_{j+1} S_{j+1} - 2 t_{j-1}T_{j-1} - 2 t_{j+1}T_{j+1} + \cdots.
\end{split}
\end{equation}
Here, $- c_{l_{j}} C_{l_{j}}$ and $- c_{r_{j}} C_{r_{j}} $ are multi-site flip terms associated with sites $j-1$ (or equivalently $j$ since $m_{j-1} = 0$) and $j+1$ respectively. As before, the ellipsis encodes Hamiltonian terms that commute with $H_{0}$. We will incorporate all the second-order perturbation theory results and some terms generated from higher order for the second line in Eq.~\eqref{eq:ti2ti} and the third line in Eq.~\eqref{eq:ti2si}. For the second-order perturbation theory formula in Eq.~\eqref{eq:2ndog}, if we consider a term with one $H_{p}$ chosen to be a flip term $- c_{l_{j}} C_{l_{j}}$, the other $H_{p}$ should be also be a flip term (either $- c_{r_{j}} C_{r_{j}}$ or $- c_{l_{j}} C_{l_{j}}$) for the matrix elements to be non-vanishing. Hence, one can separate purely flip-term contributions, and non-flip-term contributions wherein none of $H_{p}$ is chosen to be flip term. 

Consider the flip-term contribution when $m_{j-2} =m_{j+1} = 1$, where $-c_{l_{j}} C_{l_{j}}$  is simply a double-site flip term and $c_{r_{j}} C_{r_{j}}$ is a single-site flip term. One can therefore write
 \begin{equation}
 \begin{split}
  & c_{l_{j}} C_{l_{j}}  = \frac{h_{j-1}h_{j}}{|J_{j-1}|} (F_{j-1} \odot F_{j}),  \\
  & c_{r_{j}} C_{r_{j}} = h_{j+1} F_{j+1}.
  \end{split}
 \end{equation}
 Using similar logic as in deriving the decimation rule for the ferromagnetic and antiferromagnetic bonds, one can show that flip-term contributions to the second-order perturbation theory are
 \begin{equation}
 \label{eq:ti2flip2}
 \begin{split}
     H^{(2)}_{F} & = \frac{h_{j-1}h_{j}h_{j+1}}{|J_{j-1}| t_{j}}  \mathcal{P}_{T_{j}} (F_{j-1} \odot F_{j} \odot F_{j+1}) \mathcal{P}_{T_{j}} \\
     & \quad+  \frac{h_{j-1}^{2}h_{j}^{2}}{2|J_{j-1}|^{2} t_{i}} \mathcal{P}_{T_{j}} (F_{j-1} \odot F_{j})^{2} \mathcal{P}_{T_{j}} + \frac{h_{j+1}^{2}}{2t_{i}} \mathcal{P}_{T_{j}} F_{j+1}^{2} \mathcal{P}_{T_{j}} 
 \end{split}
 \end{equation}
with $\mathcal{P}_{T_{j}}$ the projection operator to the lowest-energy eigenstates of $H_{0} = -2 t_{j}T_{j}$. The first term on the right side of Eq.~\eqref{eq:ti2flip2} simultaneously flips spins at sites $j-1$, $j$, and $j+1$, and after the projection and the decimation procedure, it becomes a flip term for the super-site in which the aforementioned three sites are combined. This contribution is given in the second line of Eq.~\eqref{eq:ti2hi}, with the numerical factor $\frac{(1+2^{1/2})}{4}$ coming from the projection $\mathcal{P}_{T_{j}}$. Note that while we derived the rules in Eq.~\eqref{eq:ti2hi} for the special case where $m_{j-2}=m_{j+1}=1$, using a trick similar to that used for Eqs.~\eqref{eq:f2ndflipsub} and \eqref{eq:f2ndflip} shows that this rule applies to other cases in which $m_{j-2}$ or $m_{j+1}$ is 0, and the super-site formed after the decimation is still linked by some bond with $m_{i}=0$. The remaining two terms in Eq.~\eqref{eq:ti2flip2} do not flip any spins; rather, they can be shown to generate Ising interaction: 
\begin{equation}
    \frac{h_{j-1}^{2}h_{j}^{2}}{2|J_{j-1}|^{2} t_{i}} \mathcal{P}_{T_{j}} (F_{j-1} \odot F_{j})^{2} \mathcal{P}_{T_{j}} \equiv \frac{(\sqrt{2}+1)h_{j-1}^{2}h_{j}^{2}}{32\sqrt{2} t_{j} J_{j-1}^{2}} \sigma_{j-2}^{z}\sigma_{j-1}^{z}
\end{equation}
\begin{equation}
    \frac{h_{j+1}^{2}}{2t_{i}} \mathcal{P}_{T_{j}} F_{j+1}^{2} \mathcal{P}_{T_{j}}  \equiv  \frac{(\sqrt{2}-1) h_{j+1}^{2}}{8t_{i}} \sigma_{j+1}^{z}\sigma_{j+2}^{z}.
\end{equation}
We included these terms in the transformation rule for $J'_{i}$'s.  If $m_{j-2}=0$, the flip term containing sites $j-1$ and $j$ flips three or more spins. We will ignore this flip-term-squared contribution as mentioned at the beginning of this Appendix. Similarly, if $m_{j+1}=0$, we ignore flip-term-squared contributions associated with site $j+1$. 

Finally, we discuss second-order perturbation theory for non-flip-term contributions: 
\begin{itemize}
    \item Choosing both $H_{p}$'s in Eq.~\eqref{eq:2ndog} to be $- 2 t_{j-1}T_{j-1}$  leads to the Ising interaction
    \begin{equation}
        \frac{t_{i-1}^{2}}{2t_{i}} \sigma_{j-2}^{z} \sigma_{j-1}^{z}.
    \end{equation}
    We included this contribution in the second line of Eq.~\eqref{eq:ti2ji}, in the third argument of $\text{amax}$.  We ignore this contribution when $m_{j-2} =0$; in this case, the bond at $j-2$ is already decimated, and $|J_{j-2}|$ is likely much larger than this newly generated Ising interaction. An analogue of this term generated upon choosing $H_{p}$'s to be $- t_{j+1}T_{j+1}$ can be thrown out for similar reasons: For $t_{j+1} \neq 0$ in our scheme, $m_{j+1} =0$ always. The generated Ising interaction mediates coupling between two spins joined by the bond $j+1$, which already has been decimated, and hence is likely to have a much larger energy scale than the generated Ising interaction.
    \item Choosing one $H_{p}$ to  be $- 2 t_{j-1}T_{j-1}$ and the other to be $- 2 t_{j+1}T_{j+1}$, in either order, yields the correction
    \begin{equation}
        \frac{t_{j-1}t_{j+1}}{t_{j}} T'_{j-1}.
    \end{equation}
    This contribution is reflected in the fourth line of Eq.~\eqref{eq:ti2ti}. Note that if $m_{j-2}=m_{j+1} = 1$, this contribution vanishes because $t_{j+1} =0$, as intended.
    \item Choosing one $H_{p}$ to be $- 2 t_{j+1}T_{j+1}$ and the other to be $- 2 s_{j-2}S_{j-2}$ leads to the contribution in the second line of Eq.~\eqref{eq:ti2si}.
    \item We drop a term generated from choosing both $H_{p}$'s to be $s_{j-2}S_{j-2}$. One can show that this term corresponds to an interaction between spins at sites $j-3$, $j-2$ and $j-1$. The rationale is that $S_{j}$ is generated through employing higher-order perturbation theory in other decimations, and although it generate three-body Ising interactions, its contribution is generally much smaller than two-body nearest-neighbor Ising interactions generated through lower-order perturbation theory.
\end{itemize}

We have now exhausted the possible terms that appear in second-order perturbation theory, and explained which terms should be retained versus dropped.  Finally, we discuss higher-order contributions. Recall that second-order perturbation theory does not generate a nonzero $t'_{j-1}$ when $m_{j+1} = 1$. Similarly, second-order perturbation theory fails to generate non-zero $s'_{j-1}$ when $m_{j+1}=1$, $m_{j+2}=0$ and $m_{j-2} =0$. For the first instance, we generate nontrivial $t'_{j-1}$ at third-order in perturbation theory by choosing the first and the third $H_{p}$'s in Eq.~\eqref{eq:3rdpgen} to be $-h_{j+1}F_{j+1}$ (due to the condition on the $m_{i}=0$ mark, the flip term associated with site $j+1$ is a single-site flip term), and the second $H_{p}$ to be $-2 t_{j-1}T_{j-1}$.  To generate non-trivial $s'_{j-1}$, we go to fourth order.  The following three choices for $H_{p}$'s in Eq.~\eqref{eq:4thpgen} are relevant:
  \begin{itemize}
      \item The first and third $H_{p}$'s are set as $-h_{j+1}F_{j+1}$, the second  is chosen to be $-2 t_{j+2}T_{j+2}$, and the last one is $-2 t_{j-1}T_{j-1}$.
      \item  The second and fourth $H_{p}$'s are set as $-h_{j+1}F_{j+1}$, the third is $-2 t_{j+2}T_{j+2}$, and the first is $-2 t_{j-1}T_{j-1}$.
      \item  The first and  fourth $H_{p}$'s are $-h_{j+1}F_{j+1}$, while the second and third are chosen to be $-2 t_{j+2}T_{j+2}$ and $-2 t_{j-1}T_{j-1}$, in either order.
  \end{itemize}
  These contributions together generate the term in the third line of Eq.~\eqref{eq:ti2si}.
 
\subsubsection{Other cases}
The transformation rule when $m_{j-1}=1$ and $m_{j} =0$ is simply a spatially inverted version of the rule for $m_{j-1}=0$ and $m_{j} =1$. Finally, if $m_{j-1} = m_{j}=1$, we assume that $t_{j}=0$, and there is no decimation rule in this case.

\subsection{Triple bond decimation}
 The triple bond decimation induced by $-s_{j} S_{j}$ brings the following changes to the Hamiltonian parameters:
 
\begin{equation}
m'_{i} = \begin{cases}
m_{i} & i \leq j-2 \\
m_{i+3} & i> j-2
\end{cases}
\end{equation}
\begin{equation} 
\label{eq:sihi}
h'_{i} = \begin{cases}
h_{i} & i < j-1 \\
\frac{2^{-1/4}+2^{1/4}}{4}\frac{h_{j-1}h_{j}h_{j+1}h_{j+2}}{ s_{j} J_{j-1} J_{j+1}} & i = j-1 \\
h_{i+3} & i>j-1
\end{cases}
\end{equation}
\begin{equation}
J'_{i} = \begin{cases}
J_{i} & i \leq j-2 \\
J_{i+3} & i > j-2
\end{cases}
\end{equation}
\begin{equation}
\label{eq:siti}
t'_{i} = \begin{cases}
t_{i} & i < j-1 \\
(1-m_{j-2}m_{j+1}) \frac{t_{j-2}t_{j+1}}{s_{j}} &  i = j-1 \\
t_{i+3} & i>j-1
\end{cases}
\end{equation}
\begin{equation}
\label{eq:sisi}
s'_{i} = \begin{cases}
s_{i} &   i < j-2 \\
m_{j+2}\frac{s_{j-2}t_{j+1}}{s_{j}} & i = j-2 \\
m_{j-2}\frac{s_{j+2}t_{j-2}}{s_{j}} & i = j-1 \\
s_{i+3} & i > j-1
\end{cases}.
\end{equation}
All these results are derived from second-order perturbation theory. The procedure shares many details with the case covered in Appendix~\ref{app:tib}. We start with the setup
\begin{equation}
\begin{split}
   &  H = H_{0} + H_{p}, \, \\
   & H_{0} = -2s_{j} S_{j}, \\
   & H_{p} = - c_{l_{j}} C_{l_{j}} - c_{r_{j}} C_{r_{j}} -2 t_{j-1} T_{j-1} - 2t_{j+2} T_{j+2} \\
   & \quad \quad \, \, -2 s_{j-2} S_{j-2} - 2s_{j+1} S_{j+1} + \cdots.
\end{split}
\end{equation}
 Similarly as before, $- c_{l_j} C_{l_j}$ and $- c_{r_j} C_{r_j}$ are multi-site flip terms that cover spins on the left and right of the bond $j$ respectively. 
In our setup, if $s_{j} \neq 0$, then $m_{j-1} = m_{j+1} =0$. Hence, both flip terms are multi-site flip terms that cover at least two sites. As in Appendix~\ref{app:tib}, one can divide terms generated at second order into flip-term contributions and non-flip-term contributions.

  The flip-term contributions are given by
  \begin{equation}
 \begin{split}
     H^{(2)}_{F}= &  \frac{c_{l_{j}} c_{r_{j}}}{ s_{j}}  \mathcal{P}_{S_{j}} (C_{l_{j}} \odot C_{r_{j}}) \mathcal{P}_{S_{j}}  \\
     & \quad \quad +  \frac{c_{l_{j}}^{2}}{2s_{j}} \mathcal{P}_{S_{j}} C_{l_{j}}^{2} \mathcal{P}_{S_{j}}  +  \frac{c_{r_{j}}^{2}}{2s_{j}} \mathcal{P}_{S_{j}} C_{r_{j}}^{2}  \mathcal{P}_{S_{j}},
 \end{split}
 \end{equation}
where $\mathcal{P}_{S_{j}}$ is analogous to $\mathcal{P}_{T_{j}}$ but instead projects onto the ground states of $H_{0} = -2s_{j} S_{j}$. One can show that when $m_{j-2} = m_{j+3} = 1$, the first term is equivalent to a single-site flip term after the decimation, whose coefficient appears in the second line of Eq.~\eqref{eq:sihi}. While derived for a simple case of $m_{j-2} = m_{j+3} = 1$, with the trick we employed, this rule works even when either or both of $m_{j-2}$ or $m_{j+3}$ are $0$.  As for other pieces, upon following the argument in Appendix~\ref{app:more}, one may discard the flip-term-squared contributions in the second line. 

  For non-flip-term contributions, we have the option to choose either $-2 t_{j-1} T_{j-1}$,  $- 2t_{j+2} T_{j+2}$,  $ -2 s_{j-2} S_{j-2}$ or $- 2s_{j+1} S_{j+1}$ for each $H_{p}$ in Eq.~\eqref{eq:2ndog} to generate them. We do the following:
 \begin{itemize}
     \item We drop any contributions that are generated from choosing the same term for two $H_{p}$'s in Eq.~\eqref{eq:2ndog}. Plugging in $-2 t_{j-1} T_{j-1}$ in both $H_{p}$'s leads to a small antiferromagnetic interaction we ignore; the same is true for plugging in $-2 t_{j+2} T_{j+2}$ in both $H_{p}$'s. Similarly, choosing $ -2 s_{j-2} S_{j-2}$ or $ -2 s_{j+1} S_{j+1}$ for both $H_{p}$'s lead to a non-nearest neighbor term we chose to ignore before as well.
     \item Choosing $-2 t_{j-1} T_{j-1}$ and $- 2t_{j+2} T_{j+2}$ for the two $H_{p}$'s gives rise to the term $-2t'_{j-1}T_{j-1}$ after decimation, as in the second line of Eq.~\eqref{eq:siti}. Note that we drop this term if $m_{j-2}=m_{j+2} =0$
     \item Choosing $ -2 s_{j-2} S_{j-2}$  and $- 2t_{j+2} T_{j+2}$ leads to the correction in the second line of Eq.~\eqref{eq:sisi}, while choosing $- 2t_{j+2} T_{j+2}$ and $- 2s_{j+1} S_{j+1}$ gives the correction in the third line of Eq.~\eqref{eq:sisi}.
     \item We ignore the contribution from choosing $ -2 s_{j-2} S_{j-2}$ and $- 2s_{j+1} S_{j+1}$ due to the similar reasoning presented in Appendix.~\ref{app:tib}.
 \end{itemize}

\section{The fixed point values of $p_{i}$'s}
\label{app:fpp}

 In the main text and in Appendices~\ref{app:df} and \ref{app:rule}, we assumed that $p=2^{-1/4}$. We here show that $ p_{i}= \pm 2^{-1/4}$ are the only stable fixed-point values of $p_{i}$ in the strong-disorder limit. To this end, we let individual $F_{i}$'s have different $p_{i}$ values, as opposed to one uniform value across the whole system. We then explore how $p_{i}$'s renormalize after each decimation.
 
  Decimating the ferromagnetic bond at $j$ transforms $p_{i}$'s before the decimation to $p'_{i}$'s as follows:
\begin{equation}
    p'_{i} = \begin{cases}
    p_{i} & i < j \\
    \frac{2^{-1/2}  + p_{j} p_{j+1}}{p_{j+1} + p_{j}} & i=j \\
    p_{i+1} & i>j
    \end{cases}.
\end{equation}
 Assuming that $p_{i}$ and $p'_{i}$ are all given by the same value $p$, the only two consistent values of $p$ are $\pm 2^{-1/4}$. We investigate the stability of this solution under the decimation procedure by expanding $p_{i} = 2^{-1/4} + \delta_{i}$ and $p'_{i} = 2^{-1/4} + \delta'_{i}$, assuming $\delta_{i} \ll 1$. 
 By Taylor expanding one can show that
 \begin{equation}
     \delta'_{j} = O(\delta_{j}^{2}) + O(\delta_{j}\delta_{j+1}) + O (\delta_{j+1}^{2}).
 \end{equation}
 That is, to lowest order, $\delta'_{j}$ is \textit{quadratic} in $\delta_{i}$'s. Hence, assuming that initially $\delta_{j}$ and $\delta_{j+1}$ are sufficiently small, $\delta'_{j}$ is much smaller than $\delta_{j}$ or $\delta_{j+1}$! Hence, the fixed point value $p_{i} = p =2^{-1/4}$ is stable to small perturbations.  
 The solution $p_{i} = p = -2^{-1/4}$ can be similarly shown to be stable under ferromagnetic bond decimations.
 
 Antiferromagnetic bond decimations mark the corresponding bond $j$ with $m_{j}=0$. In our schemes, $m_{j}$ labels direct how to form multi-site flip terms with individual $F_{i}$'s but do not modify information about $F_{i}$'s. Hence, in our formulation, antiferromagnetic bond decimations do not modify $p_{i}$'s. 
 
   We conduct a similar exercise for double-bond and triple-bond decimations. The decimations associated with $T_{j}$ and $S_{j}$ respectively transform $p_{i}$'s as
 \begin{equation}
     p'_{i} = \begin{cases}
     p_{i} & i < j-1\\
     \frac{p_{j-1} + p_{j+1} + p_{j}^{*}\left(2^{-1/2} + p_{j-1}p_{j+1} \right)}{ 1 + p_{j}^{*}(p_{j-1} + p_{j+1}) + \sqrt{2} p_{j-1}p_{j+1}} & i = j-1 \\
     p_{i+2} & i > j-1
     \end{cases}
 \end{equation}
  \begin{equation}
  \begin{split}
     & p'_{i} = \begin{cases}
     p_{i} & i < j-1\\
     \frac{p_{j-1} + p_{j+2} + p_{j,j+1}^{*}\left(2^{-1/2} + p_{j-1}p_{j+2} \right)}{ 1 + p_{j,j+1}^{*}(p_{j-1} + p_{j+2})  + \sqrt{2} p_{j-1}p_{j+2}} & i = j-1 \\
     p_{i+3} & i > j-1
     \end{cases}\\
     & p_{j,j+1} =  \frac{2^{-1/2}   + p_{j} p_{j+1}}{p_{j+1} + p_{j}} .
    \end{split}
 \end{equation}
 In both cases, one can expand $p_{i} = \pm 2^{-1/4} + \delta_{i}$, $p'_{j-1} = \pm 2^{-1/4} + \delta'_{j-1}$ to show that $\delta'_{j-1}$ is, to lowest order, quadratic in $\delta_{i}$.
 
 Finally, we comment on how a single-site and double-site decimation modifies $p_{i}$. There, as we saw in the previous Appendices and the main text, $p_{i} = \pm 2^{-1/4}$ is a fixed point of the transformation. One can perform the same exercise of expanding $p_{i} = \pm 2^{-1/4} + \delta_{i}$ and investigate how $p'_{i} = \pm 2^{-1/4} + \delta_{i}'$ evolves. Unfortunately, in this case, the $\delta'_{i}$'s that non-trivially renormalize are, to the lowest order, linear in $\delta_{i}$. However, we also observed that site decimations generate a ferromagnetic interaction of order $\Omega$, and in the strong-disorder limit site decimations are immediately followed by ferromagnetic bond decimations. If one combines the effect of ferromagnetic bond decimations, once more the non-trivially renormalized $\delta'_{i}$'s are quadratic in $\delta_{i}$'s. Hence, in any strong-disorder fixed point $p_{i} =\pm 2^{-1/4}$ persist as stable fixed-point values.
 
 As a closing remark, there may be some strong-disorder fixed points with $p_{i}$ themselves randomly distributed. We reserve exploring such strong-disorder fixed points for future work.

\section{More on the irrelevancy of the multi-site flip terms}
\label{app:more}

 Here, we lay out the rationale for dropping the contributions from multi-site flip terms in our strong-disorder RG scheme. We specifically argue that $(i)$ multi-site flip terms are renormalized to a very small magnitude compared to the RG energy scale at each step and $(ii)$ non-local Majorana fermion interactions generated from multi-site flip terms are much smaller than the nearest-neighbor Majorana fermion hopping terms we keep track of in the $-t_{i} T_{i}$ and $-s_{i} S_{i}$ terms from Eq.~\eqref{eq:fullRGham} and can be ignored. Point $(i)$ justifies ignoring any possible decimation procedure coming from the multi-site flip terms, while $(ii)$ justifies ignoring terms generated due to multi-site flip terms during the decimation procedure associated with local terms we do keep track of in our RG procedure.
 
The key observation behind the argument is that antiferromagnetic bonds are typically very far from each other in the limit we are considering, and up to some energy scale $\Lambda_{FG}$ ($FG$ standing for Ferromagnetic Griffiths), the RG flow is ``unaware'' of the presence of antiferromagnetic bonds. Hence, when the RG energy scale satisfies $\Omega > \Lambda_{FG}$, the ferromagnetic Griffiths behavior characterized by the distribution of couplings given in Eq.~\eqref{eq:griffithscaling} governs the scaling behavior. In Eq.~\eqref{eq:griffithscaling}, when $\Omega$ is small enough, we have $\alpha \gg \beta$, and ferromagnetic Ising interactions are on average much larger than flip terms. Hence, when antiferromagnetic bonds appear rarely enough that $\Lambda_{FG}$ is sufficiently small, one can generally expect flip terms to be renormalized to very small values at $\Omega \sim \Lambda_{FG}$, especially compared to ferromagnetic interactions present or generated mid-RG. Meanwhile, up to $\Omega > \Lambda_{FG}$, there is only a small number of clusters connected by $m_{i}=0$ bonds whose sizes are larger than two; most sites are not flanked by $m_{i}=0$ bonds at all, or are flanked by one bond due to the presence of rare antiferromagnetic bonds.

 Recall that in the RG flow we presented in the main text, there is also an energy scale $\Lambda_{MF}$  below which spins are mostly pinned by strongly antiferromagnetic bonds with $m_{i}=0$ and the physics of the system is governed by domain-wall Majorana fermions. When $\Lambda_{MF} <\Omega < \Lambda_{FG}$, multi-site clusters form, and multi-site flip terms potentially disrupt the RG procedure we presented in the main text. The high-level idea is that since flip terms are renormalized to a very small magnitude at $\Omega \sim \Lambda_{FG}$, their effect will be suppressed during $\Lambda_{MF} < \Omega < \Lambda_{FG}$ as well; when $\Omega< \Lambda_{MF} $, all spins in the system effectively form a single cluster, and the notion of individual flip terms, whether multi-site or single-site, effectively disappears.

Let us spell out more detailed arguments in favor of points $(i)$ and $(ii)$ from the beginning of the Appendix. Even when $\Lambda_{MF} < \Omega < \Lambda_{FG}$, the majority of decimations at RG steps are ferromagnetic bond decimations. Another key observation regarding the nearest-neighbor Ising interaction is that, according to the rule we derived in Appendix~\ref{app:rule}, when the nearest-neighbor Ising interaction $J_{i}$ obeys a non-trivial RG transformation rule, the rule always involves the ``$\text{amax}$'' symbol introduced at the beginning of Appendix~\ref{app:rule}. The rule essentially chooses the dynamically generated Ising interaction only when it is larger than the Ising interaction $J_{i}$ already present.  In other words, the magnitude of the nearest-neighbor Ising interaction \textit{does not} decrease under the RG transformation rule. Hence, he relevant RG energy scales in the regime $\Lambda_{MF} < \Omega < \Lambda_{FG}$ are effectively lower-bounded by the ferromagnetic Ising interactions present/generated up to the scale $\Omega \sim \Lambda_{FG}$, and these ferromagnetic Ising interactions are much larger than $h_{i}$'s that parameterize the magnitude of spin flip terms. This conclusion implies that the possibility that any flip terms---including multi-site ones that start to appear frequently in the regime $\Lambda_{MF} < \Omega < \Lambda_{FG}$---enter as the RG energy scale is greatly suppressed in the regime of our interest, justifying $(i)$.

 To justify $(ii)$, we observe that in our RG, hopping or interaction between domain wall Majorana fermions are essentially generated by the physical processes of flipping spins back and forth. Using this observation, one may estimate and compare contributions to Majorana fermion hoppings from single-site flip terms and multi-site flip terms. To do so, assume that at the middle of an RG step with RG energy scale $\Omega \sim \Lambda_{FG}$, there are two bonds at $j$ and $j+n$ which become two neighboring bonds, say $j'$ and $j'+1$ with labels $m_{j'} =m_{j'+1}=0$ at a later RG step. Denote the amplitude for Majorana fermion hopping between bond $j$ and bond $j+n$ at the later RG step of the RG as $t_{j, j+n}$. 
 When only the single-site flip terms are utilized, the simplest way to generate $t_{j, j+n}$ is to flip spins between bond $j$ and bond $j+n$ back and forth. Then, one may estimate $t_{j,j+n}$ to be
\begin{equation}
\label{eq:tjjnappD}
    t_{j,j+n} \sim \frac{\displaystyle\prod_{i=j+1}^{j+n} h_{i}^{2} }{\displaystyle\prod_{k=1}^{2n-1} \Omega_{k}}.
\end{equation}
The fact that one flips spins back and forth is reflected in the $h_{i}^{2}$ in the numerator, while $\Omega_{k}$ in the denominator is the energy scale lying between $\Lambda_{P}$ and $\Lambda_{FG}$ and characterizes the virtual spin-flip energy cost. Note that numerator is a product of $2n$ energy scales, while the denominator is a product of $2n-1$ energy scales, so the expression on the right side has the correct dimensions. Finally, invoking the same logic used to justify $(i)$, generally $\Omega_{k} \gg h_{i}$ for any $k$ and $i$. Hence, when $n$, the number of spins involved to generate $t_{j,j+n}$, is larger, the generated $t_{j,j+n}$ will be smaller.

Multi-site flip term contributions for the same Majorana fermion hoppings appear at a later step of the RG, when the bond $j$ and $j+n$ becomes two neighboring bonds $j'$ and $j'+1$, by flipping the cluster including the bonds $j'$ and $j'+1$ back and forth. From the perspective of the renormalized system at the RG energy scale $\Omega \sim \Lambda_{FG}$, spins comprising the cluster at the later step are not only those between bond $j$ and $j+n$ -- a large number of spins to the left of the bond $j$ and to the right of the bond $j+n$ are also involved. One may write down a formula analogous to Eq.~\eqref{eq:tjjnappD} for the multi-site flip term contributions; however, it will involve a number of $h_{i}$'s much larger than $2n$ in the numerator and a number of $\Omega_{k}$'s much larger than $2n-1$ in the denominators. Hence, the generated Majorana fermion hoppings are expected to be much smaller than the single-site contributions that involve fewer spins. Furthermore, all multi-site fermion terms are presumably irrelevant (as claimed in the original paper by Fisher \cite{Fishersdrg_2}, supported by recent numerics in \cite{Chepiga2023_1}; see also \cite{Monthus2018}) near the infinite randomness fixed point of our consideration. Hence, as long as Majorana fermion hopping terms generated from multi-site flip terms are kept to be sufficiently small as we suggested here, they cannot alter the IR physics. This argument justifies cutting off terms generated from multi-site flip terms in our decimation procedure.

\bibliography{ref}

\end{document}